\documentclass[11pt]{article}
\usepackage{epsfig}
\usepackage{amsfonts}
\usepackage{mathrsfs}
\usepackage{latexsym}
\usepackage{color}
\usepackage[usenames,dvipsnames]{xcolor}
\usepackage{amsmath,amssymb}
\usepackage{verbatim}
\definecolor{red}{rgb}{1,0,0}
\usepackage{cite}
\usepackage{graphicx}
\usepackage{subcaption}
\usepackage{xcolor,colortbl}
\usepackage{authblk}
\usepackage{float}
\numberwithin{equation}{section}
\def\bea{\begin{eqnarray}} 
\def\eea{\end{eqnarray}}
\def\be{\begin{equation}} 
\def\ee{\end{equation}} 
\def\ba{\begin{array}}
\def\ea{\end{array}} 
\def\nn{\nonumber}

\title{{\Large{\textbf{Uncovering novel phase structures in $\Box^k$ scalar theories \\ with the renormalization group}}}} 

\author[1,2]{M. Safari\thanks{safari@bo.infn.it}}

\author[2]{G.P. Vacca\thanks{vacca@bo.infn.it}}
\affil[1]{\small\it Dipartimento di Fisica e Astronomia, 
Universit\`a di Bologna, via Irnerio 46, 40126 Bologna, Italy}
\affil[2]{\it INFN - Sezione di Bologna, via Irnerio 46, 40126 Bologna, Italy}

\begin{document}

\thispagestyle{empty}

\renewcommand{\thefootnote}{\fnsymbol{footnote}}
\setcounter{footnote}{0}
\setcounter{figure}{0}

\maketitle

\begin{center}


\vspace{5mm}

\begin{abstract}
We present a detailed version of our recent work on the renormalization group approach to multicritical scalar theories with higher derivative kinetic term of the form $\phi(-\Box)^k\phi$ and upper critical dimension $d_c = 2nk/(n-1)$. Depending on whether the numbers $k$ and $n$ have a common divisor two classes of theories have been distinguished which show qualitatively different features. For coprime $k$ and $n-1$ the theory admits a Wilson-Fisher type fixed point with a marginal interaction $\phi^{2n}$. We derive in this case the renormalization group equations of the potential at the functional level and compute the scaling dimensions and some OPE coefficients, mostly at leading order in $\epsilon$. While giving new results, the critical data we provide are compared, when possible, and accord with a recent alternative approach using the analytic structure of conformal blocks. Instead when $k$ and $n-1$ have a common divisor we unveil a novel interacting structure at criticality. In this case the phase diagram is more involved as other operators come into play at the scale invariant point. $\Box^2$ theories with odd $n$, which fall in this class, are analyzed in detail. Using the RG flows that are derived at quadratic level in the couplings it is shown that a derivative interaction is unavoidable at the critical point. In particular there is an infrared fixed point with a pure derivative interaction at which we compute the scaling dimensions. For the particular example of $\Box^2$ theory in $d_c=6$ we include some cubic corrections to the flow of the potential which enable us to compute some OPE coefficients as well.
\end{abstract}


\end{center}




\setcounter{page}{1}
\renewcommand{\thefootnote}{\arabic{footnote}}
\setcounter{footnote}{0}

\newpage
\tableofcontents
\newpage

\section{Introduction}

Quantum field theory (QFT) has proven remarkably effective at describing physical systems close to their critical point where the correlation length tends to infinity. 
One of the essential tools in this regard is the renormalization group (RG) which governs the effective description of physical systems in terms of the length scales at which they are probed. In particular universality of physical properties in the vicinity of a scale invariant point which had long remained a conundrum owes its resolution to this notion \cite{Wilson:1971bg}. RG has not only brought conceptual insight but through epsilon expansion has provided a perturbative framework to compute critical properties \cite{Wilson:1971dc}. 

Applicability of RG methods relies solely on scale symmetry which is an inherent 
feature of physical systems at criticality. However, scale invariance often comes along with conformal invariance. In particular unitary scale invariant theories have been shown to be conformal in two dimensions and perturbatively also in four dimensions. The presence of conformal symmetry allows to take advantage of the whole apparatus of conformal field theory (CFT). Indeed developing CFT methods and applying them in the study of critical systems has been an active 
line of research in the recent years \cite{ElShowk:2012ht,Rychkov:2015naa,Basu:2015gpa,Nii:2016lpa,Gliozzi:2016ysv,Gliozzi:2017hni,Alday:2016njk,Codello:2017qek,Gliozzi:2017gzh}.

Well-known examples that have been studied extensively with quantum field theory methods include the Ising model as a unitary theory described by a $\phi^4$ potential and the Lee-Yang edge singularity which is nonunitary and is characterized by a cubic interaction $\phi^3$ \cite{Fisher:1978pf}. Built upon these results and their extensions \cite{Gracey:2017okb} such models have been generalized to all multicritical theories with both RG \cite{ODwyer:2007brp,Codello:2017hhh,Codello:2017epp} and CFT techniques \cite{Gliozzi:2016ysv,Gliozzi:2017hni,Codello:2017qek}. This includes theories with even \cite{ODwyer:2007brp,Codello:2017hhh,Gliozzi:2016ysv,Gliozzi:2017hni} and odd potentials \cite{Codello:2017qek,Codello:2017epp}. 

Another class of scalar theories that further generalizes the above models and has recently received more attention in the literature includes those with a higher derivative kinetic term of the form $\phi(-\Box)^k\phi$ \cite{Osborn:2016bev,Brust:2016gjy,Brust:2016zns,Brust:2016xif,Gracey:2017erc,Safari:2017irw} or $\Box^k$ theories for short, where $k$ is a positive integer. Despite their nonunitary nature, theories with higher derivatives have found interesting physical applications. The theory of elasticity \cite{Nakayama:2016dby} provides such an example. Furthermore particular quartic derivative models have been shown to describe the isotropic phase of Lifshitz critical theories, and may be relevant for the physics of certain polymers \cite{polimers}. The latter has also been studied with $\epsilon$-expansion techniques \cite{Hornreich:1975zz,Diehl:2002wn} and more recently with non perturbative functional RG methods \cite{Bonanno:2014yia,Zappala:2017vjf} as well.



Apart from possible physical applications, $\Box^k$ theories are interesting also at the theoretical level as they provide a framework to build new universality classes and serve as a testing ground for methods and ideas in higher dimensional CFTs. Moreover, they demonstrate new features not present in standard QFTs and in this sense they are instructive from the point of view of perturbation theory.
 
At the free theory level $\Box^k$ theories have been previously studied in \cite{Osborn:2016bev} where the coefficient of the energy-momentum tensor two-point function for $k=2,3$ is extracted, and in \cite{Brust:2016gjy,Brust:2016zns,Brust:2016xif} where general $\Box^k$ theories have been investigated with particular global symmetries, mostly motivated by possible links to quantum gravity in de Sitter space. At the interacting level multicritical $\Box^k$ theories, with and without global symmetries, that demonstrate a generalized Wilson-Fisher fixed point have been analyzed in \cite{Gliozzi:2016ysv,Gliozzi:2017hni} using the analytic structure of conformal blocks. In \cite{Gracey:2017erc} $O(N)$ symmetric $\Box^k$ theories with quartic interaction have been explored in the large $N$ limit.
 
In a recent letter \cite{Safari:2017irw} we have reported the results of our study of general multicritical $\Box^k$ single-scalar theories with $\mathbb{Z}_2$ symmetric critical interactions using standard perturbative RG methods and extracted critical information by performing $\epsilon$ expansion below an upper critical dimension $d_c = 2nk/(n-1)$. This upper critical dimension is fixed by the requirement that $\phi^{2n}$ be a marginal operator. We have divided such theories into two classes which we have referred to as {\it first} and {\it second} type theories. The first class consists of theories in which the numbers $k$ and $n-1$ are relatively prime. These theories admit 
a generalized Wilson-Fisher fixed point with a single marginal interaction $\phi^{2n}$ as in \cite{Gliozzi:2016ysv,Gliozzi:2017hni}. In the second class $k$ and $n-1$ have a common divisor. The phase structure of these theories is more involved compared to that of the first type theories and is characterized by the presence of fixed points with derivative interactions. Contrary to theories of the first type there are no fixed points with pure potential interactions. 

In this work we present the details of our approach which relies on functional perturbative RG \cite{ODwyer:2007brp,Codello:2017hhh,Codello:2017epp,Osborn:2017ucf} but give some new results as well. In particular we extend the beta function of the potential for $\Box^2$ theories of the second type, that was reported in \cite{Safari:2017irw} at quadratic level, to cubic order in the couplings. This is necessary for the calculation of OPE coefficients beyond free theory for non derivative operators. For this purpose we need not only the counter-terms cubic in the potential but also those with one power of $Z$ couplings that parameterize two-derivative operators. The Euclidean higher-derivative scalar theories that we consider here are free of an explicit scale and therefore do not suffer from inconsistencies discussed in \cite{Aglietti:2016pwz}.

The paper is organized as follows. In the first part of Sect.\ref{s:general} we present the general setup and basic definitions. The ordering of couplings/operators according to their canonical dimension in $\Box^k$ theories is different from that of standard theories. We devote a separate subsection to clarifying this issue and determining couplings of the same dimension that can possibly mix together. In the last part of this section we resort to dimensional analysis to constrain as much as possible the structure of beta functionals before getting into perturbative loop calculations. In Sect.\ref{s:type1} we analyze theories of the first type and report the cubic beta functional of the potential as well as the beta functional of $Z$ at quadratic order. We obtain critical data such as the field and coupling anomalous dimensions and OPE coefficients in terms of $\epsilon$ in the neighbourhood of the generalized Wilson-Fisher fixed point and compare with the literature when possible. Sect.\ref{s:type2} discusses $\Box^2$ theories of the second type which correspond to odd values of $n$. The beta functional of the potential as well as the two and four-derivative couplings are computed at quadratic level in terms of $V$ and $Z$. We describe the phase diagram of these theories and present the field anomalous dimension and some critical exponents in the vicinity of the infrared fixed point. Finally, for the particular case of $n=3$ we extend the beta function of the potential in Sect.\ref{s:type2.exmpl} by including cubic corrections that are required for the computation of OPE coefficients at order $\epsilon$. We then conclude in Sect.\ref{s:conclusions} and devote several appendices to details of the computations.





\section{General $\Box^k$ scalar theories} \label{s:general}

\subsection{Setup and definitions}

Let us begin with presenting the general setup and basic definitions. Our goal is to study critical  theories with a single scalar field that are deformations of a higher-derivative free theory of the form 
\be \label{hft}
\mathcal{L}_{HFT} = {\textstyle{\frac{1}{2}}}\phi\, (-\square)^k \phi, 
\ee
where $k$ is a positive integer. We will calculate critical properties of such deformations using the renormalization group in dimensional regularization and $\overline{\mathrm{MS}}$ scheme and perform $\epsilon$-expansion in $d=d_c-\epsilon$, that is below a critical dimension $d_c$ which will be determined shortly. The canonical dimension of the field is 
\be
\delta = \frac{d}{2} - k.
\ee
The presence of interactions will induce a running for the coefficient of the kinetic term \eqref{hft}. We adopt the convention that this coefficient is implicitly absorbed into the field $\phi$ so that the kinetic term is always in the canonical form. This implies that the field $\phi$ depends on the RG scale $\mu$ and we define its anomalous dimension $\gamma_\phi$ as
\be 
\mu\frac{d}{d\mu} \phi = -\gamma_\phi\, \phi.
\ee
The propagator of the massless theory \eqref{hft} satisfies the following differential equation where $\delta^d_x$ is the $d$-dimensional Dirac delta function at the spacetime point $x$
\be 
(-\square_x)^k G_x = \delta^d_x.
\ee
Apart from a numerical coefficient, the $x$ dependence of the propagator is fixed by translation and scale invariance. The solution to this equation in coordinate space is
\be \label{G}
G_x = \frac{c}{|x|^{2\delta}}, \qquad c \equiv \frac{1}{(4\pi)^k\Gamma(k)}\,\frac{\Gamma(\delta)}{\pi^{\delta}}.
\ee
Although we will be using such coordinate space expressions as well, our perturbative analysis of the following sections rely heavily on momentum space calculations. Also, we will be generically dealing with Feynman diagrams in which a propagator is replaced with a bunch of propagators. It is therefore useful to have the momentum space expression for a bunch of $r$ propagators
\be \label{Grp}
\int_x \!e^{ip\cdot x}\, G^r_x = \frac{1}{(4\pi)^{rk}\Gamma^r(k)}\,
\frac{\Gamma^r(\delta)}{\Gamma(r\delta)}\,
\Gamma(k-(r-1)\delta)
\left(\frac{p^2}{4\pi}\right)^{(r-1)\delta-k},
\ee
which clearly reduces to $p^{-2k}$ for $r=1$. Throughout this paper we have adopted a shorthand notation for the integrals which can be found in Appendix.\ref{s:not}. This quantity has a pole when
\be 
(r-1)\delta_c -k = 0, 1, 2, \cdots
\ee
where $\delta_c$ is the field dimension at the critical point $\epsilon=0$. Let us denote by $n$ the value of $r$ at which the first pole occurs. This satisfies $(n-1)\delta_c = k$. Then $\delta_c$ and the space dimension at criticality $d_c$ will depend on $k,n$ in the following way  
\be 
\delta_c = \frac{k}{n-1}, \qquad d_c = \frac{2nk}{n-1}.
\ee
The assumption that such an $n$ exists is equivalent to the fact that leading quantum corrections induced by non-derivative couplings are quadratic. This follows from the fact that, apart from numerical factors and couplings, \eqref{Grp} shows the expression for a diagram with two non-derivative vertices. The upper critical dimension $d_c$ can also be fixed by the requirement that $\phi^{2n}$ be a marginal operator. However, as we will see, this is not necessarily the only marginal interaction.

\subsection{General pattern of coupling mixing} \label{ss:de}

The pattern of coupling mixing in $\Box^k$ theories is more involved than in the standard case with a 2-derivative kinetic term. Before getting into the perturbative calculations it is therefore very useful to sort the operators according to their canonical dimension and determine those of the same dimension which can potentially mix together.

A derivative operator in the action has at least two fields, so let us schematically denote a general $2\ell$-derivative operator as $\phi^\sigma (\partial^\ell\!\phi)^2$, where operators with the same number of fields and derivatives are not distinguished, and for simplicity of argument treat the potential, which corresponds to $\ell=0$, on the same footing. Such an operator has the canonical dimension 
\be 
[\phi^\sigma\; (\partial^\ell\!\phi)^2] = \frac{\sigma k}{n-1} + \frac{2k}{n-1} + 2\ell.
\ee
We would like to find the lowest dimensional operator of the above form of which the coupling mixes with that of an operator with higher derivatives
. In other words, we are looking for the smallest $\sigma$ for which the operator $\phi^\sigma (\partial^\ell\!\phi)^2$ has the same dimension as a higher derivative operator, say, an operator with $2(\ell+\Delta\ell)$ derivatives. Such a higher derivative operator will of course appear with only two fields and with the smallest possible $\Delta\ell$. This is found by solving the equation
%
%
\be 
[\phi^\sigma\; (\partial^\ell\!\phi)^2] = [(\partial^{\ell+\Delta\ell}\!\phi)^2] \qquad \Leftrightarrow  \qquad
\frac{\sigma k}{n-1} =  2\Delta\ell.
\ee
Let us call $\ell_*$ the smallest value of $\Delta\ell$ for which a solution for $\sigma$ exists in the above equation, and denote the corresponding solution by $\sigma_*$. For instance, for $k=1$ we have $\ell_*=1$ and $\sigma_*=2(n-1)$, while for $k=2$ we have $\ell_*=1$ and $\sigma_*=n-1$. Both $\ell_*$ and $\sigma_*$ are independent of $\ell$. This is telling us that the lowest dimensional $\ell$-derivative operator that mixes with higher derivative operators is $[\phi^{\sigma_*}\, (\partial^\ell\!\phi)^2]$ which mixes with $[(\partial^{\ell+\ell_*}\!\phi)^2]$. 

It is clear that other values of $\Delta\ell$ that admit solutions for $\sigma$ are multiples of $\ell_*$, and consequently their solutions are also (the same) multiples of $\sigma_*$. In particular $k$ must be a multiple of $\ell_*$, because $\Delta\ell = k$ admits a solution, and that is $\sigma_*=2(n-1)$. So the set of operators is partitioned into equivalence classes, where two operators belong to the same equivalence class iff their number of derivatives differ by multiples of $\ell_*$. Couplings of operators belonging to different equivalence classes do not mix together.      
Therefore, schematically, and without distinguishing different operators of the same dimension and the same number of derivatives, the mixing of operator couplings is described by the following table 
\be \label{mixgen}
\ba{|llll|lll|lll|ll}
\cline{1-10}
1& \phi & \cdots & \phi^{\sigma_*-1} & \phi^{\sigma_*} & \cdots & \phi^{2\sigma_*-1} & \phi^{2\sigma_*} & \cdots & \phi^{3\sigma_*-1} & \cdots  &\quad \times(\partial^\ell\phi)^2 \\ \cline{1-10}
\multicolumn{4}{c|}{} & 1 & \cdots & \phi^{\sigma_*-1} & \phi^{\sigma_*} & \cdots & \phi^{2\sigma_*-1} & \cdots &\quad \times(\partial^{\ell+\ell_*}\phi)^2 \\ \cline{5-10}
\multicolumn{7}{c|}{} & 1 & \cdots & \phi^{\sigma_*-1} & \cdots &\quad \times(\partial^{\ell+2\ell_*}\phi)^2 \\ \cline{8-10}
\ea
\ee
where each row is understood to be multiplied by the operator on the right and collects operators with the same number of derivatives. Operators belonging to the same column have the same dimension and their couplings can therefore potentially mix together. There are $\ell_*$ of such tables corresponding to $\ell =0,1,\cdots, \ell_*-1$. For different $\ell$ belonging to this set the couplings of operators in the tables do not mix together, and altogether the $\ell_*$ tables describe the mixing of all possible couplings. For each $\ell_*$, table \eqref{mixgen} sorts operators according to their canonical dimensions. The table for $\ell_* =0$, which includes non-derivative operators in its first row, is shown as  
\be \label{mixingv}
\ba{|llll|lll|lll|ll}
\cline{1-10}
1& \phi & \cdots & \phi^{\sigma_*+1} & \phi^{\sigma_*+2} & \cdots & \phi^{2\sigma_*+1} & \phi^{2\sigma_*+2} & \cdots & \phi^{3\sigma_*+1} & \cdots  &\quad  \\ \cline{1-10}
\multicolumn{4}{c|}{} & 1 & \cdots & \phi^{\sigma_*-1} & \phi^{\sigma_*} & \cdots & \phi^{2\sigma_*-1} & \cdots &\quad \times(\partial^{\ell_*}\phi)^2 \\ \cline{5-10}
\multicolumn{7}{c|}{} & 1 & \cdots & \phi^{\sigma_*-1} & \cdots &\quad \times(\partial^{2\ell_*}\phi)^2 \\ \cline{8-10}
\ea
\ee
Unlike the standard case of $k=1$ where couplings with positive dimension (relevant) correspond to operators with no derivatives and do not mix with other couplings, for $k>1$ this is not always the case. 

Let us consider the column of marginal operators in \eqref{mixingv}. Such operators can potentially be present at a scale invariant point. If any of the corresponding couplings happens to be non zero at a fixed point then one is forced to take into account that operator along with all operators that appear on its left in table \eqref{mixingv}, i.e. those with the same number of derivatives but less number of fields. This makes it necessary to identify all marginal operators in order to have an idea which operators must be present at lowest order in a derivative expansion. This can be done by dimensional analysis requiring $[\phi^{\sigma}(\partial^{\ell}\phi)^{2}] = d_c$ which gives the condition
\be \label{marginal}
2k-\frac{\sigma}{n-1}k = 2\ell.
\ee
Since $\ell$ is non-negative, we are interested in non-negative integer values the left hand side can take. The values $\sigma=0,2n-2$ are always a solution and correspond to the two operators $\phi^{2n}$ and $\partial^{2k}\phi^2$, but intermediate values of $\sigma$ may also solve the equation. Whether such values exist depends on whether or not $k$ and $n-1$ have a non-trivial common divisor. 

Consider first the case where $k$ and $n-1$ are relatively prime numbers. In this case if $k$ is also an odd number it is clear that no intermediate values of $\sigma$ solve \eqref{marginal}. The $\ell_*=0$ pattern of mixing, which includes non-derivative couplings as well, in this case is given by the following table
%
\be  \label{t1}
\ba{|llll|lll|lll|l}
\cline{1-10}
1& \phi & \cdots & \phi^{2n-1} & \phi^{2n} & \cdots & \phi^{4n-3} & \phi^{2(2n-1)} & \cdots & \phi^{6n-5} & \cdots    \\ \cline{1-10}
\multicolumn{4}{c|}{} & 1 & \cdots & \phi^{2n-3} & \phi^{2(n-1)} & \cdots & \phi^{4n-5} & \cdots \quad\times \!(\partial^k\phi)^2  \\ \cline{5-10}
\ea
\ee
If instead $k$ is even  then $\sigma=n-1$ will also solve the equation and this corresponds to operators of the form $\partial^k\phi^{n+1}$. In principle, such operators can therefore be present in a scale invariant deformation of \eqref{hft}. However if $k$ is even and $k$ and $n-1$ are relatively prime then $n+1$ is an odd number and the operators $\partial^k\phi^{n+1}$ are $\mathbb{Z}_2$ odd. Therefore it is consistent to set their couplings to zero and consider only deformations of the form $\phi^{2n}$. The $\ell_*=0$ table describing the mixing pattern in this case is
%
\be  \label{t2}
\ba{|llll|lll|lll|l}
\cline{1-10}
1& \phi & \cdots & \phi^n & \phi^{n+1} & \cdots & \phi^{2n-1} & \phi^{2n} & \cdots & \phi^{3n-2} & \cdots   \\ \cline{1-10}
\multicolumn{4}{c|}{} & 1 & \cdots & \phi^{n-2} & \phi^{n-1} & \cdots & \phi^{2n-3} & \cdots \quad \times\!\partial^k\phi^2 \\ \cline{5-10}
\multicolumn{7}{c|}{} & 1 & \cdots & \phi^{n-2} & \cdots \quad \times\!(\partial^k\phi)^2  \\ \cline{8-10}
\ea 
\ee
Next, consider the case where $k$ and $n-1$ have a common divisor. In this case it is easily seen that intermediate values of $m$ that solve \eqref{marginal} always exist, and among them $\mathbb{Z}_2$ even operators are always present. Consequently one cannot a priori omit such operators. In this case a truncation of the action to non-derivative interactions is inconsistent. For the special case of $k=2$, $n$ must be an odd number $n=2m+1$. The $\ell_*=0$ mixing pattern in this case is given as
%
\be  \label{t3}
\ba{|llll|lll|lll|l}
\cline{1-10}
1& \phi & \cdots & \phi^{2m+1} & \phi^{2(m+1)} & \cdots & \phi^{4m+1} & \phi^{2(2m+1)} & \cdots & \phi^{6m+1} & \cdots   \\ \cline{1-10}
\multicolumn{4}{c|}{} & 1 & \cdots & \phi^{2m-1} & \phi^{2m} & \cdots & \phi^{4m-1} & \cdots \quad \times\!(\partial\phi)^2  \\ \cline{5-10}
\multicolumn{7}{c|}{} & 1 & \cdots & \phi^{2m-1} & \cdots \quad \times\!(\partial^2\phi)^2   \\ \cline{8-10}
\ea
\ee
Based on this reasoning we distinguish theories of the {\it first type} where $k$ and $n-1$ are relatively prime from those of the {\it second type} where $k$ and $n-1$ have a common divisor. The two type of theories have qualitatively different features. In the next section we will study each type in turn, but before getting into perturbative RG calculations let us see what we can infer just from dimensional analysis about possible terms that can appear in the beta functions.

\subsection{Structure of beta functions} 

In order to perform a RG analysis of scalar $\Box^k$ theories, instead of dealing with a finite truncation of the renormalized action, following \cite{ODwyer:2007brp,Codello:2017hhh,Codello:2017epp,Osborn:2017ucf} we find it more convenient to adopt a functional framework in which at a given order in a derivative expansion infinitely many couplings are collected into a finite number of functions that parameterize operators of the given derivative order (see e.g \cite{Codello:2017hhh} for further details).

One of the advantages of using the functional framework is that it allows us to extract many critical quantities with a single computation. Apart from that the functional approach gives more insight into the structure of the flow equations as it shows how the beta functions of many couplings are related to each other. In fact dimensional analysis alone fixes the structure of the beta functions to a high extent. To see this explicitly suppose that we are interested in the beta functional of  couplings corresponding to a $2a$-derivative operator, which we call $\beta_a$. In the functional approach such couplings are collected into a function that we refer to as $W_a(\phi)$. For $a\geq 2$ there is more than one function of this sort as the basis of $2a$-derivative operators has more than one element
. However for the purpose of the argument of this section we do not need to distinguish such functions.

The beta functional $\beta_a$ depends on the functions $W_b(\phi)$, $b=0,1,2,\cdots$ and their field derivatives. We would like to see how much dimensional analysis can tell us about this dependence before getting into perturbative loop computations. 
For this purpose, consider a diagram with $N$ vertices of which the $i$th vertex represents interactions parameterized by $W_{a_i}(\phi)$ with $i=1,2,\cdots, N$. In the most general case any two vertices are connected by a bunch of propagators which we refer to as an edge. Figs.(\ref{melon},\ref{zmelon},\ref{zzmelon}) and Fig.\eqref{triangle} provide examples of such diagrams with two and three vertices. The number of edges in the diagram is $I\equiv N(N-1)/2$. If we regard each edge as a single internal line the resulting graph will have $L\equiv I -N+1$ independent loops. Let us also denote the total number of propagators by $R$. Now, a diagram gives rise to an $\epsilon$ pole only if (omitting vertex couplings, external lines and their derivatives) 
it is dimensionless. Imposing this condition gives the following constraint 
\be \label{dlc} 
\left(R - I +L\right)\frac{2nk}{n-1} - 2k R + 2 \Big(\sum_ia_i-a\Big) =0.
\ee
The expression in the parenthesis in the first term is the total number of loops in the diagram, which is multiplied by the space dimension $d_c$. The second term takes into account the propagators each of which has a mass dimension $-2k$. Finally the last term considers possible derivatives from the vertices that can act on the propagators keeping in mind that this must be such that vertices with $2a_1,2a_2,\cdots,2a_N$ derivatives contribute to a $2a$ vertex. For instance one (but not the only) possibility is that out of all the derivatives in the vertices a number $2(\sum_ia_i-a)$ act on internal lines and $2a$ remain external. The above equation can be put in a more useful form
\be  \label{constraint}
\sum_ia_i-a = \frac{(N-1)n-R}{n-1}k.
\ee
Given $k,n$ and the total number of propagators in a diagram, this equation constrains the sum of derivatives of vertices that can contribute to the $2a$-derivative vertex. This will prove very useful in determining possible terms in a beta functional and what diagrams to compute. In this work we are interested in quadratic and cubic contributions in the couplings, that is $N=2,3$. 


\section{Theories of the first type} \label{s:type1}

As argued in Sect.\ref{ss:de}, for theories of the first type it is consistent to consider only non-derivative deformations of higher-derivative free theories \eqref{hft}. These are collectively expressed as a potential function $V(\phi)$, which give the lowest order truncation in a derivative expansion
\be 
\mathcal{L} = {\textstyle{\frac{1}{2}}}\phi\, (-\square)^k \phi + V(\phi).
\ee
In the following we will calculate the beta functional of the potential at cubic order in the couplings, i.e. cubic order in $V$. To compute the functional betas it is convenient to adopt the background-field method where the the field $\phi$ is shifted as $\phi\rightarrow\phi+f$ and $f$ is considered as the quantum field. We will therefore only deal with vacuum diagrams with vertices that are functions of $\phi$. As the first step we need to evaluate the quadratic and cubic counter-terms relevant for our purpose.


\subsection{Counter-term diagrams}

The counter-term diagrams for $V(\phi)$ at quadratic and cubic order in the $V$-couplings are discussed in Appendix.\ref{s:vct} for theories with general $k,n$. Here we pick those for theories of the first type and extract their divergences. At quadratic level the only diagram that has a pole contributing to the potential is an $(n-1)$-loop diagram given by
\begin{center}
\includegraphics[width=0.35\textwidth]{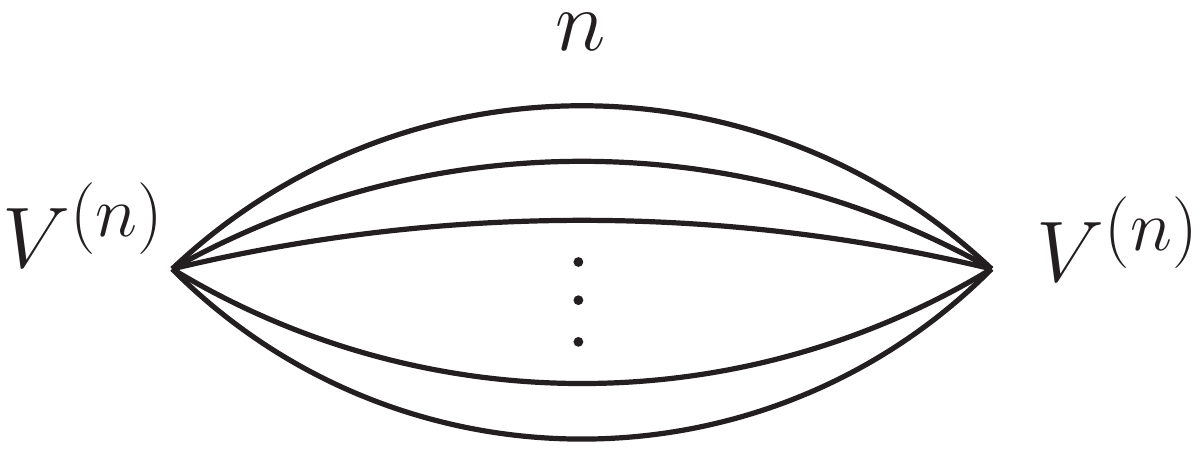}
\end{center} 
where $n$ in the superscript of $V^{(n)}$ refers to the number of field derivatives of the potential. The divergent part of this diagram is already given in the appendix and is nothing but Eq.\eqref{ul} for $r=n$, which corresponds to $l=0$. We call this $V_{c.t.2}$
\be  \label{vct2}
V_{c.t.2} \equiv \frac{1}{(n-1)\epsilon}\,\frac{1}{n!}\,\frac{1}{[(4\pi)^k\Gamma(k)]^{n}}\,
\frac{\Gamma^n(\delta_c)}{\Gamma(n\delta_c)}\,V^{(n)}\hspace{1pt}^2.
\ee 
Here, all the signs and symmetry factors have been taken into account. Let us now move to the cubic counter-terms. There are three types of cubic diagrams discussed separately in Appendices \ref{ss:tri}, \ref{ss:dmelon} and \ref{ss:cmelon}. Following the general discussion of Appendix.\ref{ss:tri} the first type of such diagrams, that is the triangle diagrams, are divided into two sets. The first are those in which the edges, which form a melon diagram, do not give rise to a subdivergence. For theories of the first type this means that $r,s,t\neq n$ in the diagram
\begin{center}
\includegraphics[width=0.4\textwidth]{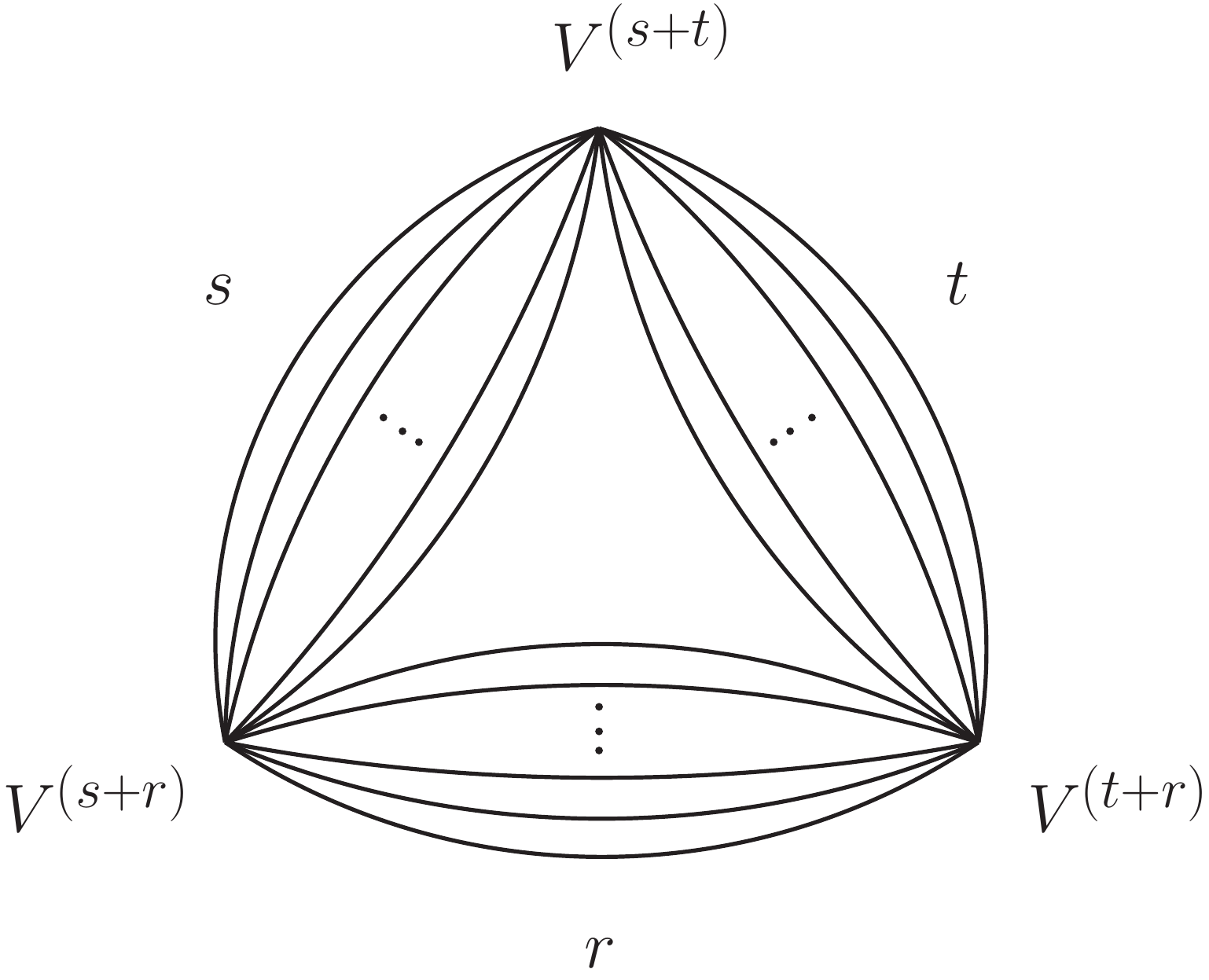} 
\end{center}
The divergence is extracted using equations \eqref{tricomp} and \eqref{gold} and picking the $y=z=0$ pole  
\be 
-\frac{1}{\epsilon}\frac{1}{[(4\pi)^k\Gamma(k)]^{2n}}\frac{\Gamma(\delta_c)^{2n}}{(n-1)\Gamma(n\delta_c)}\frac{1}{6}\hspace{-0.03\textwidth}\sum_{{\footnotesize \ba{c}r\!+\!s\!+\!t\!=\!2n\\[-5pt] r,s,t \neq n\ea}}\hspace{-0.02\textwidth}\frac{K^{k,n}_{rst}}{r!s!t!} \;V^{(r+s)}V^{(s+t)}V^{(t+r)},
\ee
where the number of propagators $r,s,t$ is positive and for compactness of the expression we have defined the following quantity
\be 
K^{k,n}_{rst} \equiv \frac{\Gamma\left((n-r)\delta_c\right)\Gamma\left((n-s)\delta_c\right)\Gamma\left((n-t)\delta_c\right)}{\Gamma\left(r\delta_c\right)\Gamma\left(s\delta_c\right)\Gamma\left(t\delta_c\right)}.
\ee
Next, consider triangle diagrams in which one edge (melon diagram) leads to a divergence. For theories of the first type this means that one of the three bunches must have $n$ propagators. We take $r=n$ so the diagram will be
\begin{center}
\includegraphics[width=0.4\textwidth]{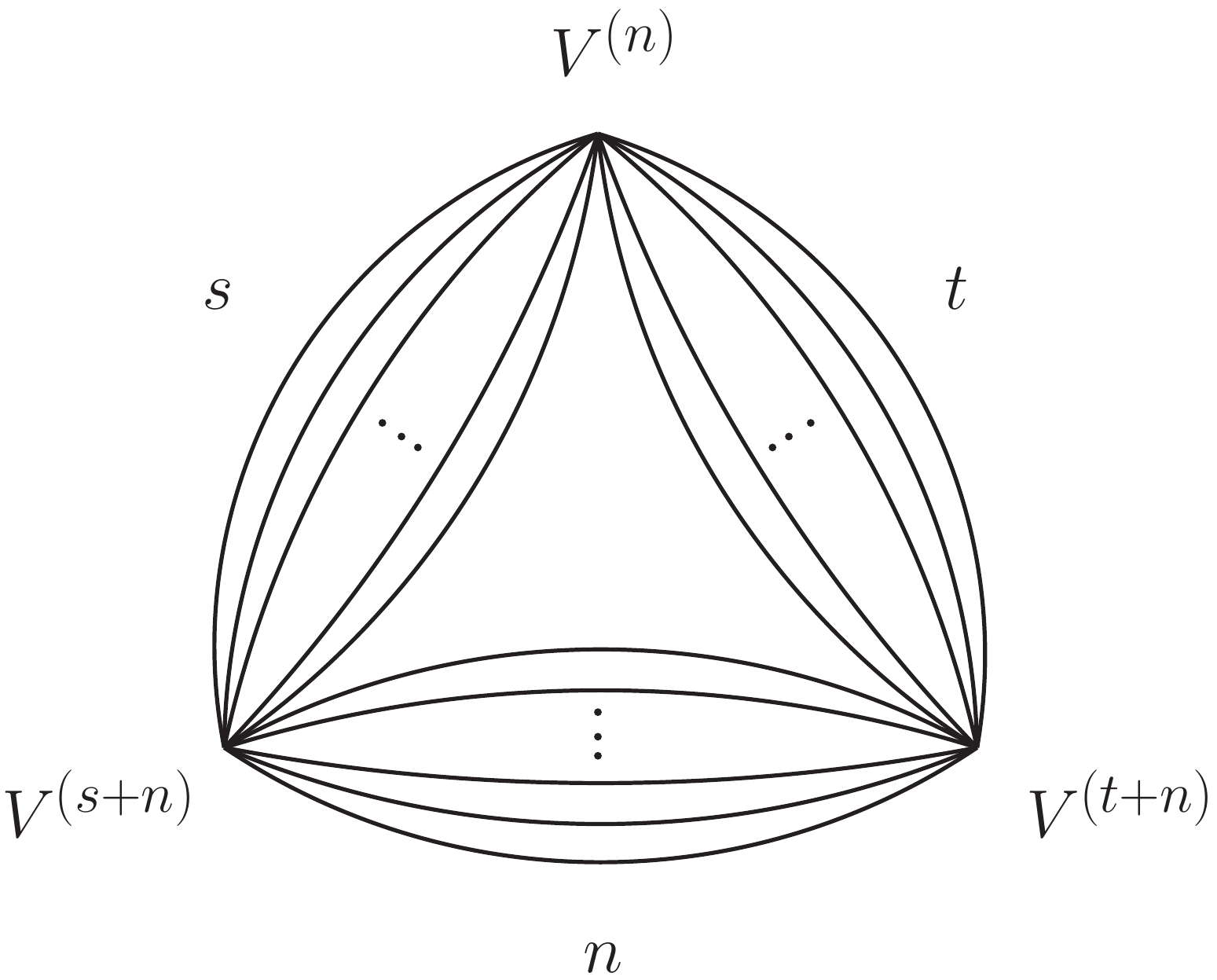}
\end{center}
One has to multiply \eqref{tricomp} by three as there are three possibilities for the subdivergent edge. We then get the following pole term
{\setlength\arraycolsep{2pt}
\bea
&-& \frac{1}{n!}\,\frac{1}{[(4\pi)^k\Gamma(k)]^{2n}}\frac{\Gamma(\delta_c)^{2n}}{\Gamma(n\delta_c)^2}\frac{1}{(n-1)^2} \\
&\times&\!\! \sum_{\footnotesize s+t=n}\left(\frac{1}{\epsilon^2} -\frac{(n-1)(\psi(s\delta_c)+\psi(t\delta_c)+3\gamma)+2n\psi(\delta_c)-(3n-1)\psi(n\delta_c)}{2\epsilon}\right) \frac{V^{(n)}V^{(n+s)}V^{(n+t)}}{s!t!}. \nn
\eea}%
The relevant double-melon diagrams consist of two melons with $n$ propagators each
\begin{center}
\includegraphics[width=0.55\textwidth]{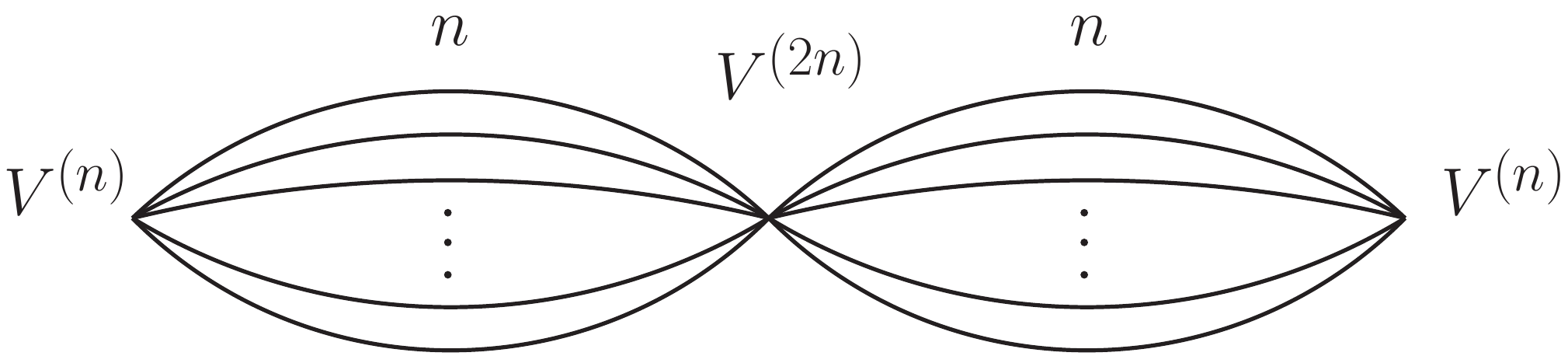}
\end{center}
The pole is easily extracted using \eqref{doubcomp} and \eqref{gg}
\be 
-\frac{2}{n!^2}\;\frac{1}{[(4\pi)^k\Gamma(k)]^{2n}}\frac{\Gamma(\delta_c)^{2n}}{\Gamma(n\delta_c)^2}\,\frac{1}{(n-1)^2}\left(\frac{1}{\epsilon^2} -\frac{(n-1)\gamma + n(\psi(\delta_c) - \psi(n\delta_c))}{\epsilon}\right) V^{(n)}\hspace{1pt}^2\,V^{(2n)}.
\ee
Finally, the melon diagram that contributes to the potential at cubic level has $n$ propagators. In this case one of the vertices is the counter-term vertex $U_{0,c.t.}$ discussed in \eqref{ss:qmelon} which is nothing but the quadratic counter-term for the potential $V_{c.t.2}$
\begin{center}
\includegraphics[width=0.35\textwidth]{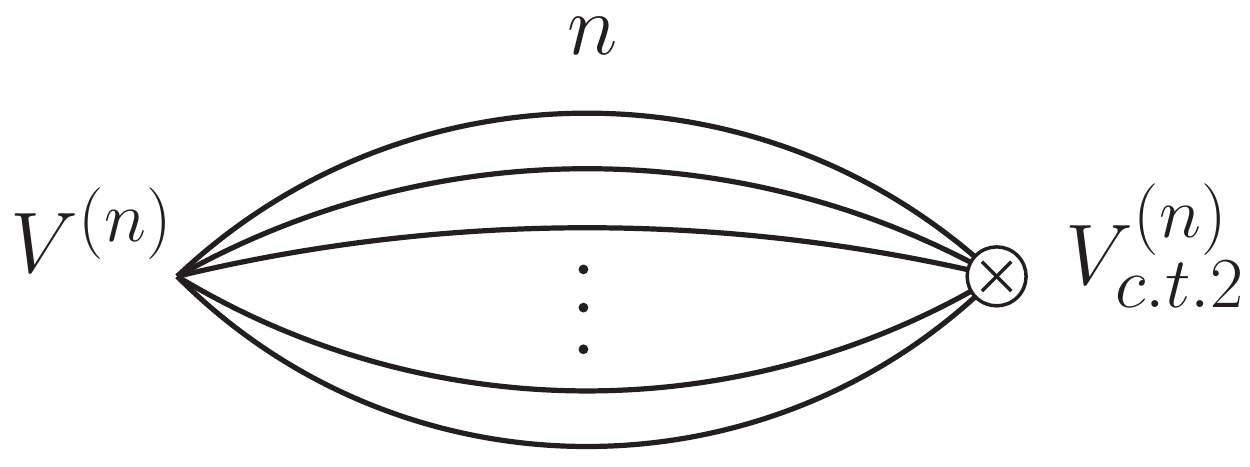}
\end{center}
The pole term for this diagram is simply extracted from \eqref{l0} 
\be
\frac{1}{n!^2}\frac{1}{[(4\pi)^k\Gamma(k)]^{2n}}\frac{\Gamma(\delta_c)^{2n}}{\Gamma(n\delta_c)^2}\,\frac{2}{(n-1)^2}\left(\frac{1}{\epsilon^2} -\frac{(n-1)\gamma + n(\psi(\delta_c) - \psi(n\delta_c))}{2\epsilon}\right)\big[V^{(n)}\hspace{1pt}^2\big]^{(n)}V^{(n)} .
\ee
Let us summarize theses results. At quadratic level the only contribution to the counter-term of the potential is \eqref{vct2}. At cubic order there are several contributions the sum of which is 
{\setlength\arraycolsep{2pt}
\bea \label{vct3}
V_{c.t.3} &=& -\frac{1}{6(n-1)\epsilon}\frac{1}{[(4\pi)^k\Gamma(k)]^{2n}}\frac{\Gamma(\delta_c)^{2n}}{\Gamma(n\delta_c)}\sum_{r,s,t}\frac{K^{k,n}_{rst}}{r!s!t!} \;V^{(r+s)}V^{(s+t)}V^{(t+r)} \nn\\
&& -\frac{1}{2(n-1)\epsilon}\,\frac{1}{n!}\,\frac{1}{[(4\pi)^k\Gamma(k)]^{2n}}\frac{\Gamma(\delta_c)^{2n}}{\Gamma(n\delta_c)^2} \sum_{s,t}\frac{J^{k,n}_{st}}{s!t!} \, V^{(n)}V^{(n+s)}V^{(n+t)} \nn\\
&& +\frac{1}{(n-1)^2\epsilon^2}\,\frac{1}{n!^2}\frac{1}{[(4\pi)^k\Gamma(k)]^{2n}}\frac{\Gamma(\delta_c)^{2n}}{\Gamma(n\delta_c)^2}\,\big[V^{(n)}\hspace{1pt}^2\big]^{(n)}V^{(n)},
\eea}%
where the following quantity is defined to make the expression compact
\be 
J^{k,n}_{st} \equiv \psi(n\delta_c) - \psi(s\delta_c)-\psi(t\delta_c) + \psi(1).
\ee

\subsection{Beta function of the potential}

From the counter-terms \eqref{vct2} and \eqref{vct3} it is straightforward to compute the beta function. Given the (canonical) normalization of the field $\phi$, there is a term proportional to the field anomalous dimension in flow of the potential. We will denote by $\beta_V$ the part that is independent of the anomalous dimension
\be 
\left.\mu \frac{\partial V(\phi)}{\partial\mu}\right|_\phi = \frac{\eta}{2}\, \phi\, V^{(1)}(\phi) + \beta_V.
\ee
At quadratic level the relation between the counter-term \eqref{vct2} and the beta function is 
\be 
\beta_{V,2} = \epsilon V_{c.t.2} - \left.\mu\frac{d}{d\mu}\right|_1\!\!V_{c.t.2}  = (n-1)\epsilon V_{c.t.2}, 
\ee
where the index $1$ in the $\mu$-derivative means that the scale derivative is taken using the tree-level flow $\beta_{V,1} = \epsilon V$. At cubic level the beta function and counter-terms are related as
\be 
\beta_{V,3} = \epsilon V_{c.t.3} - \left.\mu\frac{d}{d\mu}\right|_1\!\!V_{c.t.3} - \left.\mu\frac{d}{d\mu}\right|_2\!\!V_{c.t.2} = 2(n-1)\epsilon V_{c.t.3} - \left.\mu\frac{d}{d\mu}\right|_2\!\!V_{c.t.2}.
\ee
The index $2$ refers to the fact that in the $\mu$-derivative the quadratic flow $\beta_{V,2}$ has been used. This last term in $\beta_{V,3}$ cancels precisely the contribution form the third line of \eqref{vct3}. We will not go beyond cubic order in this work.

To address critical properties one expresses all quantities in units of the RG scale $\mu$. The dimensionless couplings are conveniently expressed as the expansion coefficients of the dimensionless potential
\be \label{vdimless}
v(\varphi) = \mu^{\!-d}\, V(\mu^\delta\,\varphi)\,.
\ee
The relation between the beta functional of $v(\varphi)$ and $\beta_V$ 
is
\be \label{vrel}
\beta_v =  -\,d v + \frac{d- 2k+\eta}{2} \,\varphi \, v^{(1)}+ \mu^{-d}\beta_V.
\ee
The normalization of couplings is not physical. We use this freedom to simplify the expression for the beta function. We make the rescaling
\be \label{vres}
v \rightarrow v\,[(4\pi)^k\Gamma(k)]^{2n}\,\Gamma(n\delta_c)^2/\Gamma(\delta_c)^{2n}.
\ee
This removes for instance all factors of $4\pi$, but the precise form is fixed by the requirement to match the CFT normalization. This will be made clear in subsequent sections. The final form of the dimensionless beta function after the rescaling \eqref{vres} is
{\setlength\arraycolsep{2pt}
\bea \label{bv1}
\beta_v &=&  -\,d v + \frac{d- 2k+\eta}{2} \,\varphi \, v^{(1)}+ \frac{1}{n!}\,v^{(n)}\hspace{1pt}^2 \nn\\
&& - \Gamma(n\delta_c)\frac{1}{3}\sum_{r,s,t}\frac{K^{k,n}_{rst}}{r!s!t!} \;v^{(r+s)}v^{(s+t)}v^{(t+r)} - \frac{1}{n!} \sum_{s,t}\frac{J^{k,n}_{st}}{s!t!}\, v^{(n)}v^{(n+s)}v^{(n+t)}.
\eea}%
where the first sum runs over positive integers $r,s,t$ subject to the conditions $r+s+t=2n$ and $r,s,t \neq n$, and the second sum runs over positive integers with a fixed sum $s+t=n$. 
The anomalous dimension $\eta$ is still unspecified at this point. This is fixed by our assumption that the field $\phi$ is always in the canonically normalized form. Let us now calculate this quantity.

\subsection{Fixed point and field anomalous dimension}

The function $V(\phi)$ induces a flow for the coefficient of the kinetic term as well. The counter-term appears at $2(n-1)$ loops and comes from the diagram
\begin{center}
\includegraphics[width=0.4\textwidth]{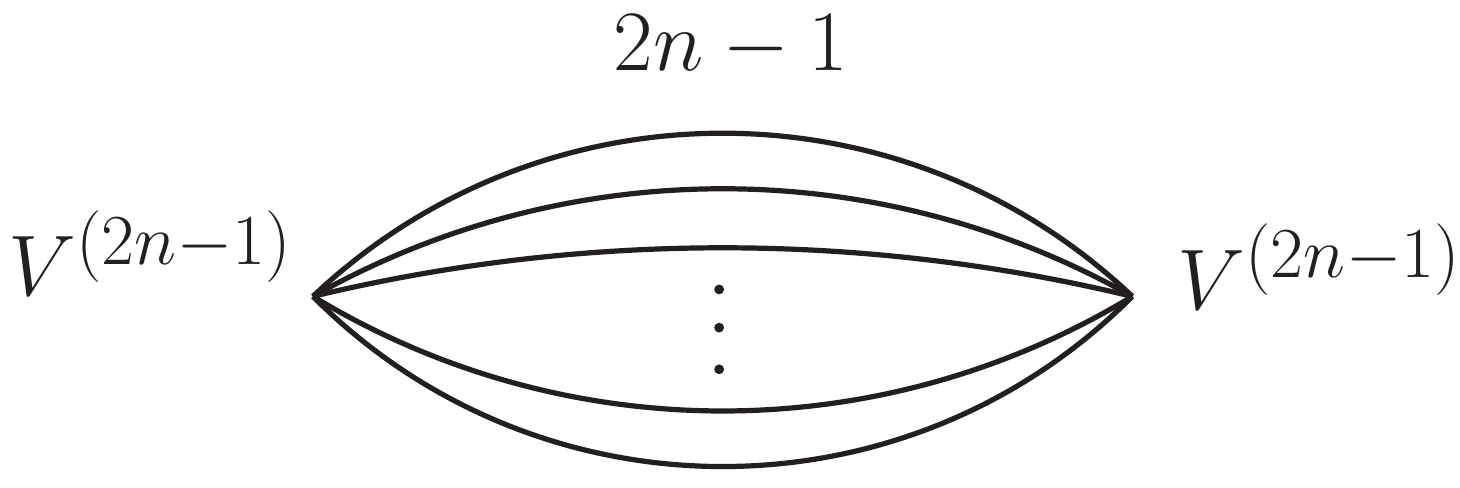}
\end{center}
The pole term of this diagram is nothing but the counter-term defined in \eqref{ul} for $l=k$
{\setlength\arraycolsep{2pt}
\bea
U_{k,c.t.} &=& \frac{1}{2(2n-1)!}\frac{1}{k(n-1)\epsilon}\,
\frac{(-1)^k}{[(4\pi)^{k}\Gamma(k)]^{2n}}\,
\frac{\Gamma^{2n-1}(\delta_c)}{\Gamma(\delta_c+2k)}\;V^{(2n-1)}(-\square)^kV^{(2n-1)}  \nn\\
&=& \frac{n}{(2n)!}\frac{2}{k(n-1)\epsilon}\,
\frac{(-1)^k}{[(4\pi)^{k}\Gamma(k)]^{2n}}\,
\frac{\Gamma^{2n-1}(\delta_c)}{\Gamma(\delta_c+2k)}\;(V^{(2n)})^2{\textstyle{\frac{1}{2}}}(\partial_{\mu_1\cdots\mu_k}\phi)^2 + \cdots \label{uk0}
\eea}%
where in the second line we have kept the term that contributes to the kinetic term coefficient. Taking the coefficient of the kinetic operator in \eqref{uk0} at $\phi=0$ and multiplying it by $2(n-1)\epsilon$ we get the flow of the kinetic term coefficient in terms of $g \equiv v^{(2n)}(0)/(2n)!$ which is simply equal to $-\eta$. Therefore we have
\be 
\eta = \frac{(-1)^{k+1}}{k}\,
\frac{{\Gamma(\delta_c+k)^2}}{\Gamma(\delta_c+2k){\Gamma(\delta_c)}}\,4n(2n)!\,g^2,
\ee
which has been written in the rescaled normalization. This rescaling of the couplings of course does not affect the anomalous dimension when written in terms of $\epsilon$. To find this we need the leading $\epsilon$ dependence of the fixed point coupling which can be obtained from \eqref{bv1}
\be
\frac{(2n)!^2}{n!^3} g =  (n-1)\epsilon +\mathcal{O}(\epsilon^2).
\ee
In terms of $\epsilon$ the anomalous dimension is a universal quantity and is given by
\be \label{etaeps1}
\eta =  (-1)^{k+1}\,
\frac{n(\delta_c)_k}{k(\delta_c+k)_k}\; \frac{4(n-1)^2n!^6}{(2n)!^3}\,\epsilon^2.
\ee
where the Pochhammer symbol $(a)_b = \Gamma(a+b)/\Gamma(a)$ has been used. This is in agreement with 
\cite{Gliozzi:2017hni}. Eq.\eqref{etaeps1} can be used in turn to find the $\epsilon$ dependence of the fixed point beyond leading order. Using \eqref{bv1} the fixed point at quadratic order in $\epsilon$ is found to be
\be  \label{fp}
\frac{(2n)!^2}{n!^3} g = \displaystyle  (n-1)\epsilon - n \,\eta  + \displaystyle \frac{n!^4(n-1)^2}{(2n)!}\bigg[\frac{1}{3}\,\Gamma(n\delta_c)\;n!^2\sum_{r,s,t}\frac{K^{k,n}_{rst}}{(r!s!t!)^2}  +\sum_{s,t}\frac{J^{k,n}_{st}}{s!^2t!^2} \bigg]\epsilon^2  +\mathcal{O}(\epsilon^3).
\ee
The $\epsilon$ dependence of the fixed point at quadratic order can be used to find some coupling anomalous dimensions beyond leading order.

\subsection{Coupling anomalous dimensions}

As an advantage of having the functional beta of the potential it is possible to calculate the anomalous dimensions of many couplings with a simple calculation and express the result for all couplings through a compact formula. This can be done by inserting the expansion 
\be \label{vexp}
v(\varphi) = \sum_i g_i\, \varphi^i,
\ee
for the potential function into the flow equation \eqref{bv1} and computing the stability matrix evaluated at the fixed point \eqref{fp}. We have chosen a coupling normalization with no factorials for later convenience, however this choice is physically irrelevant and does not affect the anomalous dimensions as functions of $\epsilon$. The coupling anomalous dimensions are generally obtained by diagonalization of the stability matrix, but in particular cases such as a triangular stability matrix they can be read off simply from the diagonal elements. If this is the case here, the coupling anomalous dimensions from \eqref{bv1} are given by
{\setlength\arraycolsep{2pt}
\bea  \label{cad} 
\tilde \gamma_i &=&  \frac{\eta}{2} i + 2n\, \eta\, \delta^{2n}_i 
+ \frac{(n-1)i!}{(i-n)!}\,\frac{2n!}{(2n)!}\left[\epsilon - \frac{n}{n-1} \,\eta\right] \nn \\[2mm]
&+& \frac{(n-1)^2i!n!^6}{(2n)!^2}\,\Gamma(n\delta_c)\sum_{r,s,t}\frac{K^{k,n}_{rst}}{(r!s!t!)^2} \left[\frac{2n!}{3(i-n)!} -\frac{r!}{(i-2n+r)!}\right]\epsilon^2 \nn \\[1mm]
&+& \frac{(n-1)^2i! n!^5}{(2n)!^2}\,\hspace{-0.01\textwidth}\sum_{s,t}\frac{J^{k,n}_{st}}{s!^2t!^2}\left[\frac{1}{(i-n)!} -\frac{2s!}{n!(i-2n+s)!}\right]\epsilon^2,
\eea}%
in which terms with negative factorials in the denominator are interpreted as zero. This formula is valid for instance when there is no mixing, that is for the range $0\le i \leq 2n$ corresponding to relevant and marginal couplings, which is still a new result. However if we downgrade the result to the leading order the range of validity will extend to infinity, including all irrelevant couplings as well. This is because despite the mixing of couplings of irrelevant non-derivative operators with those of derivative operators, the stability matrix at order $\epsilon$ is lower triangular, in the sense that in the quadratic terms in the beta function of $g_i$, couplings of higher derivative operators cannot appear together with the dimensionless coupling $g$. At this order the anomalous dimensions read
\be 
\tilde \gamma_i =  \frac{(n-1)i!}{(i-n)!}\,\frac{2n!}{(2n)!} \epsilon +\mathcal{O}(\epsilon^2),
\ee
which are independent of $k$ and in agreement with \cite{Gliozzi:2016ysv,Gliozzi:2017hni}. These anomalous dimensions vanish at order $\epsilon$ unless $i\geq n$. For instance the anomalous dimension $\tilde\gamma_2$ is of order $\epsilon^2$ for $n>2$.  
%
%
%
This can be obtained from the general formula \eqref{cad}. Notice that, independent of the value of $k$, $\phi^2$ is always a relevant operator, and hence \eqref{cad} will not be affected my mixing at any order in $\epsilon$. For theories with $n>2$ the second and third terms in the first line and the first terms in the brackets in the second and third lines vanish. Therefore $\tilde\gamma_2$ takes the following form
{\setlength\arraycolsep{2pt}
\bea  
\tilde \gamma_2 &=&  \eta 
-\frac{(n-1)^22!n!^6}{(2n)!^2}\,\Gamma(n\delta_c)\sum_{r,s,t}\frac{K^{k,n}_{rst}}{(r!s!t!)^2} \;\frac{r!}{(2-2n+r)!}\,\epsilon^2 \nn\\
&=&  \eta 
-\frac{(n-1)^22n!^6}{(2n)!^2}\,\Gamma(n\delta_c)\;\frac{K^{k,n}_{2n-2,1,1}}{(2n-2)!^2} \;(2n-2)!\,\epsilon^2 \nn\\[2mm]
&=&  \left[(-1)^{k+1}\,
\frac{n(\delta_c)_k}{k(\delta_c+k)_k} 
-\frac{n(2n-1)\Gamma(k)^2(\delta_c)_k}{\Gamma(2k)(\delta_c-k)_k}\right] \frac{4(n-1)^2n!^6}{(2n)!^3}\,\epsilon^2.
\eea}%
For the special case $k=1$ this reduces to
\be   
\tilde \gamma_2 = \left[1+\frac{n(2n-1)}{n-2}\right] \frac{4(n-1)^2n!^6}{(2n)!^3}\,\epsilon^2 = \frac{8(n+1)(n-1)^3n!^6}{(n-2)(2n)!^3}\,\epsilon^2,
\ee
in agreement with \cite{Codello:2017qek}.

\subsection{OPE coefficients}

In this section we compute what we call the $\overline{\mathrm{MS}}$ OPE coefficients, which can be calculated by expanding the coupling beta functions around the fixed point and reading off the coefficient of the quadratic terms in the coupling deformations. We distinguish them from the standard OPE coefficients by a tilde $\tilde C^l{}_{ij}$. These quantities which from now on we refer to as OPE coefficients for short depend on the renormalization scheme. However, the non zero OPE coefficients that we find with this method and using dimensional regularization are those that are dimensionless at $\epsilon=0$ \cite{Codello:2017hhh}. These OPE coefficients can be shown to be less sensitive to changes of scheme and in this sense universal (we refer the reader to section 2 of \cite{Codello:2017hhh} for further details).

Despite the anomalous dimensions which are unaffected by the coupling normalizations, the OPE coefficients do depend on this choice. To match the CFT normalization the operator couplings that are used to extract the OPE coefficients appear with the same coefficient which motivates our choice \eqref{vexp}. Using this expansion in the functional flow \eqref{bv1}, the universal OPE coefficients are extracted
%
{\setlength\arraycolsep{2pt}
\bea \label{ope}
\tilde C^l_{\;ij} &=& \frac{1}{n!} \frac{i!}{(i-n)!}\frac{j!}{(j-n)!} -\Gamma(n\delta_c) \frac{(n-1)n!^3}{(2n)!}\sum_{r,s,t}\frac{K^{k,n}_{rst}}{r!s!t!^2} \,\frac{j!}{(j-s-t)!}\frac{i!}{(i+s-2n)!}\,\epsilon \nn\\
&-& \frac{(n-1)n!^2}{(2n)!}\,\hspace{-0.01\textwidth}\sum_{s,t}\frac{J^{k,n}_{st}}{s!t!} \;\left[\frac{1}{n!}\frac{j!}{(j-n-s)!}\frac{i!}{(i-n-t)!}+\frac{1}{s!}\frac{i!}{(i-n)!}\frac{j!}{(j-n-s)!} \right. \nn\\
&+& \left. \frac{1}{s!}\frac{j!}{(j-n)!}\frac{i!}{(i-n-s)!}\right]\epsilon
+2(2n)!\left(i\delta^{2n}_j + j\delta^{2n}_i+2n\,\delta^{2n}_i\delta^{2n}_j\right)\epsilon.
\eea}%
The component $l$ is fixed by the universality condition for $\tilde C^l_{\;\,ij}$ 
\be 
[g_ig_j] = [g_l] \quad\Rightarrow\quad  i+j-l  =2n,
\ee
which is independent of $k$. These OPE coefficients have been extracted from the cubic beta functional after the rescaling \eqref{vres}. As pointed out earlier, this rescaling affects the OPE coefficients by a global overall factor and has been fixed such that the OPE coefficients are consistent with the CFT normalization. This is done by requiring any of the leading terms that is independent of $\epsilon$ in \eqref{ope} to match the value found from simple Wick counting. 

The compact expression \eqref{ope} looks rather involved but can be compared to known results for special cases. For instance 
plugging $i=n-m$, $j=n+m+1$ and hence $l=1$ into our general formula we get
{\setlength\arraycolsep{2pt}
\bea \label{univc1}
C^1_{\;n-m, n+m+1} &=&  - \Gamma(n\delta_c) \frac{(n-1)n!^3}{(2n)!}\sum_{r,s,t}\frac{K^{k,n}_{rst}}{r!s!t!^2} \,\frac{(n+m+1)!}{(n+m+1-s-t)!}\frac{(n-m)!}{(s-m-n)!}\,\epsilon \nn\\
&=& - \Gamma(n\delta_c) \frac{(n-1)n!^3}{(2n)!}\frac{K^{k,n}_{n-m-1,n+m,1}}{(n-m-1)!(n+m)!} \,(n+m+1)!(n-m)!\,\epsilon \nn\\[8pt]
%
%
&=& - \frac{\left(\delta_c\right)_k\,(n-1)\Gamma\left(k\right)}{\left(-m\delta_c\right)_k\left((m+1)\delta_c\right)_k}\,\frac{(n+m+1)!(n-m)!}{(n-m-1)!(n+m)!} \,\frac{n!^3}{(2n)!}\,\epsilon,
\eea}%
where we have used the fact that the denominators in the first line are factorials of non negative numbers
\be 
n+m+1-s-t \geq 0, \qquad
s-m-n \geq 0,
\ee
so the sum of these two expressions must also be non negative $1-t\geq 0$, and therefore we must have $t=1$. This means that the sum of the two expressions above is actually zero and so each non negative summand $n+m+1-s-t=r-n+m+1$ and $s-n-m$ must also vanish separately, giving $r=n-m-1$ and $s=n+m$. We have also used
\be  
K^{k,n}_{n-m-1,n+m,1} = \frac{\Gamma\left(k\frac{m+1}{n-1}\right)\Gamma\left(k\frac{-m}{n-1}\right)\Gamma\left(k\right)}{\Gamma\left(k\frac{-m}{n-1}+k\right)\Gamma\left(k\frac{m+1}{n-1}+k\right)\Gamma\left(\frac{k}{n-1}\right)} = \frac{\Gamma\left(k\right)}{\left(k\frac{-m}{n-1}\right)_k\left(k\frac{m+1}{n-1}\right)_k\Gamma\left(\frac{k}{n-1}\right)}.
\ee
The result \eqref{univc1} is in agreement with 
\cite{Gliozzi:2017hni}, if we take into account the difference in the normalizations. Namely, in \cite{Gliozzi:2017hni} the coefficients in $\langle\phi^l\phi^l\rangle$ are normalized to unity, while here we only normalize $\langle\phi\phi\rangle$. In other words, to compare with \cite{Gliozzi:2017hni} we have to make the additional transformation $\phi^l\rightarrow \phi^l \sqrt{l!}$, and therefore $\tilde C^k_{\;ij}\rightarrow \tilde C^k_{\;ij}\sqrt{i!}\sqrt{j!}/\sqrt{k!}$.

\subsection{Beta function of two-derivative couplings}

Let us take a step further in the derivative expansion and consider the following Lagrangian for theories of the first type
\be 
\mathcal{L} = {\textstyle{\frac{1}{2}}} \phi \square^2\phi + {\textstyle{\frac{1}{2}}}Z(\phi)(\partial\phi)^2 + V(\phi).
\ee
The function $Z$ affects the flow of the potential starting from cubic terms in the flow and leaves the quadratic flow unaltered. Let us calculate 
the contribution of the potential and the wavefunction $Z$ to the flow of $Z$ at quadratic level. The contribution that is quadratic in the potential is present only for the case $k=1$. This is seen from the argument of Sect.\ref{ss:qmelon} and the expression for the counter-term can be found from Eq.\eqref{ul} for $l=1$ and $r=2n-1$. This comes from the diagram in Fig.\ref{melon} but applied to the standard case with $k=1$. The contribution to the beta function is found by multiplying the coefficient of $(\partial\phi)^2$ by $2(n-1)\epsilon$ 
\be 
-\frac{1}{(2n-1)!}\,\frac{1}{(4\pi)^{2n}}\,\frac{\Gamma^{2n-1}(\delta_c)}{\Gamma(\delta_c+2)}\,V^{(2n)}\hspace{1pt}^2 = -\frac{2(n-1)^2}{(2n)!}\,\frac{\Gamma^{2n-2}(\delta_c)}{(4\pi)^{2n}}\,V^{(2n)}\hspace{1pt}^2 \qquad \mathrm{present\;\,only\;\,for}\;\, k=1
\ee
Let us now compute the quadratic contribution that includes both $V$ and $Z$ couplings. The relevant diagram is 
\begin{center}
\includegraphics[width=0.4\textwidth]{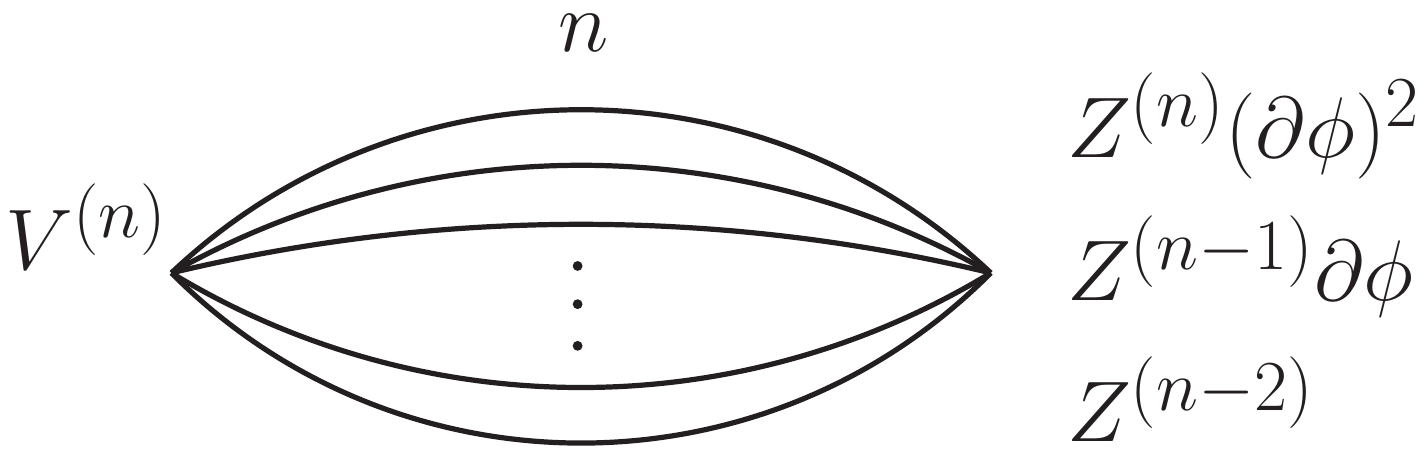}
\end{center}
At the $Z$-vertex there are three possibilities depending on whether the propagators are connected to one or both fields in $(\partial\phi)^2$ or only to fields in $Z(\phi)$. The counter-term corresponding to this diagram is given as 
{\setlength\arraycolsep{2pt}
\bea  
&& \frac{1}{(n-1)\epsilon}\frac{1}{n!}Z^{(n)}V^{(n)} (\partial\phi)^2 \; \frac{1}{[(4\pi)^k\Gamma(k)]^n}\,\frac{\Gamma(\delta_c)^n}{\Gamma(n\delta_c)}  \nn\\
&+& \frac{2-k}{(n-1)\epsilon} \frac{1}{n!}Z^{(n-1)}V^{(n+1)} (\partial\phi)^2 \; \frac{1}{[(4\pi)^{k}\Gamma(k)]^n}\,\frac{\Gamma(\delta_c)^n}{\Gamma(n\delta_c)} 
\eea}%
This has to be multiplied by $2(n-1)\epsilon$ to give the contribution to the beta function
\be 
\frac{1}{n!} \, \frac{1}{[(4\pi)^{k}\Gamma(k)]^n}\,\frac{2\Gamma(\delta_c)^n}{\Gamma(n\delta_c)}\, \left[Z^{(n)}V^{(n)} + (2-k) Z^{(n-1)}V^{(n+1)} \right]
\ee
This generalizes the result for the standard $k=1$ case given in \cite{ODwyer:2007brp}. In terms of the dimensionless fields \eqref{vdimless} and 
\be \label{zdimless}
z(\varphi) = \mu^{\!-2(k-1)}\, Z(\mu^\delta\,\varphi),
\ee
and after the rescaling \eqref{vres} the complete beta function reads
\be  \label{bz} 
\beta_z = -2(k-1)z+\frac{d-2k}{2}\varphi\, z^{(1)} + \frac{2}{n!} \left[z^{(n)}v^{(n)} + (2-k) z^{(n-1)}v^{(n+1)}\right] -\frac{2}{(2n)!}\,v^{(2n)}\hspace{1pt}^2 \delta_{k,1},
\ee
where the Kronecker delta $\delta_{k,1}$ in the last term guarantees that this term is present only for $k=1$. The term $\frac{1}{2}\eta \varphi  z^{(1)}$ proportional to the anomalous dimension is omitted as it is of cubic order in the couplings. 
From this beta function it is possible to calculate the anomalous dimensions of couplings corresponding to two-derivative operators, $\tilde\omega_i$, at order $\epsilon$. At this level of approximation the results are valid for all indices $i$ because the stability matrix is lower triangular. A simple calculation gives
\be 
\tilde\omega_i = -2(k-1) + \frac{d-2k}{2}i + \frac{2(n-1)n!}{(2n)!}\,\frac{i!(i+1-(k-1)n)}{(i-n+1)!}\,\epsilon.
\ee
This reproduces the result of \cite{Codello:2017hhh} for the special case $k=1$.

\section{Theories of the second type: The $k=2$, $n=2m+1$ example} \label{s:type2}

So far we have been dealing with theories for which $k$ and $n-1$ are relatively prime. We will now turn our attention to theories of the second type in which $k$ and $n-1$ have a common divisor. As argued in Sect.\ref{ss:de} these theories are considerably more involved. Instead of complicating the calculations, as a first encounter with such theories we find it more instructive to concentrate on the special case of $k=2$ which already shows the novel features of second-type theories. For the $k=2$ case, theories with even values of $n$ fall in the first class already discussed. We therefore consider here $\Box^2$ theories with $n=2m+1$, where $m$ is a natural number. For such theories, apart from the kinetic operator there are two marginal operators $\phi^{2(2m+1)}$ and $\phi^{2m}(\partial\phi)^2$ at the critical dimension. According to arguments of Sect. A potential approximation is therefore inconsistent as a lowest-order derivative expansion and one has to take into account 2-derivative interactions as well. Therefore, we consider the Lagrangian  
\be 
\mathcal{L} = {\textstyle{\frac{1}{2}}} \phi\square^2\phi +{\textstyle{\frac{1}{2}}}Z(\phi)(\partial\phi)^2 + V(\phi).
\ee
These theories are labelled by the number $m$ in terms of which the field dimension at criticality and the upper critical dimension are 
\be 
\delta_c = \frac{k}{n-1} =  \frac{1}{m}, \qquad
d_c = \frac{2nk}{n-1} = 4 +\frac{2}{m}. \qquad
\ee
In the following we first calculate the quadratic flow of the functions $V$ and $Z$ for general $m$. For this purpose we need the quadratic counter-terms. 

\subsection{Counter-terms for $V$ of the form $V^{(a)}\hspace{1pt}^2$} \label{ss:vv}

Let us first consider counter-terms that are quadratic in the $V$ couplings. According to the arguments of Sect.\ref{ss:qmelon} these can contribute at any order in the number of derivatives. In fact Eq.\eqref{c1} shows that the diagram of Fig.\ref{melon} with $r = (2+l)m+1$ contribute to operators with $2l$ derivatives. Here we are interested in the counter-term $U_{0,c.t.}$ which contribute to the potential $V$. We thus need to choose $l=0$. This gives a melon diagram with $r=2m+1$ propagators   
%
\begin{center}
\includegraphics[width=0.35\textwidth]{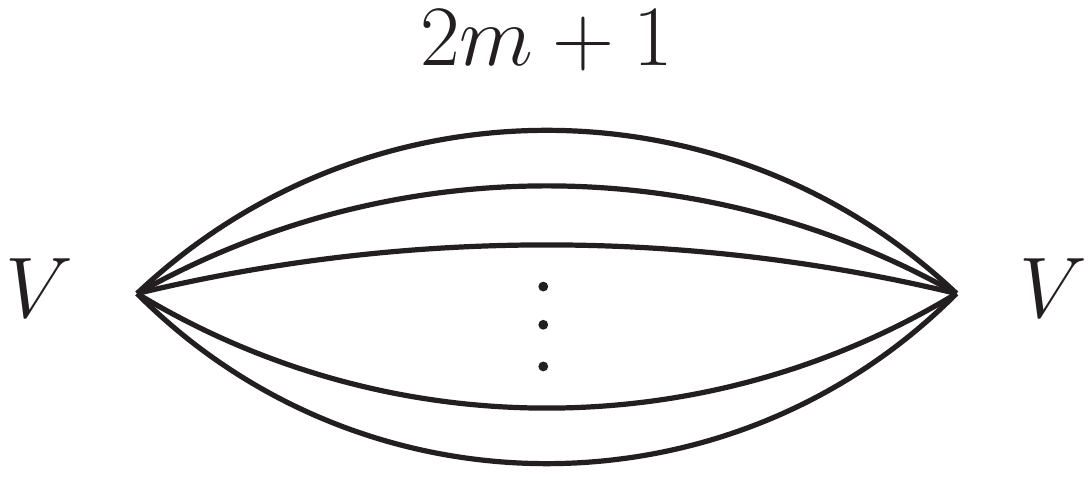}
\end{center}
The pole of this diagram is found from Eq.\eqref{ul}, setting $r=2m+1$ and $l=0$ (of course with $k=3$ and $\delta_c = m^{-1}$)
\be
U_{0,c.t.} = \frac{1}{2m\epsilon}\,\frac{1}{(2m+1)!}\,\frac{1}{(4\pi)^{2(2m+1)}}\,
\frac{\Gamma^{2m+1}(\delta_c)}{\Gamma(\delta_c+2)}\,(V^{(2m+1)})^2.
\ee
The contribution to the beta function is found by multiplying this expression by $2m\epsilon$.

\subsection{Counter-terms for $Z$ of the form $V^{(a)}\hspace{1pt}^2$} \label{ss:vv}

The contribution to the function $Z$ comes from \eqref{ul} with $l=1$. This corresponds to $r=3m+1$ giving rise to the diagram 
\begin{center}
\includegraphics[width=0.35\textwidth]{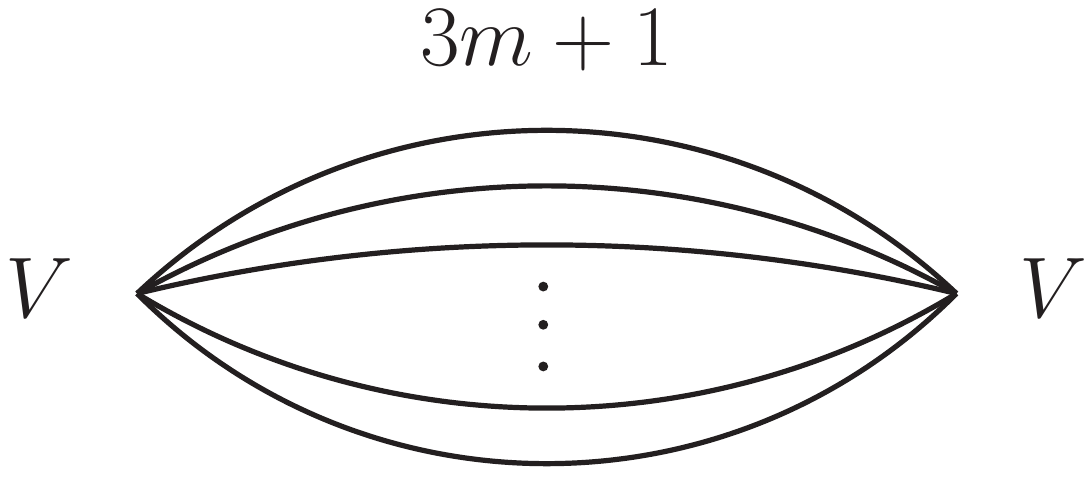}
\end{center}
Using Eq.\eqref{ul} with $r=3m+1$ and $l=1$, and doing integration by parts the pole term reads  
\be
U_{1,c.t.} = -\frac{1}{3m\epsilon}\,\frac{1}{(3m+1)!}\,\frac{1}{(4\pi)^{3(2m+1)}}\,
\frac{\Gamma^{3m+1}(\delta_c)}{\Gamma(\delta_c+3)}\,V^{(3m+2)}\hspace{1pt}^2(\partial\phi)^2.
\ee
This has to be multiplies by $2\times 3m\epsilon$ to give the contribution to the beta function.


\subsection{Counter-terms for $V$ of the form $V^{(a)}Z^{(b)}$}

Let us now consider quadratic contributions to the potential counter-term with a $V$-coupling and a $Z$-coupling. The relevant diagram is a melon diagram with $m+1$ propagators. 
\begin{center}
\includegraphics[width=0.35\textwidth]{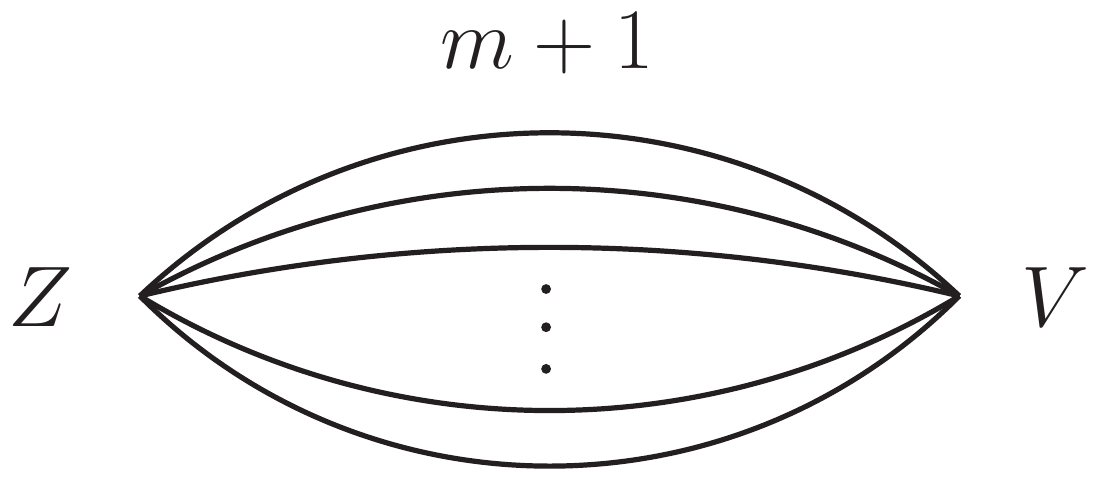}
\end{center}
The expression for such diagram for arbitrary $r$ number of propagators is found in Sect.\ref{s:zvmelon}. The diagram of relevance here is evaluated by setting $r=m+1$ in Eq.\eqref{mel}. The first two terms in this equation involve $G^{m+1}_{xy}$ which is finite for $k=2$ and $n=2m+1$. We are therefore left with the last term, the pole of which can be easily calculated. This gives
\be  \label{zv_v}
\frac{\Gamma^{m}(\delta_c)}{(4\pi)^{2m+1}}\,\frac{V^{(m+1)}Z^{(m-1)}}{(m+1)!}\; \,\frac{1}{\epsilon}
\ee

\subsection{Counter-terms for $Z$ of the form $V^{(a)}Z^{(b)}$}

A similar diagram but with $2m+1$ propagators contribute to the wavefunction $Z$
\begin{center}
\includegraphics[width=0.35\textwidth]{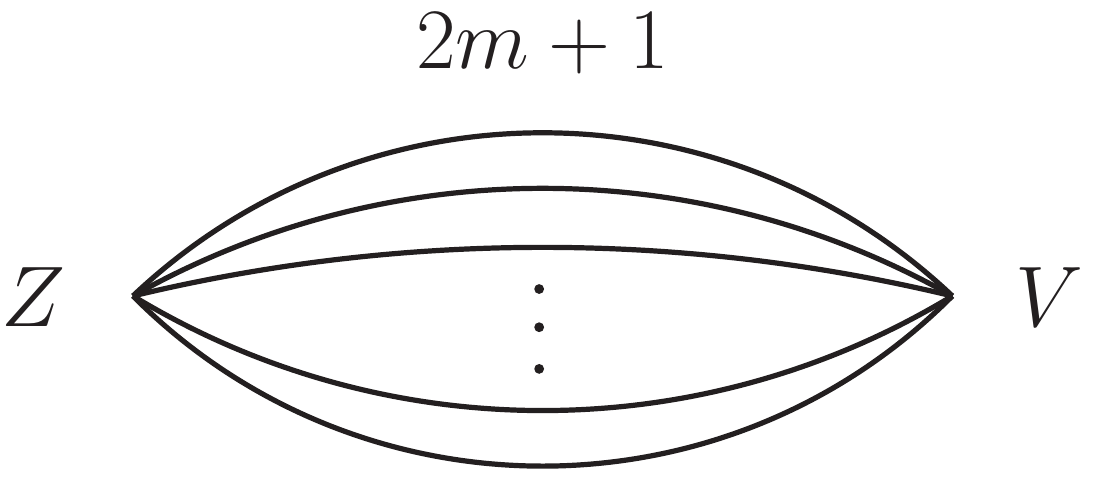}
\end{center}
This diagram can also be calculated from Eq.\eqref{mel} by setting $r=m+1$, but this time $G^{m+1}_{xy}$ is divergent and so the first two terms also contribute to the pole. In fact a simple calculation shows that the last two terms cancel together and the pole term reads 
\be  \label{zv_z}
\frac{V^{(2m+1)}}{2(4\pi)^{2(2m+1)}}\,\frac{m\Gamma(\delta_c)^{2m}}{(m+1)\epsilon} \frac{Z^{(2m+1)}}{(2m+1)!} \,(\partial\phi)^2
\ee
To obtain the beta function one has to multiply this by $2\times 2m\epsilon$. 

\subsection{Counter-terms for $Z$ of the form $Z^{(a)}Z^{(b)}$}

There is also a counter-term that is quadratic in $Z$. This contributes only to $Z$. The diagram is an $m$-loop melon diagram   
\begin{center}
\includegraphics[width=0.35\textwidth]{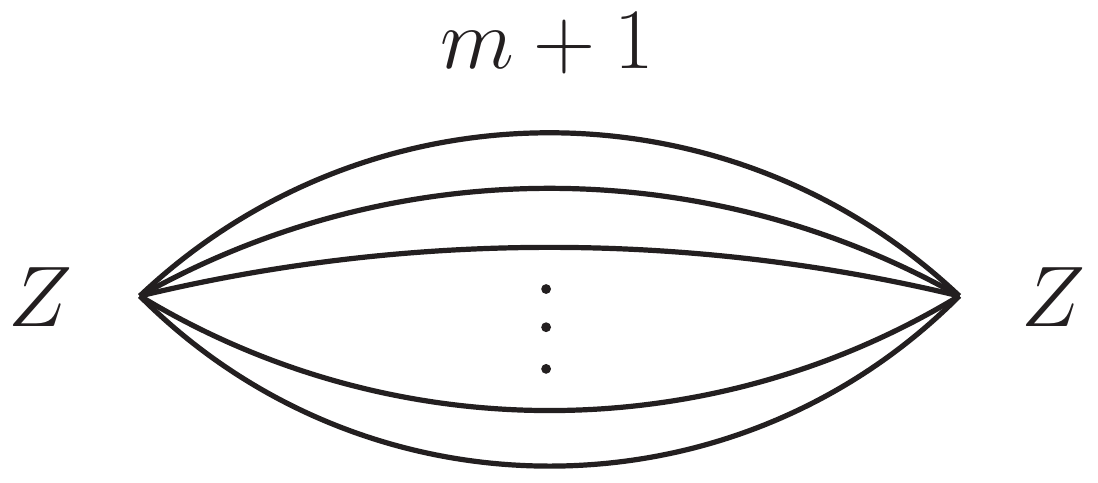}
\end{center}
Each vertex can appear in three different ways as discussed in Sect.\ref{s:zvmelon}, so all together the expression for this diagram can be written as the sum of six terms. This given for a general number of propagators $r$ in \eqref{zz}. 
To complete the calculation we need to know the divergences of the combinations of propagators and their derivatives in \eqref{zz} for $r=m+1$. The quantity $G^{m+1}$ is finite so the first three terms do not contribute.  For the last three terms we need to extract the divergences of \eqref{r-2}, \eqref{r-3}, \eqref{r-4} and \eqref{mel}. Setting $r=m+1$ these are found to be
\be  
\int_x \!e^{ip\cdot x} \; (\partial_\mu\partial_\nu G_x)^2G^{m-1} = -  \frac{1}{(4\pi)^{2m+1}}\,\frac{(m(2m+3)+2)\Gamma(\delta_c)^m}{m^2(m+1)(2m+1)\epsilon} \,p^2 +\mathrm{finite} 
\ee
\be   
\int_x \!e^{ip\cdot x} \; \partial_\mu\partial_\nu G_x\,\partial^\mu G\,\partial^\nu G_x \, G^{m-2} = -\frac{1}{(4\pi)^{2m+1}}\,\;\frac{(m+2)\Gamma(\delta_c)^{m}}{m^2(m+1)(2m+1)\epsilon}\,p^2 + \mathrm{finite}
\ee
\be  
\int_x \!e^{ip\cdot x} \; (\partial G)^4 G^{m-3} =  -\frac{1}{(4\pi)^{2m+1}}\;\frac{2\Gamma(\delta_c)^{m}}{m^2(m+1)(2m+1)\epsilon}\,p^2 + \mathrm{finite}
\ee
\be  
\int_x \!e^{ip\cdot x}\, (\partial G)^2 G^{m-1} = \frac{1}{(4\pi)^{2m+1}}\,\frac{2\Gamma(\delta_c)^{m}}{m(m+1)\epsilon} +\mathrm{finite}
\ee
Using the first three of these pole terms, the combination of propagators and their derivatives in the fourth term of \eqref{zz} becomes
\be 
\ba{cc}
 2(\partial_\mu\partial_\nu G_x)^2G^{m-1}+4(m-1)\partial_\mu\partial_\nu G_x\,\partial^\mu G\,\partial^\nu G_x \, G^{m-2}+(m-1)(m-2)(\partial G)^4 G^{m-3} \\[2mm]
\displaystyle = \frac{1}{(4\pi)^{2m+1}}\;\frac{2(5m+2)\Gamma(\delta_c)^{m}}{m(m+1)(2m+1)\epsilon}\,\square_x\delta_{xy} + \mathrm{finite}
\ea
\ee
These poles can now be inserted into \eqref{zz} to obtain the final counter-term quadratic in $Z$ 
\be 
\frac{1}{(4\pi)^{2m+1}}\left[\frac{(3m+2)\Gamma(\delta_c)^{m}}{4(2m+1)(m+1)!\epsilon}\,(Z^{(m)})^2 +\frac{\Gamma(\delta_c)^{m}}{2(m+1)!\epsilon}\,Z^{(m+1)} Z^{(m-1)}\right] (\partial\phi)^2.
\ee
This must be multiplied by $2\times m\epsilon$ to give the corresponding beta function contribution.

\subsection{Beta functions at quadratic order}

Summing up all the contributions and simplifying a bit we get the dimensionful beta functions for $V$ and $Z$ at quadratic order 
{\setlength\arraycolsep{2pt}
\bea
\beta_V &=& \frac{m\Gamma^{m}(\delta_c)}{(4\pi)^{2m+1}} \; \frac{V^{(m+1)}Z^{(m-1)}}{(m+1)!} + \frac{1}{(2m+1)!}\,\frac{m^2\Gamma^{2m}(\delta_c)}{(4\pi)^{2(2m+1)}}\,
\frac{1}{m+1}\,(V^{(2m+1)})^2, \label{bvq} \\[10pt]
\beta_Z &=& -\frac{1}{(3m+1)!}\,\frac{m^3\Gamma^{3m}(\delta_c)}{(4\pi)^{3(2m+1)}}\,
\frac{2}{(2m+1)(m+1)}\,(V^{(3m+2)})^2 + \frac{m^2\Gamma(\delta_c)^{2m}}{(4\pi)^{2(2m+1)}}\,\frac{2}{(m+1)} \frac{V^{(2m+1)}Z^{(2m+1)}}{(2m+1)!} \nn\\[5pt]
&& +\frac{m\Gamma(\delta_c)^{m}}{(4\pi)^{2m+1}}\,\frac{(3m+2)}{2(2m+1)(m+1)!}\,(Z^{(m)})^2 +\frac{m\Gamma(\delta_c)^{m}}{(4\pi)^{2m+1}}\,\frac{1}{(m+1)!}\,Z^{(m+1)}Z^{(m-1)}. \label{bzq} 
\eea}%
where (just like $\beta_V$) $\beta_Z$ refers to the $\eta$-independent term in the flow of $Z$. Let us move to dimensionless variables 
\be \label{vzdimless}
v(\varphi) = \mu^{\!-d}\, V(\mu^\delta\,\varphi), \qquad z(\varphi) = \mu^{\!-2}\, Z(\mu^\delta\,\varphi),
\ee
and perform the rescalings   
\be  \label{vzres}
v\rightarrow (m+1)\,\frac{(4\pi)^{2(2m+1)}}{m^2\Gamma^{2m}(\delta_c)}\, v, \qquad
z\rightarrow \frac{(4\pi)^{2m+1}}{m\Gamma^{m}(\delta_c)}\, z.
\ee
The rescaling for $v$ is compatible with the $v$-rescaling introduced earlier in \eqref{vres} for the first-type theories, which is made to match the CFT normalization. For the dimensionless variables the beta functions reduce to 
{\setlength\arraycolsep{2pt}
\bea
\beta_v &=& -d\,v + \frac{d-4}{2}\varphi \,v^{(1)} + \frac{(v^{(2m+1)})^2}{(2m+1)!} +  \frac{v^{(m+1)}z^{(m-1)}}{(m+1)!}, \label{bvqdl} \\[3mm]
\beta_z &=& -2\,z + \frac{d-4}{2}\varphi \,z^{(1)} -\frac{2(m+1)}{(2m+1)}\,\frac{(v^{(3m+2)})^2}{(3m+1)!} + 2 \frac{v^{(2m+1)}z^{(2m+1)}}{(2m+1)!} \nn\\[2mm]
&& +\frac{3m+2}{2(2m+1)}\,\frac{(z^{(m)})^2}{(m+1)!} +\frac{z^{(m+1)}z^{(m-1)}}{(m+1)!}. \label{bzqdl}
\eea}%
Similar to the case of Eq.\eqref{bz}, the terms proportional to the anomalous dimension in $\beta_v$ and $\beta_z$ which are given by $\frac{1}{2}\eta \varphi  v^{(1)}$ and $\frac{1}{2}\eta \varphi z^{(1)}$ respectively are of cubic order in the couplings and are therefore omitted in this approximation.

\subsection{Fixed points}

\begin{figure}[h] 
\begin{center}
\includegraphics[width=0.5\textwidth]{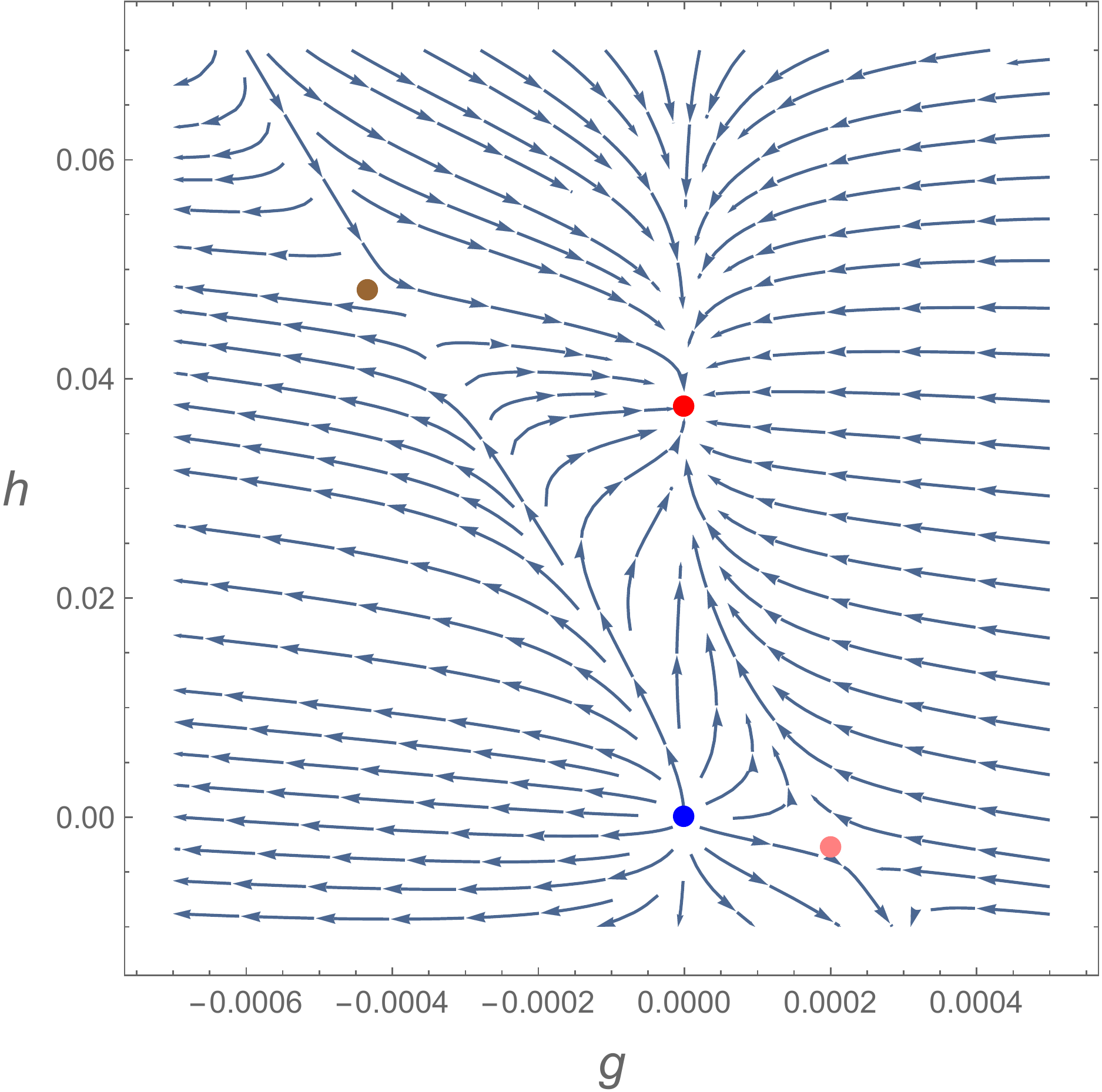}
\end{center}
\caption{Phase diagram for $\Box^2$ theory with $n=3$ for $\epsilon=0.1$. Four fixed points are present: a Gaussian fixed point, a fixed point with a derivative interaction $\phi^2 (\partial\phi)^2$, and two fixed points with a mixture of $\phi^2(\partial\phi)^2$ and $\phi^6$ interactions.} \label{fd} 
\end{figure}

Equipped with the beta functions we are now able to find the fixed points and spectrum of the couplings. For this purpose let us parameterize the $v,z$ functions as
\be \label{vzexp}
v(\varphi) = \sum_i g_i\, \varphi^i, \qquad z(\varphi) = \sum_i h_i\, \varphi^i
\ee
where $g\equiv g_{2(2m+1)}$ and $h\equiv h_{2m}$ are the marginal couplings. With this parametrization the fixed point equations for the marginal couplings at quadratic level are 
{\setlength\arraycolsep{2pt}
\bea
2m\epsilon g &=& \frac{(2(2m+1))!(2m)!}{(3m+1)!(m+1)!^2}\,g\,h + \frac{(2(2m+1))!^2}{(2m+1)!^3}\,g^2  \nn\\
m\epsilon h &=& \left[\frac{3m+2}{2(2m+1)}+\frac{m}{m+1}\right]\frac{(2m)!^2}{(m+1)!m!^2}\,h^2  - \frac{2(m+1)}{2m+1} \,\frac{(2(2m+1))!^2}{(3m+1)!m!^2}\,g^2  \label{fpe}
\eea}%
An interesting feature of these equations is that there are no non trivial fixed points with $h=0$. This means in particular that a pure potential deformation of the higher-derivative free theory \eqref{hft} is never scale invariant. Instead there exists a fixed point with a pure derivative interaction, that is with $g=0$ and 
\be \label{h}
h=\frac{2m(m+1)(2m+1)}{2+7m(m+1)} \,\frac{(m+1)!m!^2}{(2m)!^2}\,\epsilon .
\ee 
Interestingly this is also an infrared fixed point. Apart from this and the Gaussian fixed point there are two other fixed points which are a mixture of derivative and non-derivative interactions. These are made more clear in Fig.\eqref{fd} where the flow diagram is shown for the particular case of $m=1$. The picture is qualitatively the same for higher $m$.

\subsection{Spectrum of couplings} \label{ss:spectrum}

Having identified the fixed points, we will now proceed with the calculation of coupling anomalous dimensions. Table.\eqref{t3} shows that couplings of non-derivative operators up to $\phi^{2m+1}$ are not affected by mixing. This makes diagonal the $2(m+1)$-dimensional upper left block in the stability matrix, i.e. $\mathcal{M}_0^{(0)}$ in the notation of \cite{Codello:2017hhh}. The diagonal elements give the scaling dimensions of the potential couplings corresponding to $\phi^i$. The $i$th element is given, for $i=1,\cdots, 2m+1$, as  
\be 
i\frac{d-4}{2} -d + \frac{i!}{(i-m-1)!}\,\frac{(2m)!}{(m+1)!^2}\,h + 2\frac{i!}{(i-2m-1)!}\,\frac{(2(2m+1))!}{(2m+1)!^2}\,g.
\ee
Notice that the last term proportional to $g$ is present only in the last component $i=2m+1$. The anomalous dimension for the relevant coupling $g_i$ is therefore
\be \label{ad1} 
\tilde \gamma_i = \frac{i!}{(i-m-1)!}\,\frac{(2m)!}{(m+1)!^2}\,h + \frac{i!}{(i-2m-1)!}\,\frac{2(2(2m+1))!}{(2m+1)!^2}\,g, \qquad i=1,\cdots, 2m+1.
\ee 
It is clear that for $i=1,\cdots, m$ the anomalous dimension is zero at this order. 
Starting from $\phi^{2(m+1)}$ the couplings of non-derivative operators mix with those of two-derivative operators. This continues up to $\phi^{4m+1}$ after which couplings of four-derivative operators also come into play. Following the notation of \cite{Codello:2017hhh} the $i$th block of the matrix $\mathcal{M}_0^{(2)}$ governs the $\phi^i - \phi^{i-2m-2}(\partial\phi)^2$ mixing, i.e. the mixing between the couplings $g_i$ and $h_{i-2m-2}$, for $i=2m+2,\cdots, 4m+1$. These two by two matrices are given by 
\be \label{sm}
\mathcal{M}_{0,i}^{(2)} = \frac{i-2}{m}-4 - \frac{i-2}{2}\epsilon + \left(\ba{cc}  {}[\mathcal{M}_{0,i}^{(2)}]_{gg} & {}[\mathcal{M}_{0,i}^{(2)}]_{gh} \\ {}[\mathcal{M}_{0,i}^{(2)}]_{hg} & m\epsilon +  {}[\mathcal{M}_{0,i}^{(2)}]_{hh} \ea\right)
\ee
where the first three terms are understood to be proportional to the two by two identity matrix and the four elements in the last term are
{\setlength\arraycolsep{2pt}
\bea \label{m}
{}[\mathcal{M}_{0,i}^{(2)}]_{gg} &=& \frac{i!}{(i-m-1)!}\,\frac{(2m)!}{(m+1)!^2}\,h + \frac{i!}{(i-2m-1)!}\,\frac{2(2(2m+1))!}{(2m+1)!^2}\,g \nn\\
{}[\mathcal{M}_{0,i}^{(2)}]_{gh} &=& \frac{(i-2m-2)!}{(i-3m-1)!}\,\frac{(2(2m+1))!}{(m+1)!(3m+1)!}\,g \nn\\
{}[\mathcal{M}_{0,i}^{(2)}]_{hg} &=& - \frac{4(m+1)}{2m+1}\,\frac{i!}{(i-3m-2)!}\,\frac{(2(2m+1))!}{(3m+1)!m!}\,g  \\
{}[\mathcal{M}_{0,i}^{(2)}]_{hh} &=& \frac{(2m)!(i-2m-2)!}{(m+1)!^2} \left[\frac{3m+2}{2m+1}\,\frac{m+1}{(i-3m-2)!} +  \frac{1}{(i-3m-3)!} +\frac{m(m+1)}{(i-3m-1)!}\right]h \nn
\eea}%
For $i=2m+2,\cdots, 3m$ this is still diagonal because $[\mathcal{M}_{0,i}^{(2)}]_{gh}=[\mathcal{M}_{0,i}^{(2)}]_{hg}=[\mathcal{M}_{0,i}^{(2)}]_{hh}=0$. For $i=3m+1$ it is triangular because $[\mathcal{M}_{0,i}^{(2)}]_{hg}=0$. So the anomalous dimensions can be read off from the diagonal elements. The matrix element $[\mathcal{M}_{0,i}^{(2)}]_{gg}$ gives the anomalous dimensions of the $g_i$ couplings, so one can extend the range of $g_i$ anomalous dimensions \eqref{ad1}, given above, to
\be  \label{ad2}
\tilde \gamma_i = \frac{i!}{(i-m-1)!}\,\frac{(2m)!}{(m+1)!^2}\,h + \frac{i!}{(i-2m-1)!}\,\frac{2(2(2m+1))!}{(2m+1)!^2}\,g, \qquad i=0,\cdots, 3m+1.
\ee 
Furthermore, the quantity $[\mathcal{M}_{0,i}^{(2)}]_{hh}$ gives the anomalous dimensions of the $h_i$ couplings $\tilde \omega_i$. The only nonzero element that is never affected my mixing is the case $i=3m+1$ which gives the anomalous dimension of $h_{m-1}$ 
\be 
\tilde\omega_{m-1} = \frac{(2m)!}{(m+1)!}h.
\ee 
It is straightforward to calculate the anomalous dimensions also for $i=3m+2,\cdots, 4m+1$, but one has to solve the characteristic equation and the expressions are more involved. For the particular fixed point with $g=0$, the matrix \eqref{m} is diagonal all the way up to $i=4m+1$. So the coupling anomalous dimensions are easily read off from diagonal elements. In this case, using \eqref{h} it is straightforward to express the anomalous dimensions in terms of $\epsilon$ 
\be \label{cad2ep}
\tilde\gamma_i = \frac{i!}{(i\!-\!m\!-\!1)!}\;\frac{2m(2m+1)}{2+7m(m+1)} \,\frac{m!}{(2m)!}\,\epsilon,
\ee 
\be  \label{omega}
\tilde\omega_i = i! \left[\frac{3m+2}{2m+1}\frac{m+1}{(i\!-\!m\!)!} + \frac{1}{(i\!-\!m\!-\!1)!} +\frac{m(m+1)}{(i-m+1)!}\right]\frac{2m(2m+1)}{2+7m(m+1)} \frac{m!}{(2m)!}\epsilon.
\ee
At this fixed point, the range of validity for these coupling anomalous dimensions extends to infinity. This is because for $g=0$ the leading terms that contribute to the stability matrix come from quadratic terms in the beta functions with an $h$ coupling, which corresponds to a two-derivative operator. Then, from \eqref{constraint} it is seen that in a quadratic term in the beta function of $g_i$, the coupling $h$ can be coupled only to non-derivative couplings $g_j$, and in a quadratic term in the beta function of $h_i$, the coupling $h$ can be coupled either to non-derivative couplings $g_j$ or to two-derivative couplings $h_j$. Therefore the stability matrix governing the mixing in the $g_i$-$h_j$ sector is lower triangular. The above argument is in fact general: If we collectively denote the coupling of $2l$-derivative operators by $c_l$, then the beta function $\beta_{c_i}$ can include a quadratic term of the form $h c_j$ only for $j\leq i$. This constraint on $i,j$ indeed follows from \eqref{constraint} which in this case translates \nolinebreak to 
\be 
j+1-i = \frac{2m+1-r}{m}, \qquad\Rightarrow\qquad j-i = 1+ \frac{1-r}{m} \leq 0.
\ee



\subsubsection{Some scaling relations}

Let us make a small digression here to check some simple scaling relations. On general grounds one can identify two scaling directions in theory space with simple critical exponents. This can be seen most easily by working with a general action $S(\phi)$ rather than the set of functions $V(\phi)$, $Z(\phi)$, \ldots, as was instead discussed in Appendix B of~\cite{Codello:2017hhh}. 
In terms of the dimensionless field $\varphi$ we denote the local action by $s(\varphi)$ defined through 
\be \label{s}
S(\phi) = \int_x \mathcal{L}(\phi,\partial\phi,\cdots) = \int_{\tilde x} \tilde{\mathcal{L}}(\varphi,\partial\varphi,\cdots) = s(\varphi),
\ee
where $\phi = \mu^\delta\varphi$, the dimensionless spacetime variable in the second integral is $\tilde x = \mu x$ and the dots stand for higher derivatives of the field. The spacetime dependence of the fields are omitted for simplicity. Referring to the $\eta$-independent part of the flow of $S$ as $\beta_S$
\be 
\mu\, \partial_\mu S = {\textstyle{\frac{1}{2}}} \eta\, \phi \!\cdot\! S' + \beta_S,
\ee
where $S'$ is the functional derivative of the action with respect to the field $\phi$ and the dot implies integration over the spacetime variable $x$, it follows from the definition \eqref{s} that the beta function of $s$ is related to $\beta_S$ through 
\be \label{bs}
\beta_s = -d\,s(\varphi) +  {\textstyle{\frac{1}{2}}}(d-2k+\eta)\, \varphi \!\cdot\! s'(\varphi) + \beta_S.
\ee
Here again $s'(\varphi)$ is the functional derivative of the action with respect to its argument and this time the dot implies integration over the spacetime variable $\tilde x$. This equation is simply a generalization of relations such as \eqref{vrel} and its analogue for the wavefunction. The beta function $\beta_s$ depends on $s$ and its field derivatives, but there is also an explicit $\varphi$ dependence in \eqref{bs}. Let us denote the scaling solution by $s_*(\varphi)$ so that $\beta_{s_*}=0$. Taking the field derivative of the fixed point equation $\beta_{s_*}=0$ we find
\be 
0 = \frac{d\beta_{s_*}}{d\varphi} = \frac{\partial\beta_{s_*}}{\partial\varphi} + \left.\frac{\delta\beta_{s}}{\delta s}\right|_{s_*}\hspace{-7pt}\!\cdot \! s'_* = {\textstyle{\frac{1}{2}}}(d-2k+\eta) s'_* + \left.\frac{\delta\beta_{s}}{\delta s}\right|_{s_*}\hspace{-7pt}\!\cdot \! s'_*,
\ee
where the first term on the r.h.s follows from \eqref{bs}. This is simply an eigenvalue equation with the eigenvector $\delta_r s$ and eigenvalue $-\theta_r$ given by  
\be \label{sr}
\delta_r s = s'_* \qquad \theta_r = {\textstyle{\frac{1}{2}}}(d-2k+\eta).
\ee 
Let us now linearize the general equation \eqref{bs} around its fixed point
\be \label{lbs}
\left.\frac{\delta\beta_{s}}{\delta s}\right|_{s_*} \hspace{-7pt}\!\cdot \!\delta s = -d\, \delta s +  {\textstyle{\frac{1}{2}}}(d-2k+\eta)\, \varphi \!\cdot \!\delta s' + \left.\frac{\delta\beta_{S}}{\delta s}\right|_{s_*}\hspace{-7pt} \!\cdot \!\delta s.
\ee
It is straightforward to check that a perturbation of the form $\delta\tilde{\mathcal{L}} = \tilde{g}_1 \varphi$ which  depends on a single dimensionless coupling $\tilde g_1$ is in fact an eigenperturbation. This is because the last term on the r.h.s of \eqref{lbs} vanishes for this variation as it contains at least second derivatives of $\delta s$, and because for variations linear in $\varphi$ we have $\varphi \!\cdot \!\delta s' = \delta s$. So one can identify the eigenvector $\delta_1 s$ and its corresponding eigenvalue $-\theta_1$ as
\be \label{s1}
\delta_1 s = \tilde{g}_1 \int_{\tilde x} \varphi \qquad \theta_1 = {\textstyle{\frac{1}{2}}}(d+2k-\eta).
\ee
One notices that quite generally $\theta_1+\theta_r=d$. 

Let us now verify these relations in the example that we just studied, i.e. $k=2$ theory with $n=2m+1$. The anomalous dimension $\eta$ is beyond the approximation made in this analysis and can therefore be ignored here. It is easy to see that the upper left entry in the stability matrix \eqref{sm} for $i=1$, which gives the scaling dimension of the coupling $g_1$ (defined in \eqref{vzexp}), is equal to $\theta_1 = \frac{1}{2}(d+4)$. The other scaling direction $s'_*$ translated to coupling space, has nonzero components only along $g_{4m+1}$ and $h_{2m-1}$ and in this two-dimensional space it is given by the vector $(2(2m+1)g, 2m h)$. One can use the fixed point equations \eqref{fpe} to verify explicitly that
\be 
\mathcal{M}_{0,4m+1}^{(2)}\cdot \left(\!\!\ba{c} 2(2m+1)g \\ 2m h \ea\!\!\right) = -{\textstyle{\frac{1}{2}}}(d-4) \left(\!\!\ba{c} 2(2m+1)g \\ 2m h \ea\!\!\right)
\ee 
where $g,h$ are understood to take their fixed point values. This is consistent with the eigenvalue $\theta_r$ in \eqref{sr}. For the particular fixed point \eqref{h} with $g=0$, this eigenvalue equation reduces to $\tilde \omega _{2m-1} = 2m\epsilon$ which can be verified independently using the general formula \eqref{omega}.

\subsection{Counter-terms of four-derivative operators} \label{ss:rg4der}

To compute the field anomalous dimension we need the counter-term for the kinetic term coefficient. In fact the diagrams of Figs \ref{melon}, \ref{zmelon} and \ref{zzmelon} for suitable values of $r$ (to be determined shortly) give the complete counter-terms, quadratic in the potential and the wavefunction, for the four-derivative operators. As a byproduct we can find apart from the field anomalous dimension the quadratic flows of all four derivative couplings. These are introduced into the Lagrangian as
\be 
\mathcal{L} = {\textstyle{\frac{1}{2}}}W_1(\phi) (\square\phi)^2 + {\textstyle{\frac{1}{2}}}W_2(\phi)(\partial\phi)^2 \square\phi + {\textstyle{\frac{1}{2}}}W_3(\phi)(\partial\phi)^4 +{\textstyle{\frac{1}{2}}}Z(\phi)(\partial\phi)^2 + V(\phi)
\ee
which specifies our choice of basis for the four-derivative operators. The one-half factors are simply a matter of convention. The kinetic term coefficient is fixed to $W_1(0) = 1$ so that the field is always in the canonical form. We will now proceed with the computation of the flow of these functions induced by $V,Z$.

\subsubsection{Counter-terms of the form $V^{(a)}\hspace{1pt}^2$}

Let us start with the quadratic contributions in the potential. This is found immediately from \eqref{ul} by setting $l=2$ and hence $r=4m+1$. The relevant diagram is
\begin{center}
\includegraphics[width=0.35\textwidth]{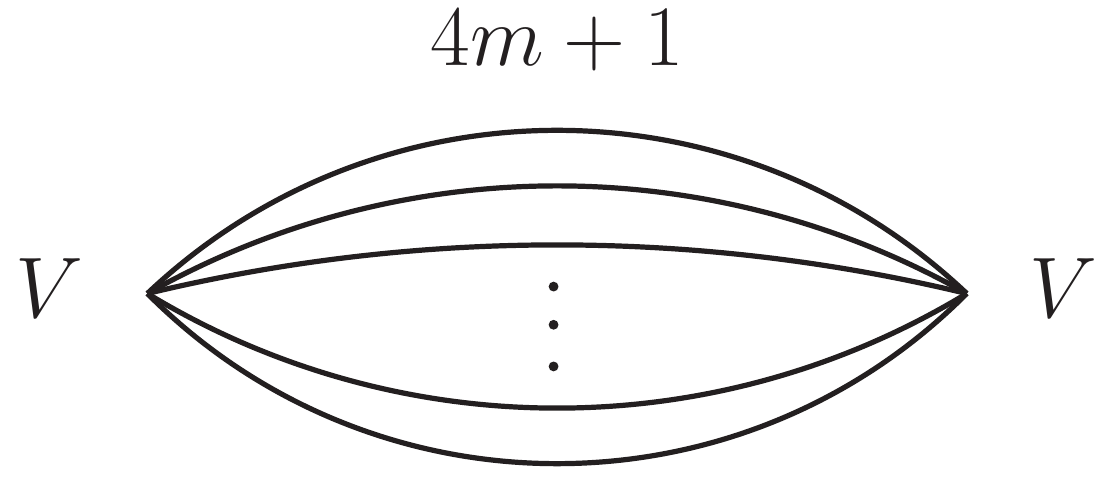}
\end{center}
which has $4m$-loops. After an integration by parts in $U_{2,c.t.}$ this evaluates to
\be \label{ct_vv_w}
\frac{1}{(4\pi)^{4(2m+1)}}\,
\frac{\Gamma^{4m+1}(\delta_c)}{\Gamma(4+\delta_c)}\,\frac{1}{8m\epsilon}\,\frac{(\square V^{(4m+1)})^2}{(4m+1)!}. 
\ee
To write this in the basis of four-derivative operators we use
\be 
\square V^{(4m+1)} =  V^{(4m+2)}\square\phi + V^{(4m+3)}(\partial\phi)^2.
\ee
This gives the following counter-term which has to be multiplied by $2\times 4m\epsilon$ to give the dimensionful beta functions
{\setlength\arraycolsep{2pt}
\bea 
&& \frac{1}{(4\pi)^{4(2m+1)}}\,\frac{\Gamma^{4m+1}(\delta_c)}{\Gamma(4+\delta_c)}\,\frac{1}{8m\epsilon}\,\frac{(V^{(4m+2)})^2}{(4m+1)!}\,(\square\phi)^2   \nn\\
&+& \frac{1}{(4\pi)^{4(2m+1)}}\,\frac{\Gamma^{4m+1}(\delta_c)}{\Gamma(4+\delta_c)}\,\frac{1}{4m\epsilon}\,\frac{V^{(4m+2)}V^{(4m+3)}}{(4m+1)!}\,\square\phi(\partial\phi)^2 \nn\\
&+& \frac{1}{(4\pi)^{4(2m+1)}}\,\frac{\Gamma^{4m+1}(\delta_c)}{\Gamma(4+\delta_c)}\,\frac{1}{8m\epsilon}\,\frac{(V^{(4m+3)})^2}{(4m+1)!}\,(\partial\phi)^4 
\eea}%

\subsubsection{Counter-terms of the form $V^{(a)}Z^{(b)}$}

The contributions of the form $V^{(a)}Z^{(b)}$ come from the diagram of Fig.\ref{zmelon} with $r=3m+1$
\begin{center}
\includegraphics[width=0.35\textwidth]{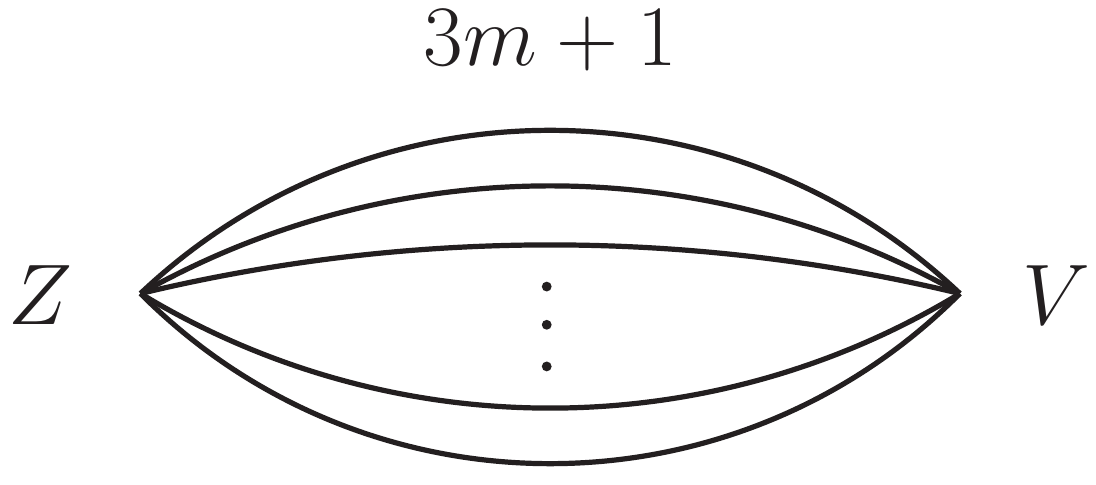}
\end{center}
This value of $r$ can easily be found from the constraint relation \eqref{constraint} for $N=2$, $a_1+a_2=1$, and $a=2$. This $3m$-loop diagram is directly calculated from \eqref{vz} by setting $r=3m+1$
{\setlength\arraycolsep{2pt}
\bea 
&& \frac{1}{(4\pi)^{3(2m+1)}}\,\frac{\Gamma^{3m+1}(\delta_c)}{\Gamma(3+\delta_c)} \; \frac{1}{(3m+1)!} Z^{(3m+1)}(\partial\phi)^2 \,\frac{1}{3m\epsilon}\, \square V^{(3m+1)}  \nn\\
&-& \frac{1}{(4\pi)^{3(2m+1)}}\,\frac{\Gamma^{3m+1}(\delta_c)}{\Gamma(3+\delta_c)} \; \frac{1}{(3m+1)!} \square Z^{(3m-1)} \,\frac{1}{6m\epsilon}\,\square V^{(3m+1)} 
\eea}%
To express this in the basis chosen for the four-derivative operators we use
{\setlength\arraycolsep{2pt}
\bea 
\square V^{(3m+1)} &=& V^{(3m+2)} \square \phi + V^{(3m+3)} (\partial\phi)^2 \nn\\
\square Z^{(3m-1)} &=& Z^{(3m)} \square \phi + Z^{(3m+1)} (\partial\phi)^2.
\eea}%
This gives the counter-terms for the three functions $W_1$, $W_2$ and $W_3$ 
{\setlength\arraycolsep{2pt}
\bea 
&-& \frac{1}{(4\pi)^{3(2m+1)}}\,\frac{\Gamma^{3m+1}(\delta_c)}{\Gamma(3+\delta_c)} \; \frac{1}{(3m+1)!}\,\frac{1}{6m\epsilon}  Z^{(3m)} \, V^{(3m+2)} (\square \phi)^2 \nn\\
&+& \frac{1}{(4\pi)^{3(2m+1)}}\,\frac{\Gamma^{3m+1}(\delta_c)}{\Gamma(3+\delta_c)} \;\frac{1}{(3m+1)!} \,\frac{1}{6m\epsilon}\, (Z^{(3m+1)} \, V^{(3m+2)}-Z^{(3m)} \, V^{(3m+3)} )\, (\partial\phi)^2\square\phi \nn\\
&+&  \frac{1}{(4\pi)^{3(2m+1)}}\,\frac{\Gamma^{3m+1}(\delta_c)}{\Gamma(3+\delta_c)} \;\frac{1}{(3m+1)!} \,\frac{1}{6m\epsilon}\, Z^{(3m+1)} \, V^{(3m+3)}\, (\partial\phi)^4 
\eea}%
which have to be multiplied by $2\times 3m\epsilon$ to give their flows.

\subsubsection{Counter-terms of the form $Z^{(a)}Z^{(b)}$}

Counter-terms quadratic in $Z$ are also present for $W_i$. These come from the $2m$-loop diagram
\begin{center}
\includegraphics[width=0.35\textwidth]{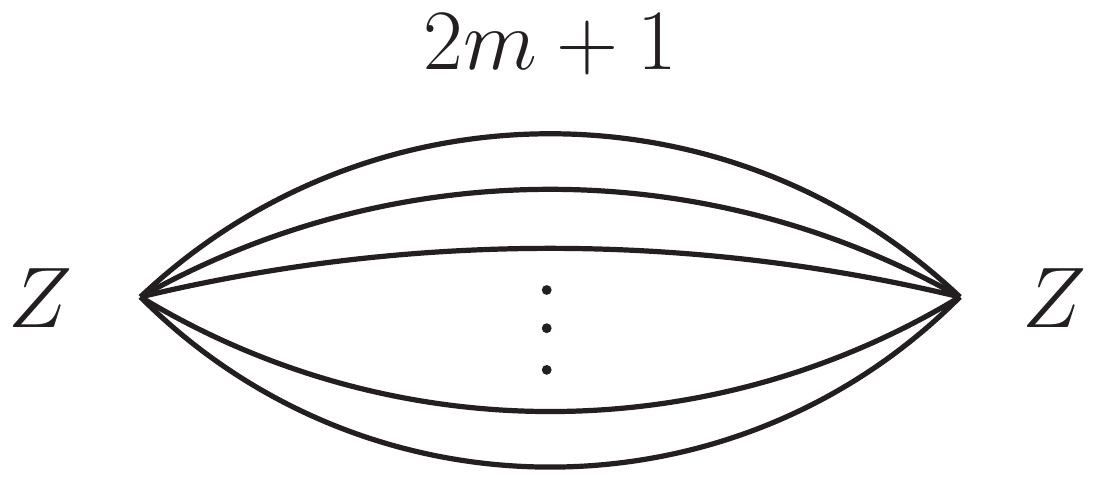}
\end{center}
This can be computed using Eq.\eqref{zz} for $r=2m+1$. Before that we need the pole terms 
\be 
\int_x \!e^{ip\cdot x}\, G^{2m+1}_x = \frac{1}{(4\pi)^{2(2m+1)}}\,
\frac{\Gamma^{2m+1}(\delta_c)}{\Gamma(\delta_c+2)}\,\frac{1}{m\epsilon} + \mathrm{finite}
\ee
\be   
\int_x \!e^{ip\cdot x} \; (\partial_\mu\partial_\nu G_x)^2G^{2m-1} = \frac{1}{(4\pi)^{2(2m+1)}}\,\frac{m(2m+3)+2}{4m^5\epsilon}\;\frac{\Gamma(\delta_c)^{2m+1}}{\Gamma(4+\delta_c)}\,p^4 + \mathrm{finite}
\ee
\be   
\int_x \!e^{ip\cdot x} \; \partial_\mu\partial_\nu G_x\,\partial^\mu G\,\partial^\nu G_x \, G^{2m-2} = \frac{1}{(4\pi)^{2(2m+1)}}\,\frac{m+2}{4m^5\epsilon}\;\frac{\Gamma(\delta_c)^{2m+1}}{\Gamma(4+\delta_c)}\,p^4 + \mathrm{finite}
\ee
\be   
\int_x \!e^{ip\cdot x} \; (\partial G)^4 G^{2m-3} = \frac{1}{(4\pi)^{2(2m+1)}}\,\frac{1}{2m^5\epsilon}\;\frac{\Gamma(\delta_c)^{2m+1}}{\Gamma(4+\delta_c)}\,p^4 + \mathrm{finite}
\ee
\be  
\int_x \!e^{ip\cdot x}\, (\partial_x G_x)^2 G^{2m-1}_x = -\frac{1}{(4\pi)^{2(2m+1)}}\,\frac{\delta_c}{2m+1}\,\frac{\Gamma(\delta_c)^{2m+1}}{\Gamma(\delta_c+2)}\,\frac{1}{m\epsilon}\,p^2 + \mathrm{finite}
\ee
which can be obtained form \eqref{Grp}, \eqref{r-2}, \eqref{r-3}, \eqref{r-4} and \eqref{mel} at $r=2m+1$. For instance these are used to obtain the singular factor in the fourth term of \eqref{zz}
\be
\ba{c}
2(\partial_\mu\partial_\nu G_x)^2G^{2m-1}+4(2m-1)\partial_\mu\partial_\nu G_x\,\partial^\mu G\,\partial^\nu G_x \, G^{2m-2}+(2m-1)(2m-2)(\partial G)^4 G^{2m-3} \\[2mm]
\displaystyle =  \frac{1}{(4\pi)^{2(2m+1)}}\, \frac{\Gamma(\delta_c)^{2m+1}}{\Gamma(4+\delta_c)}\,\frac{10m+3}{2m^4\epsilon}\,\square^2_x\delta_{xy} + \mathrm{finite}.
\ea
\ee
The final form of the four-derivative counter-term is 
{\setlength\arraycolsep{2pt}
\bea
&-& \frac{1}{8m(3m+1)(2m)!\epsilon}
\;\frac{1}{(4\pi)^{2(2m+1)}}\,\frac{\Gamma(\delta_c)^{2m+1}}{\Gamma(2+\delta_c)}
\,(Z^{(2m)})^2(\square\phi)^2   \nn\\
&-& \frac{1}{4m(3m+1)(2m)!\epsilon}
\;\frac{1}{(4\pi)^{2(2m+1)}}\,\frac{\Gamma(\delta_c)^{2m+1}}{\Gamma(2+\delta_c)}
\,Z^{(2m)}Z^{(2m+1)}(\partial\phi)^2\square\phi \nn\\
&+& \frac{1}{8(3m+1)(2m+1)!\epsilon}\;\frac{1}{(4\pi)^{2(2m+1)}}\;\frac{\Gamma(\delta_c)^{2m+1}}{\Gamma(2+\delta_c)}\,(Z^{(2m+1)})^2 (\partial\phi)^4  
\eea}%
These are multiplied by $2 \times 2m\epsilon$ to give the beta function contribution.

\subsection{Beta functions of four-derivative couplings and $\gamma_\phi$} 

Summing up all the contributions and reading off the coefficients of the basis operators we find the dimensionful beta functions for $W_i$. After rescaling the potential and the wavefunction according to \eqref{vzres} these beta functions, which refer to the $\eta$-independent parts of the corresponding flows, are
{\setlength\arraycolsep{2pt}
\bea 
\beta_{W_1} &=& \frac{(m+1)^2\Gamma(\delta_c)}{m^4\Gamma(4+\delta_c)}\,\frac{V^{(4m+2)}\hspace{1pt}^2}{(4m+1)!} -\frac{(m+1)\Gamma(\delta_c)}{m^3\Gamma(3+\delta_c)} \; \frac{Z^{(3m)} \, V^{(3m+2)}}{(3m+1)!} \nn\\
&-& \frac{\Gamma(\delta_c)}{m^2\Gamma(2+\delta_c)}\, \frac{Z^{(2m)}\hspace{1pt}^2}{2(3m+1)(2m)!}
\eea}%

{\setlength\arraycolsep{2pt}
\bea 
\beta_{W_2} &=& \frac{2(m+1)^2\Gamma(\delta_c)}{m^4\Gamma(4+\delta_c)}\,\frac{V^{(4m+2)}V^{(4m+3)}}{(4m+1)!} + \frac{(m+1)\Gamma(\delta_c)}{m^3\Gamma(3+\delta_c)} \;\frac{Z^{(3m+1)} \, V^{(3m+2)}-Z^{(3m)} \, V^{(3m+3)}}{(3m+1)!} \nn\\
&-& \frac{\Gamma(\delta_c)}{m^2\Gamma(2+\delta_c)}\, \frac{Z^{(2m)}Z^{(2m+1)}}{(3m+1)(2m)!} 
\eea}%

{\setlength\arraycolsep{2pt}
\bea 
\beta_{W_3} &=& \frac{(m+1)^2\Gamma(\delta_c)}{m^4\Gamma(4+\delta_c)}\,\frac{V^{(4m+3)}\hspace{1pt}^2}{(4m+1)!} + \frac{(m+1)\Gamma(\delta_c)}{m^3\Gamma(3+\delta_c)} \;\frac{Z^{(3m+1)}\,V^{(3m+3)}}{(3m+1)!} \nn\\
&+& \frac{\Gamma(\delta_c)}{\Gamma(2+\delta_c)}\, \frac{Z^{(2m+1)}\hspace{1pt}^2}{2m(3m+1)(2m+1)!}
\eea}%
The dimensionless versions of these functions are defined as
\be \label{wdimless}
w_1(\varphi) = W_1(\mu^\delta\,\varphi), \qquad w_2(\varphi) = \mu^{\!-\delta}\, W_2(\mu^\delta\,\varphi), \qquad w_3(\varphi) = \mu^{\!-2\delta}\, W_3(\mu^\delta\,\varphi).
\ee
Their beta functions are related to those of the dimensionful ones in the following way
{\setlength\arraycolsep{2pt}
\bea  \label{bwdl}
\beta_{w_1} &=& 2\gamma_\phi\, w_1 + (\delta + \gamma_\phi)\,\varphi\, w^{(1)}_1 + \beta_{W_1}, \nn\\
\beta_{w_2} &=& (\delta +3\gamma_\phi)\,w_2 + (\delta + \gamma_\phi)\,\varphi\, w^{(1)}_2 + \mu^\delta\,\beta_{W_2}, \nn\\
\beta_{w_3} &=& (2\delta +4\gamma_\phi)\, w_3 + (\delta + \gamma_\phi)\,\varphi\, w^{(1)}_3 + \mu^{2\delta}\,\beta_{W_3}.
\eea}%
The field anomalous dimension is found by requiring that $\beta_{w_1}(\varphi=0)=0$
\be 
\gamma_\phi = \frac{\Gamma(\delta_c)}{m^2\Gamma(2+\delta_c)}\, \frac{(2m)!}{4(3m+1)}h^2 -\frac{\Gamma(\delta_c)}{m^4\Gamma(4+\delta_c)}2(m+1)^2(2m+1)^2(4m+1)!\,g^2.
\ee
Except for the first appearance of $\gamma_\phi$ as the coefficient of $w_1$ the terms involving $\gamma_\phi$ in \eqref{bwdl} are cubic order in the couplings and must therefore be omitted at the quadratic approximation considered here. 
At the infrared fixed point where $g=0$ the anomalous dimension $\eta = 2\gamma_\phi$ takes the following form in terms of $\epsilon$ 
\be   \label{eta2ep}
\eta = \frac{\Gamma(\delta_c)}{\Gamma(2\!+\!\delta_c)}\,\frac{2(m\!+\!1)^2(2m\!+\!1)^2}{(3m\!+\!1)(2\!+\!7m(m\!+\!1))^2}\frac{(m\!+\!1)!^2m!^4}{(2m)!^3}\epsilon^2.
\ee

\section{The $k=2$, $n=3$ example: including higher order corrections} \label{s:type2.exmpl}

The quadratic beta functions calculated in the previous section have been used to obtain the field and coupling anomalous dimensions at leading order in $\epsilon$. For the OPE coefficients instead they are capable of giving only the free theory value which can be obtained by Wick counting. If we wish to extend this and find some OPE coefficients beyond free theory we need some cubic corrections to the beta functions as well. For instance if we are interested in computing the OPE coefficients of non-derivative operators we must find cubic corrections to $\beta_V$ that are at least quadratic in the potential. These include corrections of the form $V^{(a)}V^{(b)}V^{(c)}$ which are cubic in the potential. However as we saw earlier the function $Z$ is non zero at the scale invariant point so corrections of the form $V^{(a)}V^{(b)}Z^{(c)}$ will also contribute to such OPE coefficients. In this section we extend the results of the previous section by calculating cubic corrections to the beta function of the potential that are of the two forms mentioned above. Although straightforward to extend to general values of $m$, for this calculation we find it more instructive to restrict ourselves to the special case of $m=1$.

\subsection{Counter-terms of the form $V^{(a)}V^{(b)}V^{(c)}$}

Here we present the diagrams that contribute to cubic terms of the form $V^{(a)}V^{(b)}V^{(c)}$ in $\beta_V$. Each diagram is followed by the expression for its UV divergence. We do not enter into the details of the calculations here and instead as an example we provide in the appendix the computational steps for one of the diagrams. For cubic contributions to the counter-term of the potential the number of propagators in the relevant diagram is $2n=6$. There are three triangle diagrams with six propagators which we will now consider. The first diagram has two propagators in each edge. This falls in the first class described in App.\eqref{ss:tri} for which non of the edges is divergent and therefore it has a simple pole in $\epsilon$. It can be evaluated using equations \eqref{tricomp} and \eqref{gold}
\begin{center}
\includegraphics[width=0.34\textwidth]{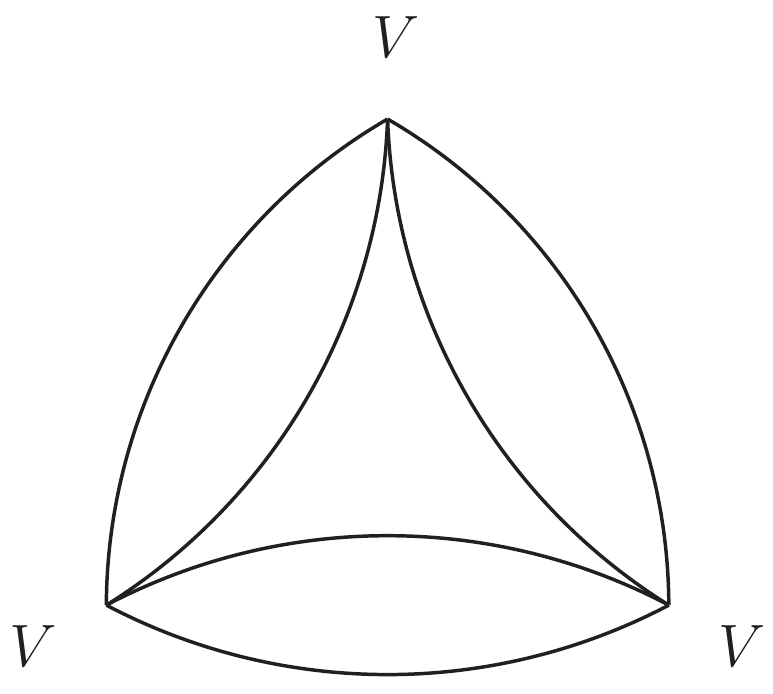}
\end{center} 
\be 
-\frac{1}{(4\pi)^{12}}\frac{1}{192\epsilon}\,V^{(4)}\hspace{1pt}^3.
\ee
The poles in $y,z$ in the Mellin-Barnes representation \eqref{tricomp} that give rise to divergences in this case are $(y,z)=(0,0)$. The next triangle diagram has an edge with three propagators. This leads to a subdivergence and a double $\epsilon$-pole. Again the relevant poles in the Mellin-Barnes integral are $(y,z)=(0,0)$ which give 
\begin{center}
\includegraphics[width=0.34\textwidth]{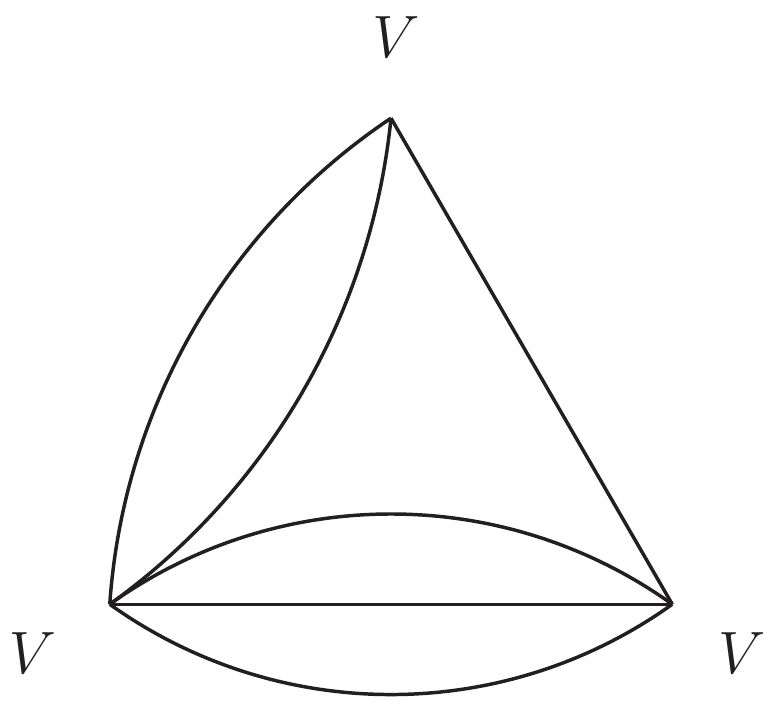} \nn
\end{center}
\be 
-\frac{1}{(4\pi)^{12}}\frac{1}{96}\left[\frac{1}{\epsilon^2}+\frac{5-2\gamma}{\epsilon}\right]  V^{(3)}V^{(4)}V^{(5)}.
\ee
The last diagram of the triangle type, shown below, has a subdivergence which comes from a melon subdiagram with four propagators. Such a subdiagram has a singularity that contributes to two-derivative operators quadratic in the potential. This has to be canceled by inverse powers of the derivative and therefore leads to non-local UV divergences in the diagram. This nonlocal contribution comes from the poles $(y,z)=(1,0)$ and $(y,z)=(0,1)$ in the Mellin-Barnes integral while the pole $(y,z)=(0,0)$ is responsible for the local part similar to the previous two cases
\begin{center}
\includegraphics[width=0.35\textwidth]{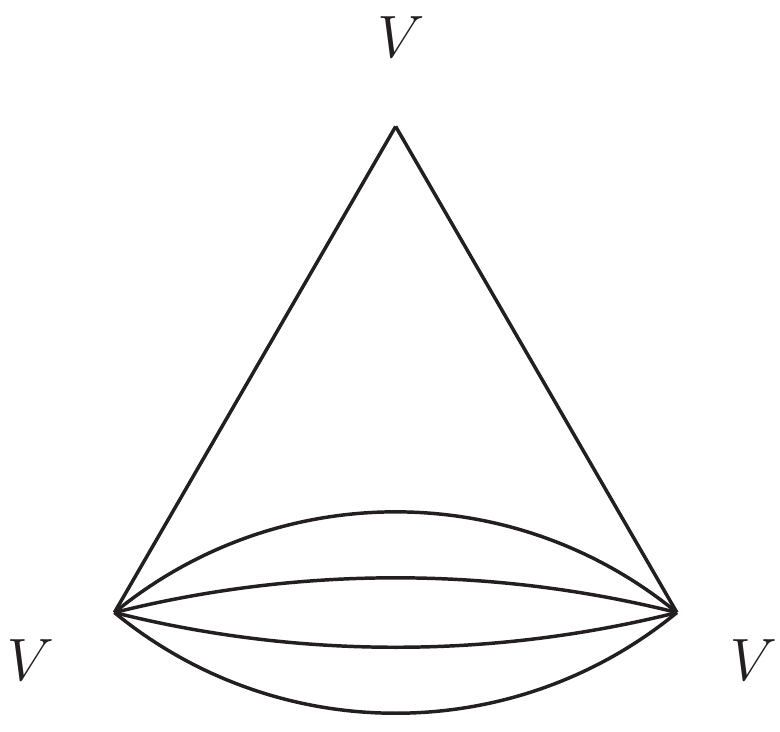}
\end{center}
\be 
\frac{1}{(4\pi)^{12}}\frac{1}{1728}\left[\frac{1}{\epsilon^2}+\left(\frac{59}{12}-2\gamma\right)\frac{1}{\epsilon}\right]  V^{(2)}V^{(5)}\hspace{1pt}^2 \,+\, \frac{1}{(4\pi)^{12}}\frac{1}{432\epsilon}\, V^{(5)}\, \Box V^{(5)} \,\frac{1}{\Box} V^{(2)} .
\ee
In addition to triangle diagrams there are cubic diagrams of the double-melon type discussed in Sect.\ref{ss:dmelon}. In the present case of $n=3$ there are two diagrams of this kind. The first consists of two melons with three propagators each, connected together. A melon with three propagators is singular and therefore such a double-melon diagram has an $\epsilon^2$ pole
\begin{center}
\includegraphics[width=0.56\textwidth]{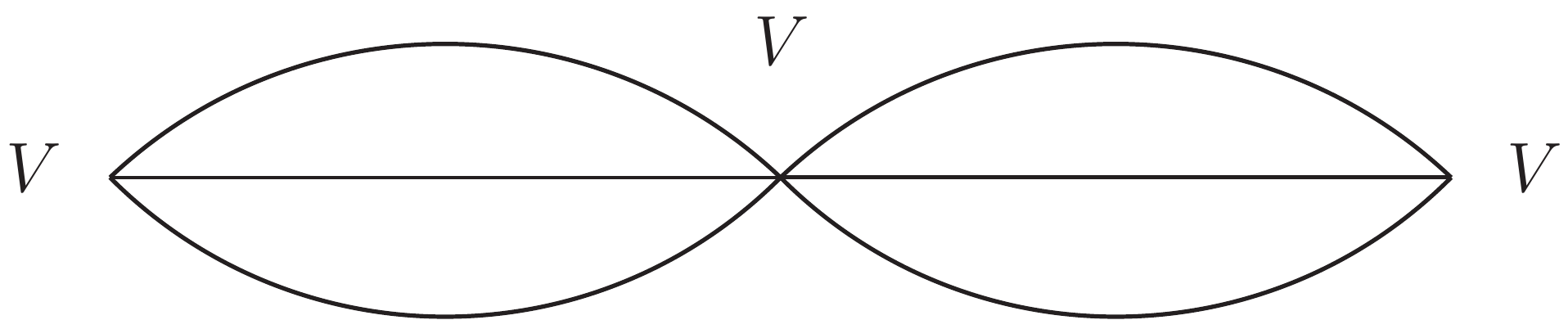}
\end{center}
\be 
-\frac{1}{(4\pi)^{12}}\frac{1}{288}\left[\frac{1}{\epsilon^2}+\left(\frac{9}{2}-2\gamma\right)\frac{1}{\epsilon}\right]  V^{(3)}\hspace{1pt}^2\,V^{(6)}
\ee
The second double-melon diagram has a melon subdiagram with two propagators and a melon subdiagram with four propagators. The one with four propagators has a singularity that contributes to two-derivative operators while that with two propagators is finite but involves two inverse derivatives. This diagram therefore gives a completely non-local term
\begin{center}
\includegraphics[width=0.56\textwidth]{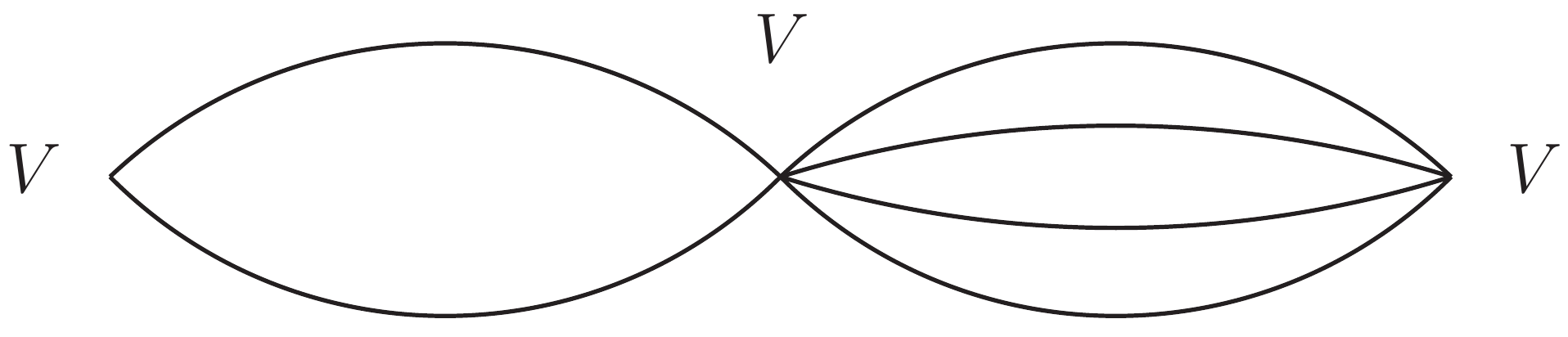}
\end{center}
\be
\frac{1}{(4\pi)^{12}}\frac{1}{432\epsilon}\, V^{(6)}\, \Box V^{(4)} \,\frac{1}{\Box} V^{(2)} 
\ee
Finally let us come to the melon diagrams that give rise to cubic corrections in the beta functions. These diagrams include a counter-term vertex. The counter-term vertices are those discussed in Sect.\ref{ss:vv}, which in this example must be evaluated for $m=1$. They read
{\setlength\arraycolsep{2pt}
\bea
U_{0,c.t.} &=& \frac{1}{(4\pi)^{6}}\,\frac{1}{24\epsilon}\,(V^{(3)})^2, \label{u0} \\ 
U_{1,c.t.} &=& -\frac{1}{(4\pi)^{9}}\,\frac{1}{432\epsilon}\,V^{(5)}\hspace{1pt}^2(\partial\phi)^2, \label{u1} 
%
\eea}%
The first diagram we will consider has three propagators and its counter-term vertex is $U_{0,c.t.}$ given in Eq.\eqref{u0}. The divergence part of this is evaluated as
\begin{center}
\includegraphics[width=0.35\textwidth]{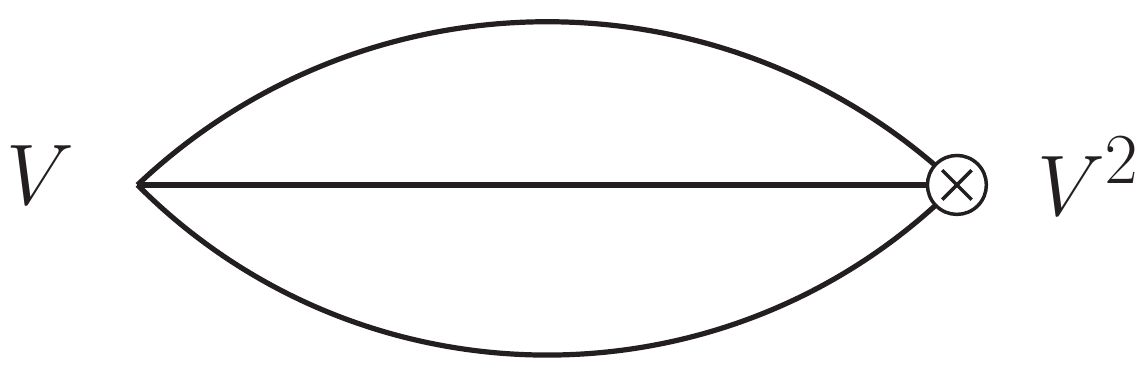}
\end{center}
\be 
\frac{1}{(4\pi)^{12}}\frac{1}{288}\left[\frac{1}{\epsilon^2}+\left(\frac{9}{4}-\gamma\right)\frac{1}{\epsilon}\right]  \big[V^{(3)}\hspace{1pt}^2\big]^{(3)}\,V^{(3)}
\ee
The second diagram involves the counter-term vertex $U_{1,c.t.}$ given in Eq.\eqref{u1} and has two propagators. Its divergence is given by the following expression which includes two non-local terms
\begin{center}
\includegraphics[width=0.35\textwidth]{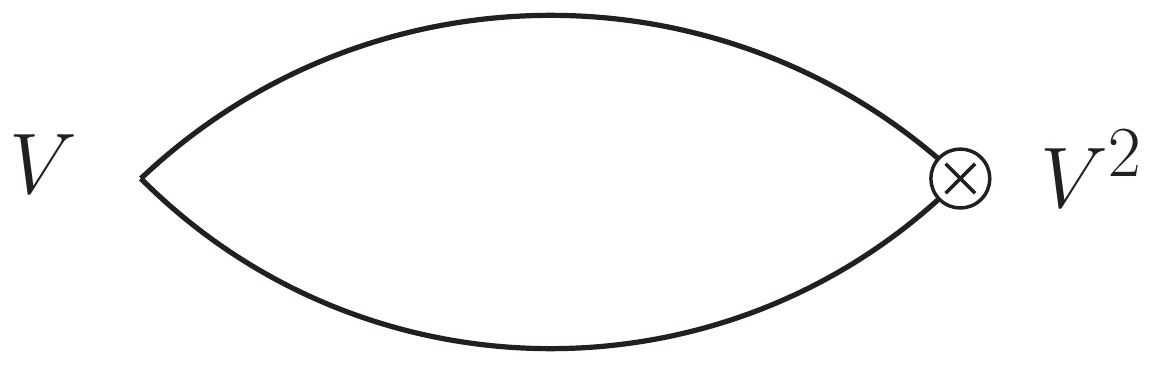}
\end{center}
{\setlength\arraycolsep{2pt}
\bea
&-& \frac{1}{(4\pi)^{12}}\frac{1}{432}\left[\frac{1}{\epsilon^2}+\frac{1-\gamma}{2\epsilon}\right]  V^{(5)}\hspace{1pt}^2\,V^{(2)} \nn\\
&-& \frac{1}{(4\pi)^{12}}\frac{1}{432\epsilon} \,V^{(5)} \,\Box V^{(5)} \,\frac{1}{\Box}V^{(2)} -\frac{1}{(4\pi)^{12}}\frac{1}{432\epsilon} \,V^{(6)} \,\Box V^{(4)} \,\frac{1}{\Box}V^{(2)}.
\eea}%
As expected no non-local terms must be present in the total counter-term. Indeed one can check that the non-local terms cancel among the diagrams. Moreover the terms proportional to $\gamma$ are also seen to cancel out. Summing up all these contributions and simplifying a bit the total counter-term that is cubic in the potential reads
{\setlength\arraycolsep{2pt}
\bea
V^{v^3}_{c.t.}  &=&- \frac{1}{(4\pi)^{12}}\frac{1}{192\epsilon}\,V^{(4)}\hspace{1pt}^3 
+ \frac{1}{(4\pi)^{12}}\frac{1}{96}\left(\frac{1}{\epsilon^2}-\frac{1}{2\epsilon}\right)  V^{(3)}V^{(4)}V^{(5)} \nn\\
&& - \frac{1}{(4\pi)^{12}}\frac{1}{576}\left(\frac{1}{\epsilon^2}-\frac{35}{36\epsilon}\right)  V^{(2)} \,V^{(5)}\hspace{1pt}^2
+\frac{1}{(4\pi)^{12}}\frac{1}{288\epsilon^2}  V^{(3)}\hspace{1pt}^2\,V^{(6)}.
\eea}%

\subsection{Counter-terms of the form $V^{(a)}V^{(b)}Z^{(c)}$}

Apart from cubic corrections in the potential it is necessary for the computation of OPE coefficients to obtain also cubic corrections of the form $V^{(a)}V^{(b)}Z^{(c)}$ in $\beta_V$. This is because for $\Box^2$ theories of the second type the function $Z$ does not vanish at the fixed point and therefore these terms can also contribute to the OPE coefficients. Here we present the diagrams leading to such terms along with the expression for their UV divergence. Computational details for a sample diagram are presented in the Appendix \ref{s:vvz.sample}. To extract the divergences we continue using the Mellin-Barnes representation. Compared to the cubic case in the potential the computation is more involved as the function $Z$ can appear in three different ways in each vertex. In order not to increase the number of diagrams these are not distinguished here. Furthermore for a diagram with given topology the function $Z$ can appear on different vertices which leads to different diagrams. By dimensional analysis a diagram with two $V$-vertices and a $Z$-vertex contributes to the flow of the potential if its number of propagators is $2n-1=5$. There are four triangle diagrams of this sort, which come in two different topologies. In the first diagram the $Z$ function appears at a four-vertex. In this case it turns out that only the cases in which both derivatives at the $Z$-vertex act on the propagator contribute to the divergence. The results is
\begin{center}
\includegraphics[width=0.34\textwidth]{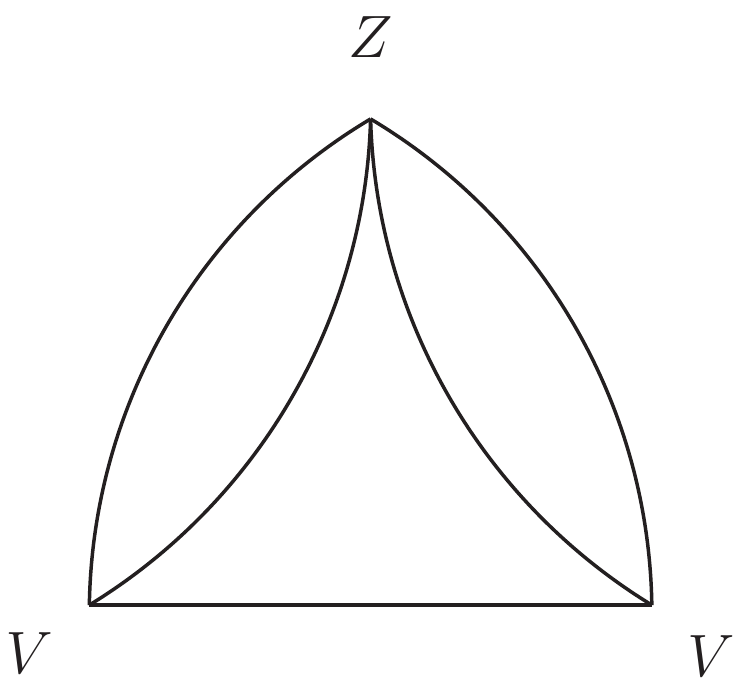}
\end{center}
\be 
-\frac{1}{(4\pi)^9}\frac{1}{4}\left[\frac{1}{3\epsilon^2}+\left(\frac{7}{6}-\frac{\gamma}{2}\right)\frac{1}{\epsilon}\right] Z^{(2)} V^{(3)}\hspace{1pt}^2
\ee
The second diagram is topologically the same as the first one except that the $Z$ function appears at a three-vertex. Similar to the previous case, the divergence receives contributions only when derivatives at the $Z$-vertex  act on the propagators. We get
\begin{center}
\includegraphics[width=0.34\textwidth]{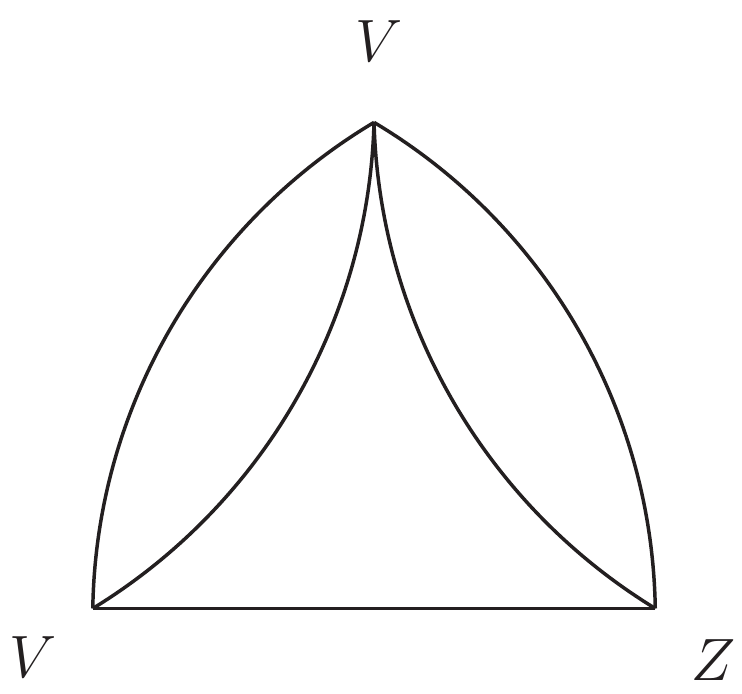}
\end{center}
\be 
-\frac{1}{(4\pi)^9}\frac{1}{4}\left[\frac{1}{3\epsilon^2}+\left(\frac{4}{3}-\frac{\gamma}{2}\right)\frac{1}{\epsilon}\right] Z^{(1)} V^{(3)}V^{(4)}
\ee
The other two triangle diagrams are also topologically the same except that in one the function $Z$ appears on the two-vertex while in the other it appears on the four-vertex. In both cases the diagram has a non-local divergence. In the first case the pole term is
\begin{center}
\includegraphics[width=0.35\textwidth]{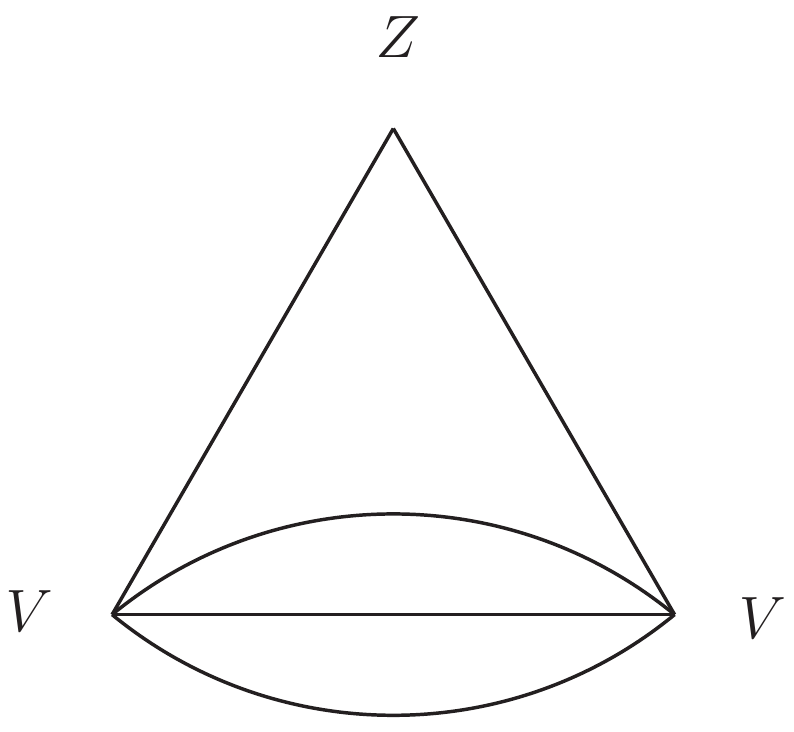} 
\end{center}
\be 
-\frac{1}{(4\pi)^9}\frac{1}{24}\left[\frac{1}{3\epsilon^2}+\left(\frac{7}{4}-\frac{\gamma}{2}\right)\frac{1}{\epsilon}\right] Z\,V^{(4)}\hspace{1pt}^2 + \frac{1}{(4\pi)^9}\frac{1}{48\epsilon}\, \frac{1}{\Box} \left(Z^{(2)}\,(\partial\phi)^2\right)V^{(4)}\hspace{1pt}^2
\ee
which includes a double pole. This is because of the subdivergence in the melon with three propagators and the overall divergence. In the second diagram there is only a simple pole from the melon subdiagram and again there are non-local divergences as well
\begin{center}
\includegraphics[width=0.35\textwidth]{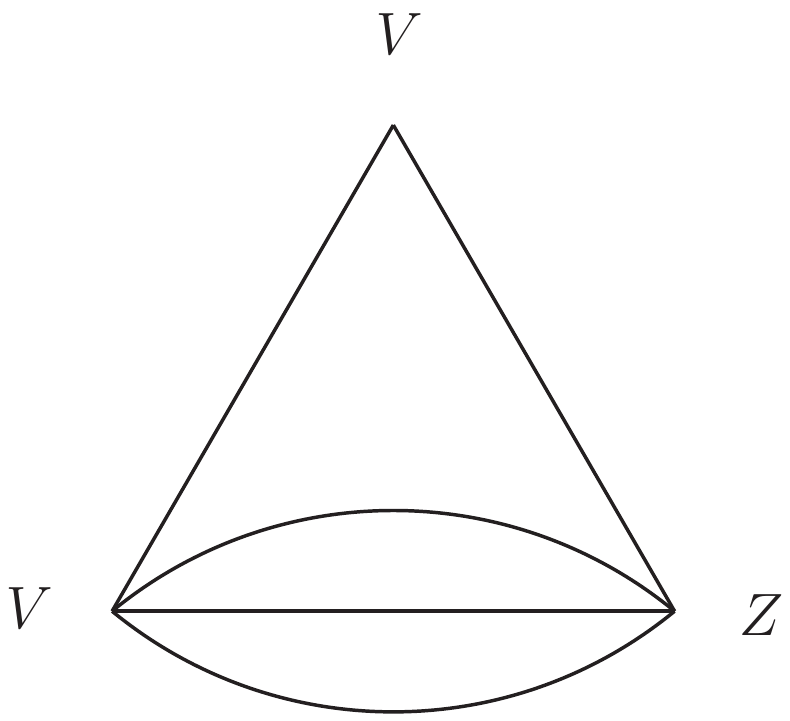}
\end{center}
{\setlength\arraycolsep{2pt}
\bea
-\frac{1}{(4\pi)^9}\frac{1}{144\epsilon}\, Z^{(2)} V^{(2)}V^{(4)} &+& \frac{1}{(4\pi)^9}\frac{1}{48\epsilon}\, Z^{(2)} \Box V^{(4)} \frac{1}{\Box} V^{(2)} \nn\\
&-& \frac{1}{(4\pi)^9}\frac{1}{48\epsilon}\, \Box Z^{(2)} V^{(4)} \frac{1}{\Box} V^{(2)} \nn\\
&+& \frac{1}{(4\pi)^9}\frac{1}{24\epsilon}\,  Z^{(4)} V^{(4)}\,(\partial\phi)^2 \, \frac{1}{\Box} V^{(2)} \label{113z4uv}
\eea}%
This diagram is analyzed in detail in Appendix \ref{s:vvz.sample}. Let us now consider double-melon diagrams. With five propagators such diagrams can take only one topologically. This consists of two melon diagrams one with two propagators and the other with three propagators connected together through a vertex. Given this topology, the function $Z$ can appear in either of the three vertices, hence there are three diagrams of this type altogether. 
When the $Z$ function appears on the two-vertex the pole terms is given as
\begin{center}
\includegraphics[width=0.55\textwidth]{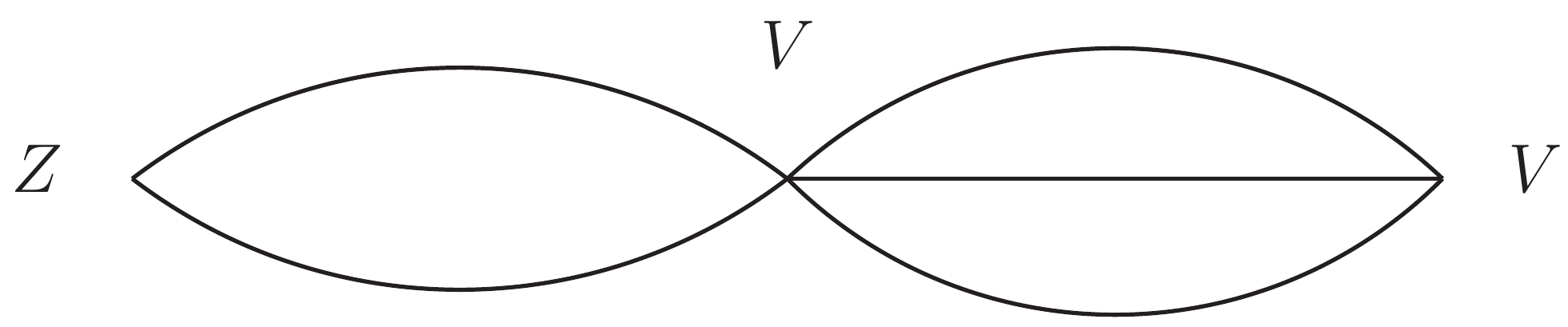}
\end{center}
\be 
-\frac{1}{(4\pi)^9}\frac{1}{24}\left[\frac{1}{\epsilon^2}+\left(\frac{15}{4}-\frac{3\gamma}{2}\right)\frac{1}{\epsilon}\right] Z\,V^{(3)}V^{(5)} + \frac{1}{(4\pi)^9}\frac{1}{48\epsilon}\, \frac{1}{\Box} \left(Z^{(2)}\,(\partial\phi)^2\right)V^{(3)}V^{(5)}
\ee
The function $Z$ can also appear on the five-vertex which leads to the $\epsilon$ pole
\begin{center}
\includegraphics[width=0.55\textwidth]{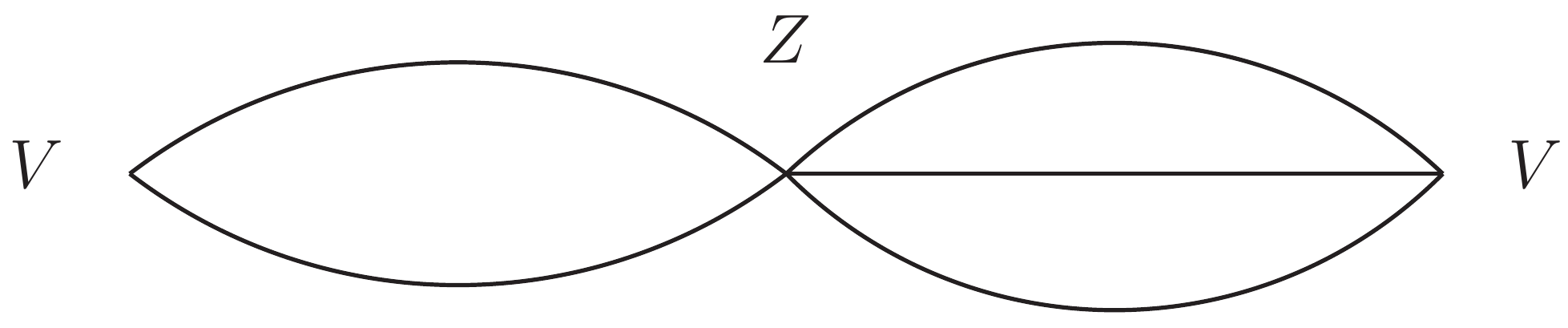}
\end{center}
{\setlength\arraycolsep{2pt}
\bea
-\frac{1}{(4\pi)^9}\frac{1}{24}\left[\frac{1}{\epsilon^2}+\left(\frac{11}{4}-\frac{3\gamma}{2}\right)\frac{1}{\epsilon}\right] Z^{(3)} V^{(2)}V^{(3)} &+& \frac{1}{(4\pi)^9}\frac{1}{48\epsilon}\,  Z^{(5)} V^{(3)}\,(\partial\phi)^2 \, \frac{1}{\Box} V^{(2)} \nn\\
&-& \frac{1}{(4\pi)^9}\frac{1}{24\epsilon}\, \Box Z^{(3)} V^{(3)} \frac{1}{\Box} V^{(2)} \nn\\
&-& \frac{1}{(4\pi)^9}\frac{1}{24\epsilon}\,Z^{(4)} V^{(4)}\,(\partial\phi)^2 \, \frac{1}{\Box} V^{(2)} 
\eea}%
Finally, if the three-vertex comes from two-derivative interactions parameterized by the $Z$ function then the pole term will be
\begin{center}
\includegraphics[width=0.55\textwidth]{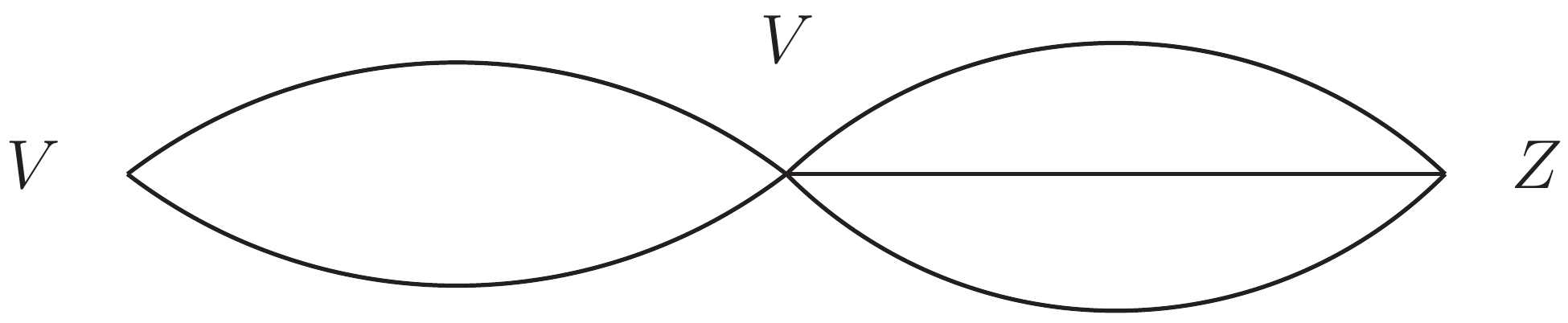}
\end{center}
\be 
\frac{1}{(4\pi)^9}\frac{1}{48\epsilon}\,  Z^{(3)} V^{(5)}\,(\partial\phi)^2 \, \frac{1}{\Box} V^{(2)}
\ee
What remains to consider is the non-trivial melon diagrams with a counter-term vertex. Let us take a diagram with a $Z$ vertex. The counter-term vertex will then have to be quadratic in the potential. Apart from the case with a single propagator just considered, To provide a counter-term for the potential the only possibility is that the diagram have two propagators and there be no derivatives on the counter-term vertex. The counter-term quadratic in $V$ and with no derivatives is given in \eqref{u0}. The pole term for such a diagram is then given as
\begin{center}
\includegraphics[width=0.35\textwidth]{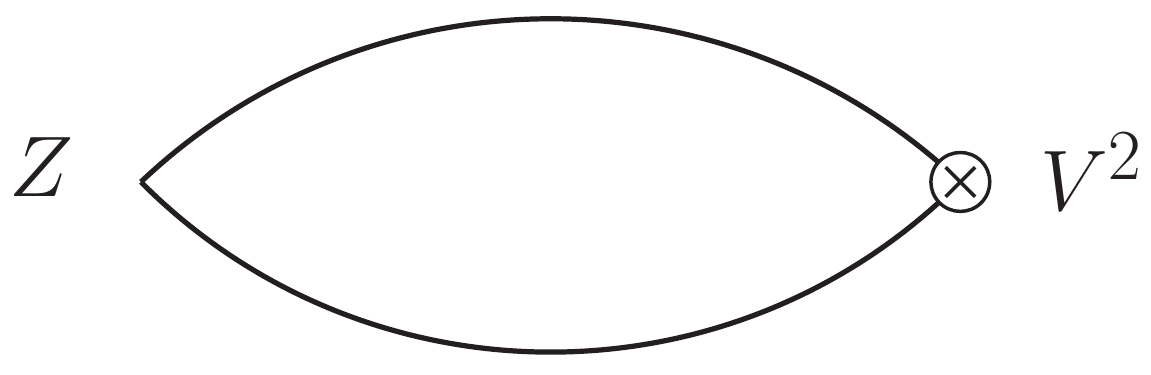}
\end{center}
\be 
\frac{1}{(4\pi)^9}\frac{1}{48}\left[\frac{1}{\epsilon^2}+\left(\frac{3}{2}-\frac{\gamma}{2}\right)\frac{1}{\epsilon}\right] Z\, \big[V^{(3)}\hspace{1pt}^2\big]^{(2)} - \frac{1}{(4\pi)^9}\frac{1}{96\epsilon}\, Z^{(2)}\,(\partial\phi)^2\, \frac{1}{\Box}\big[V^{(3)}\hspace{1pt}^2\big]^{(2)}
\ee
Next, consider a melon diagram with a $V$ vertex. Its quadratic counter-term vertex will then be of the form $V^{(a)}Z^{(b)}$. Apart from the trivial single propagator case considered above which can be omitted there are two other cases. The first has two propagators and a counter-term vertex with two derivatives. The counter-term vertex can be obtained from Eq.\eqref{zv_z} setting $m=1$
\be
\frac{1}{(4\pi)^{6}}\,\frac{1}{24\epsilon} \,V^{(3)} Z^{(3)} \,(\partial\phi)^2.
\ee
The pole of this diagram then reads
\begin{center}
\includegraphics[width=0.35\textwidth]{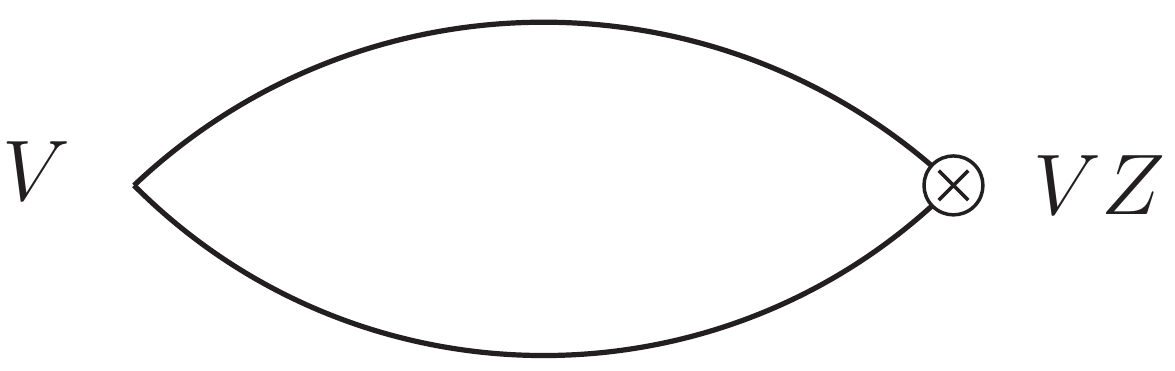}
\end{center}
\be 
\frac{1}{(4\pi)^9}\frac{1}{24}\left[\frac{1}{\epsilon^2}+\left(\frac{3}{2}-\frac{\gamma}{2}\right)\frac{1}{\epsilon}\right] Z^{(3)}\,V^{(2)}\,V^{(3)} - \frac{1}{(4\pi)^9}\frac{1}{48\epsilon}\, \big[Z^{(3)}V^{(3)}\big]^{(2)}(\partial\phi)^2\, \frac{1}{\Box}V^{(2)}.
\ee
In the second case the diagram has three propagators. The counter-term vertex then has no derivatives and is given by Eq.\eqref{zv_v} for $m=1$
\be  
\frac{1}{(4\pi)^{3}}\,\frac{1}{2\epsilon}\,V^{(2)}Z.
\ee
Using this the divergence of the diagram is found to be
\begin{center}
\includegraphics[width=0.35\textwidth]{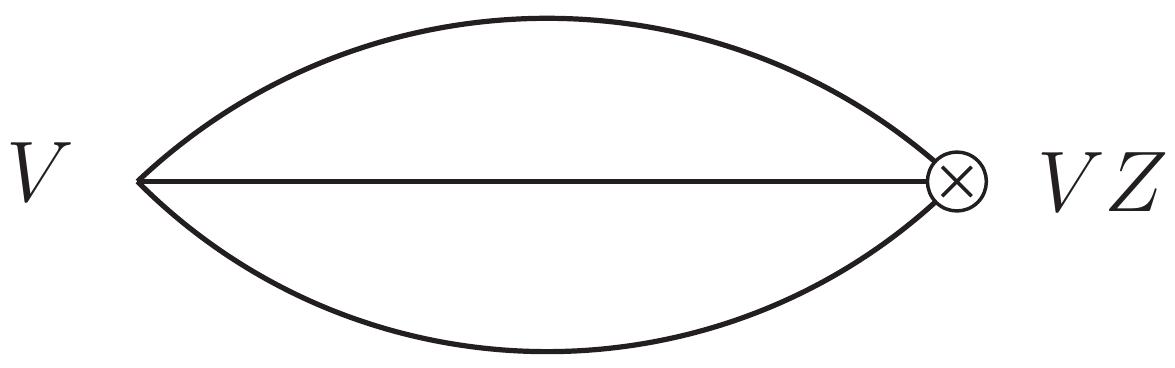}
\end{center}
\be 
\frac{1}{(4\pi)^9}\frac{1}{24}\left[\frac{1}{\epsilon^2}+\left(\frac{9}{4}-\gamma\right)\frac{1}{\epsilon}\right] \big[Z\,V^{(2)}\big]^{(3)}V^{(3)}.
\ee
Finally let us sum up these contributions to get the total counter-term of the form $V^2Z$. Just like the previous case of cubic counter-terms in $V$, we expect here that no non-local term be present in the final result. Although this is indeed the case, manipulating these terms a bit shows that the situation is slightly more involved in this case. In fact the non-local terms of the $V^2Z$ diagrams do not cancel out among the diagrams but sum up to local terms instead. The cancellation of terms proportional to $\gamma$ can be verified here as well, similar to the $V^3$ case. Summing up different contributions and simplifying, the total counter-term of the form $V^2Z$ reads 
{\setlength\arraycolsep{2pt}
\bea
V^{v^2\!z}_{c.t.}  &=& \frac{1}{(4\pi)^{9}}\frac{1}{72\epsilon}\,Z^{(2)}V^{(2)}V^{(4)}
+ \frac{1}{(4\pi)^{9}}\frac{1}{36}\left(\frac{1}{\epsilon^2}-\frac{3}{8\epsilon}\right)  Z\,V^{(4)}\hspace{1pt}^2 \nn\\
&+&  \frac{1}{(4\pi)^{9}}\frac{1}{24\epsilon^2}\, Z^{(3)} V^{(2)} V^{(3)}
+\frac{1}{(4\pi)^9}\frac{1}{24}\left(\frac{1}{\epsilon^2} - \frac{1}{4\epsilon}\right) Z^{(2)} V^{(3)}\hspace{1pt}^2 \nn\\
&+& \frac{1}{(4\pi)^9}\frac{1}{24}\left(\frac{1}{\epsilon^2} - \frac{5}{4\epsilon}\right) Z^{(1)} V^{(3)}V^{(4)} + \frac{1}{(4\pi)^9}\frac{1}{24\epsilon^2}\, Z\, V^{(3)}V^{(5)}.
\eea}%

\subsection{Beta functions}

Before proceeding with the computation of the cubic beta function for the potential, let us summarize what we have found at the quadratic level. In Sect.\ref{s:type2} we have computed the counter-terms and beta functions for $V$ and $Z$ at quadratic level for general values of $m$. Here we only need the $m=1$ results. For $m=1$ the quadratic counter-terms are
{\setlength\arraycolsep{2pt}
\bea
V_{c.t.2} &=& \frac{1}{(4\pi)^{6}}\,\frac{1}{24\epsilon}\,V^{(3)}\hspace{1pt}^2 + \frac{1}{(4\pi)^{3}}\,\frac{1}{2\epsilon}\,V^{(2)}Z,  \\[5pt]
Z_{c.t.2} &=& - \frac{1}{(4\pi)^{9}}\,\frac{1}{216\epsilon}\,V^{(5)}\hspace{1pt}^2 + \frac{1}{(4\pi)^{6}}\,\frac{1}{12\epsilon}\,V^{(3)} Z^{(3)} + \frac{1}{(4\pi)^{3}}\frac{5}{12\epsilon}\,Z^{(1)}\hspace{1pt}^2 +\frac{1}{(4\pi)^{3}}\frac{1}{2\epsilon}\,Z^{(2)} Z.
\eea}%
The $m=1$ dimensionful beta functions at quadratic level are simply obtained from the general equations \eqref{bvq} and \eqref{bzq} 
{\setlength\arraycolsep{2pt}
\bea 
\beta_{V,2} &=& \frac{1}{(4\pi)^{6}}\,
\frac{1}{12}\,V^{(3)}\hspace{1pt}^2 + \frac{1}{(4\pi)^{3}} \,\frac{1}{2}\, V^{(2)}Z, \label{bvqm1} \\[5pt]
\beta_{Z,2} &=& -\frac{1}{(4\pi)^{9}}\,
\frac{1}{72}\,V^{(5)}\hspace{1pt}^2 + \frac{1}{(4\pi)^{6}}\,\frac{1}{6}\, V^{(3)}Z^{(3)} +\frac{1}{(4\pi)^{3}}\,\frac{5}{12}\,Z^{(1)}\hspace{1pt}^2 +\frac{1}{(4\pi)^{3}}\,\frac{1}{2}\,Z^{(2)}Z. \label{bzqm1} 
\eea}%
%
%
%
%
We are now in a position to compute the cubic beta function in $V$. This is related to the quadratic and cubic counter-terms $V_{c.t.2}$ and $V_{c.t.2}$ as
\be \label{b3}
\beta_{V,3} = \epsilon V_{c.t.3} - \left.\mu\frac{d}{d\mu}\right|_1\!\!V_{c.t.3} - \left.\mu\frac{d}{d\mu}\right|_2\!\!V_{c.t.2}.
\ee
However, we are interested only in contributions of the form $V^{(a)}V^{(b)}V^{(c)}$ and $V^{(a)}V^{(b)}Z^{(c)}$. We therefore take $V_{c.t.3} = V^{v^3}_{c.t.} + V^{v^2z}_{c.t.}$ and in the last term of \eqref{b3} we pick the relevant contributions. This means that in 
\be \label{v2dot}
\left.\mu\frac{d}{d\mu}\right|_2\!\!V_{c.t.2} = \frac{1}{(4\pi)^{6}}\,\frac{1}{12\epsilon}\,V^{(3)}\beta^{(3)}_{V,2} + \frac{1}{(4\pi)^{3}}\,\frac{1}{2\epsilon}\,Z\beta^{(2)}_{V,2} + \frac{1}{(4\pi)^{3}}\,\frac{1}{2\epsilon}\,V^{(2)}\beta_{Z,2}
\ee
we only take the first term of \eqref{bvqm1} in $\beta^{(2)}_{V,2}$ and the first two terms of \eqref{bzqm1} in $\beta_{Z,2}$, while $\beta^{(3)}_{V,2}$ is used with both its terms \eqref{bvqm1}. Doing this and using
\be 
\epsilon V^{v^3}_{c.t.} - \left.\mu\frac{d}{d\mu}\right|_1\!\!V^{v^3}_{c.t.} = 4\epsilon V^{v^3}_{c.t.}, \qquad
\epsilon V^{v^2z}_{c.t.} - \left.\mu\frac{d}{d\mu}\right|_1\!\!V^{v^2z}_{c.t.} = 3\epsilon V^{v^2z}_{c.t.},
\ee
we find the desired result, i.e. the cubic beta function of the potential with at most linear dependence in $Z$.
After the rescalings \eqref{vzres} the beta function reads
{\setlength\arraycolsep{2pt}
\bea \label{dfbv}
\beta_V &=&  \,\frac{1}{6}\,V^{(3)}\hspace{1pt}{}^2  + \frac{1}{2} Z\,V^{(2)} \nn\\[2mm]
&-& \frac{1}{12} V^{(4)}\hspace{1pt}{}^3 - \frac{1}{12} V^{(3)}V^{(4)}V^{(5)} + \frac{35}{1296} V^{(2)}V^{(5)}\hspace{1pt}{}^2 \nn\\[2mm]
&+&  \frac{1}{12} Z^{(2)}V^{(2)}V^{(4)}  -\frac{1}{16}Z\,V^{(4)}\hspace{1pt}{}^2 -\frac{1}{16}Z^{(2)}\,V^{(3)}\hspace{1pt}{}^2 -\frac{5}{16}Z^{(1)}\,V^{(3)}  V^{(4)}.
\eea}%
A novel feature of this theory is that to obtain the flow of the potential at cubic order in $V$ we need the quadratic flow of $Z$, given in the last term of \eqref{v2dot}, and this is crucial for the cancellation of $\epsilon$ poles in the beta function \eqref{b3}. This is a general feature of second type theories, that the potential contributions to the flow of the potential is affected by the flow of the couplings of higher derivative operators. For completeness we report here also the $m=1$ version of the quadratic flows of $Z$ and $W_i$ functions after the rescaling \eqref{vzres}. For the $Z$ function this is
\be  \label{dfbz}
\beta_Z = -\frac{1}{18}V^{(5)}\hspace{1pt}{}^2 + \frac{1}{3}Z^{(3)}V^{(3)}+\frac{5}{12}Z^{(1)}\hspace{1pt}{}^2 +\frac{1}{2}Z\,Z^{(2)}.
\ee
For the functions $W_i$ corresponding to four-derivative operators the rescaled flows are given as  
{\setlength\arraycolsep{2pt}
\bea
\beta_{W_1} &=& \frac{1}{720}V^{(6)}\hspace{1pt}{}^2 - \frac{1}{72}Z^{(3)}V^{(5)}-\frac{1}{32}Z^{(2)}\hspace{1pt}{}^2, \label{dfbw1} \\[2mm]
\beta_{W_2} &=& \frac{1}{360}V^{(6)}V^{(7)} + \frac{1}{72}Z^{(4)}V^{(5)}-\frac{1}{72}Z^{(3)}V^{(6)}-\frac{1}{16}Z^{(2)}Z^{(3)}, \label{dfbw2} \\[2mm]
\beta_{W_3} &=& \frac{1}{720}V^{(7)}\hspace{1pt}{}^2 + \frac{1}{72}Z^{(4)}V^{(6)}+\frac{1}{96}Z^{(3)}\hspace{1pt}{}^2. \label{dfbw3}
\eea}%
The dimensionful beta functions \eqref{dfbv}, \eqref{dfbz} and \eqref{dfbw1} are related to the beta functions of the dimensionless variables defined in \eqref{vzdimless} and \eqref{wdimless} as
{\setlength\arraycolsep{2pt}
\bea 
\beta_v &=& -2(\delta+2) v + (\delta + \gamma_\phi)\,\varphi\, v^{(1)} + \mu^{-d}\beta_V \\
\beta_z &=& -2z+(\delta + \gamma_\phi)\,\varphi\, z^{(1)} +\mu^{-2}\beta_Z,
\eea}%
and \eqref{bwdl} respectively.

\subsection{Critical data}

In this section we report some of the critical data that can be extracted with a knowledge of the beta functions. In Sect.\ref{s:type2} we computed for general $m$ the beta functions of $V$ and $Z$ at quadratic level and obtained the field and coupling anomalous dimensions in terms of the couplings and at the non-trivial IR fixed point also in terms of $\epsilon$. Here for the particular case of $m=1$ we first present the anomalous dimensions for all three non-trivial fixed points in terms of $\epsilon$, which requires \eqref{dfbz} and \eqref{dfbw1} and only the quadratic terms in the beta function \eqref{dfbv}. Next we take advantage of the cubic terms in \eqref{dfbv} to extract some OPE coefficients as well. For $m=1$ the general leading order anomalous dimensions given in sections \eqref{ss:spectrum} and \eqref{ss:rg4der} reduce to 
\be 
\tilde\gamma_i = 40\,i(i-1)(i-2)\,g+\frac{1}{2}i(i-1)\,h, \qquad i\leq 4
\ee
\be 
\tilde\gamma_1 = -360\,g^2+\frac{1}{16}\,h^2.
\ee 
Notice that the couplings have been rescaled according to $g\rightarrow 2(4\pi)^6 g$ and $h\rightarrow (4\pi)^3 h$. The three fixed points shown in Fig.\ref{fd} depend on $\epsilon$ in the following way
\be 
\ba{lll}
\displaystyle g = 0 &\qquad \displaystyle g = \frac{\left(3 \sqrt{138}-13\right) \epsilon }{22200} &\qquad \displaystyle  g = -\frac{\left(13+3 \sqrt{138}\right) \epsilon }{22200} \\[3mm]
h = \displaystyle \frac{3\epsilon}{8} &\qquad \displaystyle h= \frac{1}{185} \left(42-4 \sqrt{138}\right) \epsilon  &\qquad \displaystyle h= \frac{2}{185} \left(21+2 \sqrt{138}\right) \epsilon.
\ea
\ee
The corresponding relevant spectrum can then be calculated at each fixed point. Those that are of order $\epsilon$ are given as
\be 
\ba{lll}
\displaystyle \gamma_2 = \frac{3\epsilon}{8} &\qquad \gamma_2 = \displaystyle \frac{1}{185} \left(42-4 \sqrt{138}\right) \epsilon &\qquad \gamma_2 = \displaystyle \frac{2}{185} \left(21+2 \sqrt{138}\right) \epsilon \\[2mm]
\displaystyle \gamma_3 = \frac{9\epsilon}{8} &\qquad \displaystyle \gamma_3 = \frac{2}{185} \left(50-3 \sqrt{138}\right) \epsilon &\qquad \gamma_3 = \displaystyle \frac{2}{185} \left(50+3 \sqrt{138}\right) \epsilon \\[3mm]
\displaystyle \gamma_4 = \frac{9\epsilon}{4} &\qquad \displaystyle \gamma_4= \frac{4\epsilon}{5} &\qquad \gamma_4 = \displaystyle \frac{4\epsilon}{5},
\ea
\ee
which appear from left to right in the same order as the fixed points. For all three cases, as expected $\gamma_5 = 2\epsilon$. The value of $\gamma_1$ for the three fixed points is respectively:
\be 
\gamma_1 = \frac{9 \epsilon^2}{1024}, \qquad
\gamma_1 = \frac{\left(8519-762 \sqrt{138}\right) \epsilon ^2}{1369000}, \qquad
\gamma_1 = \frac{\left(8519+762 \sqrt{138}\right) \epsilon ^2}{1369000}.
\ee
It is a straightforward task to extract the dimensionless OPE coefficients from \eqref{dfbv} in terms of the fixed point values of the couplings $g$ and $h$. Instead of giving the complete formula which is slightly involved, for simplicity we report here the result only for the infrared fixed point with $g=0$
{\setlength\arraycolsep{2pt}
\bea  \label{opeh}
\tilde C^l{}_{ij} &=&  \,\frac{1}{6}\,\frac{i!}{(i-3)!}\frac{j!}{(j-3)!} \nn\\
&-& \frac{1}{16}\frac{i!}{(i-4)!}\frac{j!}{(j-4)!}\,h -\frac{1}{8}\frac{i!}{(i-3)!}\frac{j!}{(j-3)!}\,h \nn\\
&+& \frac{1}{12} \frac{i!}{(i-2)!}\frac{j!}{(j-4)!}\,h - \frac{5}{16}\frac{i!}{(i-3)!}\frac{j!}{(j-4)!}\,h + (i\leftrightarrow j),
\eea}%
where the terms in the last line are symmetrized in $i,j$ and $h$ is understood to take its fixed point value $3\epsilon/8$. Also the dimensionless condition fixes $l=i+j-6$. It may still be useful to give a few explicit examples for the OPE coefficients in terms of $\epsilon$ at the infrared fixed point
\be 
\tilde C^1{}_{34} = 24-\frac{153}{8}\epsilon, \qquad
\tilde C^1{}_{25} = \frac{15}{2}\epsilon, \qquad
\tilde C^2{}_{26} = \frac{45}{2}\epsilon, \qquad
\tilde C^3{}_{27} = \frac{105}{2}\epsilon.
\ee
Notice that the OPE coefficients \eqref{opeh} are valid either at leading order where mixing effects do not show up, or for the components that are not affected by mixing. The first case includes an infinite number of nontrivial, i.e. order $\epsilon$, instances when the free theory contribution, which is given by the first line, vanishes. 

%
%

\section{Conclusions} \label{s:conclusions}

In this work we have shown from one side that renormalization group techniques are still fundamental tools to investigate novel critical properties of QFTs, which in recent years has been addressed with alternative powerful CFT techniques, either analytical or numerical. On
the other hand, agreement of CFT results (for instance \cite{Gliozzi:2016ysv,Gliozzi:2017hni,Safari:2017irw}) with our RG analysis, which relies neither on unitarity nor conformal symmetry, can provide further evidence for the enhancement of scale invariance to conformal invariance for nonunitary single-scalar higher dimensional theories with higher derivative kinetic terms investigated here. Indeed we show how some non trivial results obtained both with RG and CFT techniques, such as
scaling dimensions and a class of OPE coefficients, match in the first non trivial order in $\epsilon$ expansion.


We have taken advantage of the functional form of the flow equations to extract critical information and give compact formulas encompassing (sometimes infinitely) many quantities. The general pattern of coupling mixing is presented which is different from the standard case and depends on the number of derivatives in the kinetic term and the 
space dimension, or equivalently $k,n$. We have given a constraint formula that is a result of dimensional analysis and determines to a high extent the structure of terms, i.e combination of functions and their field derivatives, that appear in the beta functionals. 
Based on the possible marginal operators 
we have distinguished two types of theories and argued that they have qualitatively different features. 

Theories of the first type have a generalized Wilson-Fisher fixed point. We have extracted critical information around this fixed point and shown that when available from both approaches our results for these theories agree with those of \cite{Gliozzi:2016ysv,Gliozzi:2017hni}, which is based on the structure of conformal blocks in a CFT framework. Beyond known results we give the next to leading order values for the coupling anomalous dimensions corresponding to the relevant and marginal non-derivative operators and the leading order anomalous dimensions for the couplings of all two-derivative operators. We also give an infinite set of OPE coefficients at order $\epsilon$. 

The phase diagram for second-type theories has a richer structure. We have analyzed those with a $\Box^2$ kinetic term which correspond to odd values of $n=2m+1$ and discovered 
that apart from the Gaussian fixed point there are three nontrivial fixed points all of which include derivative interactions. In particular there is a fixed point with a pure derivative interaction that is infrared attractive. We have computed the quadratic flow, in terms of $V$ and $Z$, of couplings corresponding to operators up to four derivatives for general $m$. The field anomalous dimension and some critical exponents for the potential and $Z$ couplings are given at leading order in terms of the fixed point couplings, and for the infrared fixed point in terms of $\epsilon$ in which case the range of validity extends to all $V,Z$ couplings. 

In a previous letter \cite{Safari:2017irw}, where some of the results of this paper were anticipated, we have shown how constraints from conformal symmetry together with a knowledge of the Schwinger-Dyson equations, can be used to obtain some of these results, providing in this way indications, at least at the leading order $\epsilon$-expansion, that scale invariance could be enhanced to conformal invariance, also for second-type theories. A more comprehensive investigation along these lines is in progress \cite{Safari.Vacca}.

For the special case of $\Box^2$ theories in $6\!-\!\epsilon$ dimensions the beta functional of the potential is extended to cubic order. This enables us to compute some OPE coefficients of non-derivative operators at order $\epsilon$. For this purpose we need not only cubic contributions in the potential of the form $V^{(a)}V^{(b)}V^{(c)}$ but also cubic terms linear in the $Z$ coupling, i.e. contributions of the form $V^{(a)}V^{(b)}Z^{(c)}$. 

Several novel features appear in this computation. For instance the computation of the pure potential contribution to the flow of $V$ at cubic level requires the quadratic flow of $Z$. This is another indication 
of the fact that derivative couplings cannot be ignored in theories of the second type even at leading order in derivative expansion. Another particular feature is that nonlocal ultraviolet divergences show up in separate diagrams that finally combine to a local term in physical amplitudes. This signals the fact that it would be too naive to simply drop the nonlocal divergences in the diagrams.

A higher derivative theory of the first type may be considered as a possible construction to describe the isotropic phase of Lifshitz critical theories which is relevant for physics of certain
polymers. It is not yet clear if multicritical theories of the novel second type can play a role in understanding new phases. We also note that these theories and in particular the second type
ones have critical points, characterized by a finite dimensional UV critical surface, spanned by the UV attractive directions, and may be seen as a non unitary realization of asymptotically safe scalar theories \cite{Wilson:1971bg,Weinberg,Litim:2014uca,Bond:2016dvk}. $\Box^2$ theories are non trivial in dimensions less than 6 but it is likely that also critical theories with higher number of derivatives ($k > 2$) demonstrate at least in some cases a similar behavior.

The results of this work can be extended in several directions. We have concentrated here on $\Box^k$ theories with $\mathbb{Z}_2$-even critical interactions. It would be interesting to extend these results to the case where the scale invariant theory includes $\mathbb{Z}_2$-odd marginal operators. Another line of investigation would be including multiple scalar fields and studying various global symmetries \cite{Osborn:2017ucf}, or possibly considering fields of higher spin such as $\frac{1}{2}$ and $1$. Theories of the second type in particular have not been studied with other methods. They therefore provide a setting to apply and test alternative approaches that do not rely on RG, especially those that are solely based on conformal symmetry, which may possibly also have the potential to improve these findings.

\vspace{10pt}

\noindent
{\bf Acknowledgments}\\
We would like to thank A. Petkou for discussions.

\appendix

\section{Counter-terms induced by the potential} \label{s:vct} 

\subsection{Quadratic counter-terms} \label{ss:qmelon}

The most general diagram that contains two powers of the potential has the following melon-type structure, where here and throughout the paper the number of field derivatives on a function is denoted by a superscript enclosed in a parenthesis 
\begin{figure}[H]
\begin{center}
\includegraphics[width=0.35\textwidth]{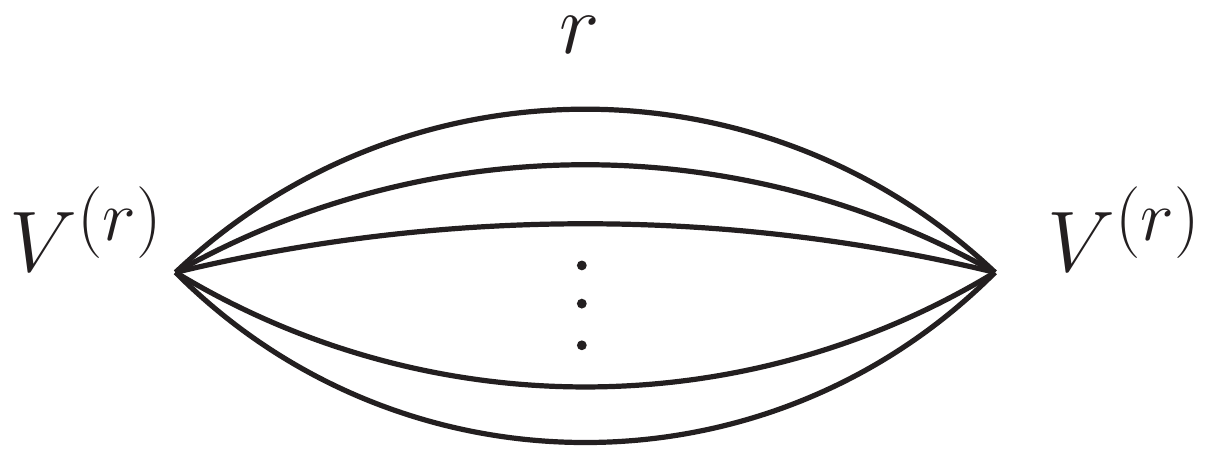}
\end{center}
\caption{General melon diagram with $r$ propagators and vertices coming from $V$.}
\label{melon}
\end{figure}
\noindent
Let $r$ be the number of propagators. Using the momentum space expression \eqref{Grp}, the coordinate-space expression for a bunch of $r$ propagators is
\be \label{Grx} 
G^r_x = \frac{1}{(4\pi)^{rk}\Gamma^r(k)}\,
\frac{\Gamma^r(\delta)}{\Gamma(r\delta)}\,
\Gamma(k-(r-1)\delta)
\left(\frac{\!-\square_x}{4\pi}\right)^{(r-1)\delta-k}.
\ee
This is singular when the argument of the rightmost Gamma function is a non-positive integer, say $-l$, in which case the box appears with the non-negative integer power $l$. In this case the singular piece of the diagram, including the $1/2$ symmetry factor, will be 
\be \label{ul}
U_{l,c.t.} = \frac{1}{2\,r!}\,\frac{1}{(4\pi)^{(r-1)(k+\delta_c)}\Gamma^r(k)}\,
\frac{\Gamma^r(\delta_c)}{\Gamma(r\delta_c)}\,\frac{(-1)^l}{l!}\, \frac{2}{(r-1)\epsilon}\,V^{(r)}(\phi)(-\square)^lV^{(r)}(\phi) 
\ee
where the number of propagators $r$ satisfies
\be \label{c1} 
\frac{n-r}{n-1}k = -l \qquad \mathrm{for} \qquad l\geq 0. 
\ee
This equation is nothing but the constraint \eqref{constraint} from dimensional analysis, for the special case of $N=2$, $a=l$ and $a_1=a_2=0$, which is found here with a different argument. It is telling us for instance that values of $l$ that are multiples of $k$ will always give rise to an integer. More specifically $l=ik$ has a solution $r=n+i(n-1)$ for $i=0,1,\cdots$. But depending on $k,n$ there can be intermediate integer values of $l,r$ that lead to a singularity, i.e. satisfy \eqref{c1}. Whether or not such intermediate values are admitted depends on whether $k$ and $n-1$ are relatively prime or not. If they are coprime numbers then such intermediate values cannot exist. Instead if they have a non-trivial common divisor then intermediate values of $l$ can arise for suitable $r$ which do not belong to the above-mentioned set. We have therefore seen in a different way a clear distinction between the two type of theories. 

We will now move to the calculation of cubic corrections in $V$-couplings. The pole term \eqref{ul} in the above melon diagram will serve as a counter-term vertex for the higher order calculations of the following section.

\subsection{Cubic counter-terms from triangle diagrams} \label{ss:tri}

The cubic counter-terms come either from diagrams with three vertices or from diagrams with two vertices of which one vertex is the quadratic counter-term vertex of the previous section. Let us divide the diagrams with three vertices into two groups. The first group consists of triangle diagrams where all three vertices are connected to each other by at least one propagator. The second group, which we call double-melon diagrams include those in which two of the vertices are not connected. Let us begin by extracting the pole term in a triangle diagram. Consider the following triangle diagram where the edges consist of bunches of $r$, $s$ and $t$ propagators 
\begin{figure}[H]
\begin{center}
\includegraphics[width=0.4\textwidth]{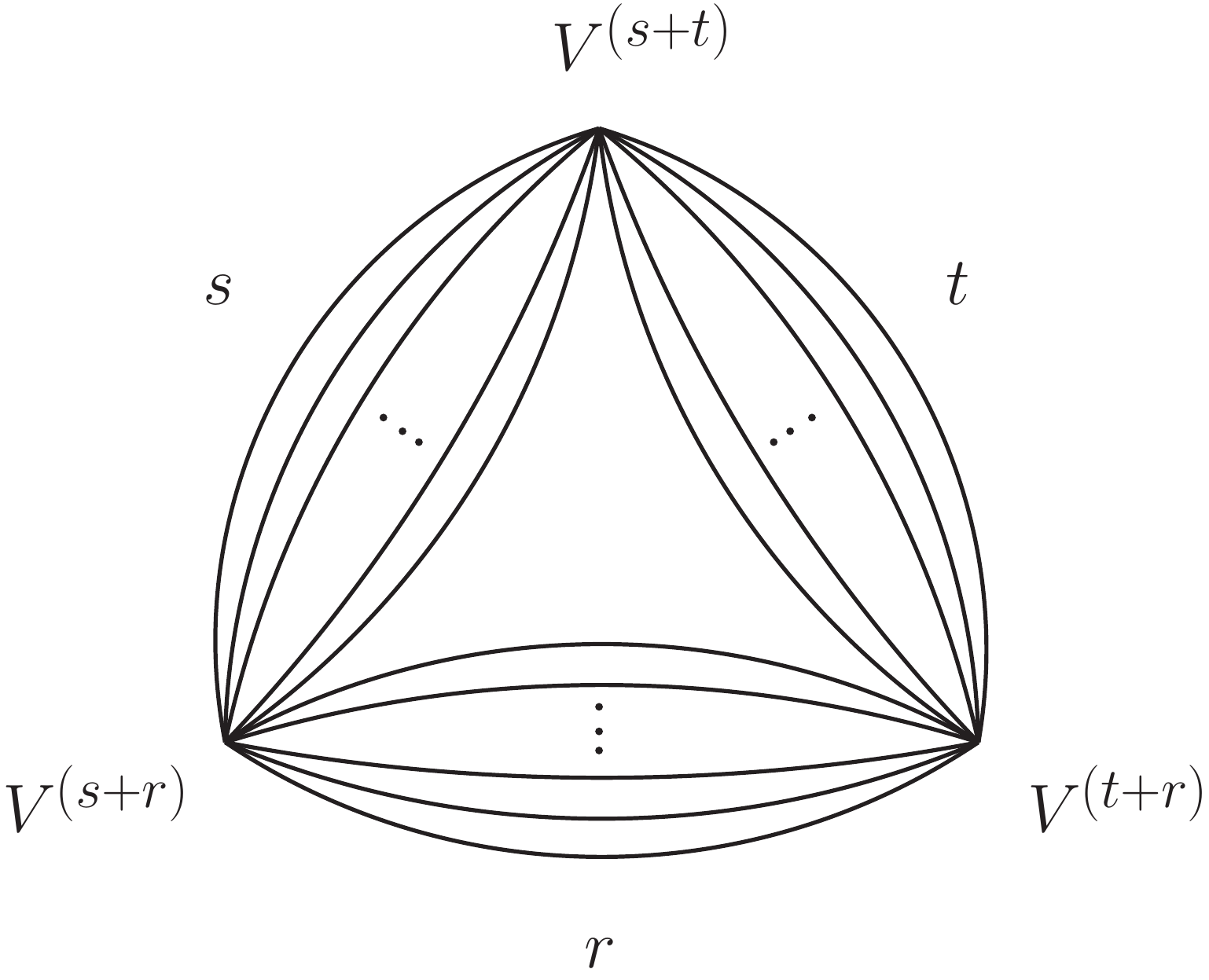} 
\end{center}
\caption{General triangle diagram with vertices from $V$ and a number of $r$, $s$ and $t$ propagators in the edges with $r,s,t\geq 1$.}
\label{triangle}
\end{figure}
\noindent
Taking into account all symmetry factors the diagram in momentum space evaluates to
\be \label{tricomp}
-\frac{1}{6}\sum_{r,s,t}\frac{1}{r!s!t!} \, V_{p_1}^{(t+s)}V_{p_2}^{(s+r)}V_{p_3}^{(r+t)}\int_p \prod_i G^{r_i}_{p+k_i},
\ee
where $i$ runs from 1 to 3, $p$ is the loop momentum integrated over according to \eqref{shn}, and $p+k_i$ is the momentum running in the edge with $r_i$ propagators which corresponds to $t,s,r$ for $i=1,2,3$ respectively. Also the derivatives of the potentials are expressed in momentum space and $p_1 = k_2 - k_1$, $p_2 = k_3 - k_2$ and $p_3 = k_1 - k_3$.
To evaluate the divergences we find it convenient to use the Mellin-Barnes representation which in this case is given by the following expression 
{\setlength\arraycolsep{2pt}
\bea \label{gold} 
\int_p \prod_i G^{r_i}_{p+k_i} &=& \frac{1}{[(4\pi)^k\Gamma(k)]^{2n}} \frac{\Gamma(\delta)^{2n}}{\Gamma(r\delta)\Gamma(s\delta)\Gamma(t\delta)}\frac{1}{\Gamma((2n -1)\delta-k)}\left(\frac{p_1}{4\pi}\right)^{(r+s+t -2)\delta-2k} \nn\\
&\times & \int_{y,z} \Gamma(-y)\Gamma(-z)\; \left(\frac{p_2^2}{p_1^2}\right)^y\left(\frac{p_3^2}{p_1^2}\right)^z \, \Gamma(y+z+2k-2(n-1)\delta) \nn\\
&\times &  \Gamma((r+t-1)\delta-k-y)\Gamma((r+s-1)\delta-k-z)\Gamma(y+z+k-(r-1)\delta).
\eea}%
It is important that the contour integrals for $y,z$ separate the poles of the first two gamma functions in the second line of \eqref{gold} from those of the third one on the right. In this case they can be taken to coincide with the imaginary axis. In order for this diagram to contribute to the counter-term of the potential the power of $p_1$ must vanish and so $r+s+t=2n$.

An edge of the triangle which forms a bunch of propagators can lead to a subdivergence if the number of propagators satisfies the condition described in the previous section, i.e. Eq.\eqref{c1}. For one of the edges to satisfy the condition \eqref{c1}, it is necessary to have its number of propagators $\geq n$. Because of the constraints $r+s+t=2n$, $r,s,t\geq 1$ on the number of propagators in each edge, one cannot have two of them $\geq n$ at the same time. So the condition \eqref{c1} can be satisfied at most by one of the edges. 

It is convenient to divide triangle diagrams into two classes, one for which non of $r,s,t$ satisfy \eqref{c1}, and one where one of $r,s,t$ satisfies \eqref{c1}. The first class has a single $\epsilon$-pole, while the second class has a double pole. For theories of the first type, an edge can be divergent only if it has $n$ propagators. Instead for second-type theories, apart from $n$ some bigger number of propagators below or equal to $2n-2$ will also lead to divergences.

\subsection{Cubic counter-terms from double-melon diagrams} \label{ss:dmelon}

A general double-melon diagram consists of two connected melon diagrams with $s$ and $t$ propagators as in the following figure
\begin{figure}[H]
\begin{center}
\includegraphics[width=0.55\textwidth]{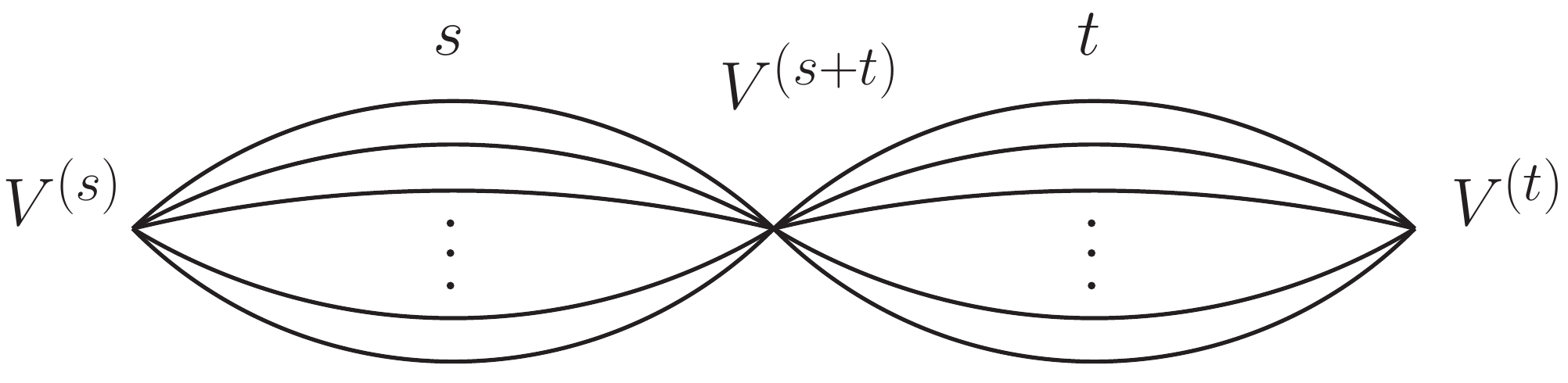}
\end{center}
\caption{General double-melon diagram with vertices from $V$ that consists of two melons, one with $s$ and the other with $t$ propagators, where $s,t\geq 2$.}
\label{dmelon} 
\end{figure}
In order for this to contribute to the counter-term of the potential the number of propagators in the two bunches $s,t$ take all positive values subject to the condition $s+t=2n$, which can be inferred from dimensional analysis, and such that at least one of the numbers 
\be 
\frac{n-s}{n-1}k,  \quad\qquad \frac{n-t}{n-1}k,
\ee
is a non-positive integer, i.e. satisfies the condition \eqref{c1} so that at least one of the melons is divergent. They can both be a non-positive integer only for $s=t=n$. Using \eqref{Grp} the total contribution of these diagrams in momentum space is
\be  \label{doubcomp}
-\frac{1}{2} \sum_{s,t} \frac{1}{s!t!}\,V^{(s)}_{p_1}V^{(s+t)}_{p_2}V^{(t)}_{p_3} \; G^s_{p_1} G^t_{p_3},
\ee
where the sum runs over positive integer $s,t$ subject to the condition $s+t=2n$. The field derivatives of the potentials are expressed in momentum space and $p_1$,$p_2$ and $p_3$ are the incoming momenta for the left, middle and the right vertex respectively. The product of the two bunch of propagators in momentum space is
\be \label{gg}
G^s_{p_1} G^t_{p_3} = \frac{(p_3^2)^{2(n-1)\delta-2k}}{(4\pi)^{2(n-1)(\delta+k)}\Gamma^{2n}(k)}\,\frac{\Gamma^{2n}(\delta)}{\Gamma(s\delta)\Gamma(t\delta)}\,
\Gamma(k-(s-1)\delta)\Gamma(k-(t-1)\delta)\left(\frac{p_3
^2}{p_1^2}\right)^{k-(s-1)\delta}.
\ee
The only case which gives rise to a double $\epsilon$-pole is $s=t=n$, while the rest lead to a single pole.

\subsection{Cubic counter-terms from melon diagrams} \label{ss:cmelon}

The final set of diagrams are melon-type diagrams of which one of the vertices is the quadratic counter-term vertex discussed in Sect.\ref{ss:qmelon} 
\begin{figure}[H]
\begin{center}
\includegraphics[width=0.35\textwidth]{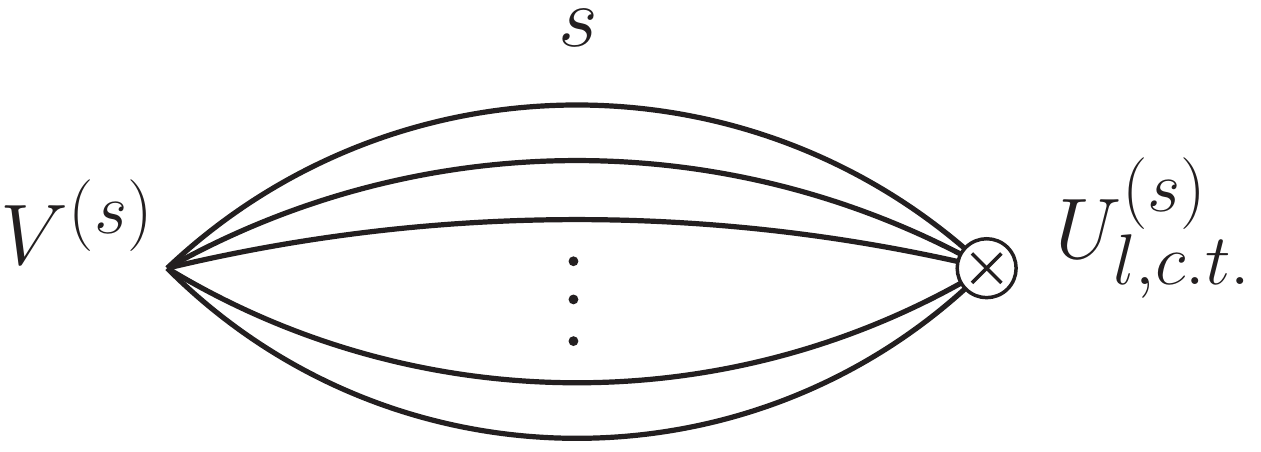}
\end{center}
\caption{General melon diagram with $s\geq 2$ propagators, that has a vertex from $V$ and a vertex from the counter-terms $U_{l,c.t.}$.}
\label{umelon}
\end{figure}\noindent
Since $U_{l,\mathrm{c.t.}}$ has $2l$ space-time derivatives, dimensional analysis shows that these diagrams can contribute to the potential if $s$ satisfies
\be \label{c3}
l = \frac{n-s}{n-1} k.
\ee
The number of space-time derivatives in $U_{l,\mathrm{c.t.}}$ is non-negative $l\geq 0$ therefore $s\leq n$. The condition \eqref{c3} can be viewed as a dual condition to \eqref{c1}. This can be seen by choosing $r,s$ to be related by $r+s=2n$, which is consistent with their range $1\leq s\leq n$ and $n\leq r\leq 2n-1$ (first interval). In other words, if $s$ satisfies \eqref{c3}, then $r=2n-s$ satisfies \eqref{c1} and vice versa. The above condition \eqref{c3} is always valid for $s=1,n$ which gives $l=k,0$. As before, intermediate values of $l$ can occur if $k$ and $n-1$ are not coprime. In momentum space this diagram is given as
\be \label{cmelonexp}
\frac{1}{r!}\,\frac{1}{(4\pi)^{(r-1)(k+\delta_c)}\Gamma^r(k)}\,
\frac{\Gamma^r(\delta_c)}{\Gamma(r\delta_c)}\,\frac{(-1)^l}{l!}\, \frac{1}{(r-1)\epsilon}V^{(s)}_{p}\sum_{a=0}^s\frac{V^{(r+s-a)}_{p-\tilde p}V^{(r+a)}_{\tilde p}}{(s-a)!a!}A^a_l(\tilde p,p)
\ee
where one must sum such expressions for all possible values of $l$. The positive integers $r,s$ are related to $l$ through
\be 
\frac{n-r}{n-1} k = -l, \quad\qquad 
\frac{n-s}{n-1} k = l. 
\ee
It is clear that $s$ can take values in $\{1,\cdots,n\}$ and $r$ can take values in $\{n,\cdots,2n-1\}$. $p$ is the total momentum running through the diagram and the quantity $A^a_l(\tilde p,p)$ is 
{\setlength\arraycolsep{2pt}
\bea  
A^a_l(\tilde p,p) &=& \frac{1}{(4\pi)^{(s-2)(\delta+k)}\Gamma^s(k)}\,
\frac{\Gamma^s(\delta)}{\Gamma(a\delta)\Gamma((s-a)\delta)}\,
\Gamma(k-(a-1)\delta)\Gamma(k-(s-a-1)\delta) \times \nn\\
&& \int_q  \frac{[(q-\tilde p)^2]^l}{(q^2)^{k-(a-1)\delta}((p-q)^2)^{k-(s-a-1)\delta}}.
\eea}%
We report here the values of this quantity for $l=0,1$ which are the only ones used in this work. For $l=0$ there is no $a$-dependence and so we drop the index 
\be   
A_0(\tilde p,p) = \frac{1}{(4\pi)^{(s-1)(\delta+k)}\Gamma^s(k)}\,
\frac{\Gamma^s(\delta)}{\Gamma(s\delta)}\,\Gamma(k-(s-1)\delta)\, \frac{1}{(p^2)^{k-(s-1)\delta}} 
\ee
The result for $l=1$ is 
{\setlength\arraycolsep{2pt}
\bea    
A_1^a(\tilde p,p) &=& \frac{1}{(4\pi)^{(s-1)(\delta+k)}\Gamma^s(k)}\,
\frac{\Gamma^s(\delta)}{\Gamma(s\delta)}\,
\Gamma(k-(s-1)\delta)  \nn\\
&\times & \left(\frac{(s-a)\delta}{k-(s-1)\delta-1}\,\frac{a}{s}+\frac{\tilde p^2}{p^2}\frac{s-a}{s}+ \frac{(p-\tilde p)^2}{p^2}\,\frac{a}{s}\right)\frac{1}{(p^2)^{k-(s-1)\delta-1}}
\eea}%
The expression \eqref{cmelonexp} for $l=0$, which corresponds to $r=s=n$ is
\be \label{l0}
\frac{1}{n!^2(n-1)\epsilon}\frac{1}{[(4\pi)^k\Gamma(k)]^{2n}}\,
\frac{\Gamma^{n}(\delta)}{\Gamma(n\delta)}\frac{\Gamma^{n}(\delta_c)}{\Gamma(n\delta_c)}\,\Gamma(k-(n-1)\delta)\, \left(\!\frac{p^2}{4\pi}\!\right)^{\!(n-1)\delta-k}\!\!\! V^{(n)}\big[V^{(n)}\hspace{1pt}^2\big]^{(n)},
\ee
while for $l=1$ is it given by
{\setlength\arraycolsep{2pt}
\bea  \label{l1}
&-& \frac{1}{r!}\,\frac{1}{(4\pi)^{(r-1)(k+\delta_c)}\Gamma^r(k)}\,
\frac{\Gamma^r(\delta_c)}{\Gamma(r\delta_c)}\, \frac{1}{(r-1)\epsilon}\; \frac{1}{(4\pi)^{(s-1)(\delta+k)}\Gamma^s(k)}\,
\frac{\Gamma^s(\delta)}{\Gamma(s\delta)}\,
\Gamma(k-(s-1)\delta) \nn\\
&\times &  V^{(s)}_p \sum_{a=0}^s \frac{V^{(r+s-a)}_{p-\tilde p} V^{(r+a)}_{\tilde p}}{(s-a)!a!}\; \left(\frac{(s-a)\delta}{k-(s-1)\delta-1}\,\frac{a}{s}+\frac{\tilde p^2}{p^2}\frac{s-a}{s}+ \frac{(p-\tilde p)^2}{p^2}\,\frac{a}{s}\right)
\eea}%
in which 
\be 
r = n+ \frac{n-1}{k} \qquad s = n- \frac{n-1}{k}. 
\ee

\section{Quadratic counter-terms of the form $V^{(a)}Z^{(b)}$} \label{s:zvmelon}

In this section we consider a general quadratic contribution with one $V$-coupling and one $Z$-coupling, i.e. a contribution of the form $V^{(a)}Z^{(b)}$. The relevant diagram is 
\begin{figure}[H]
\begin{center}
\includegraphics[width=0.41\textwidth]{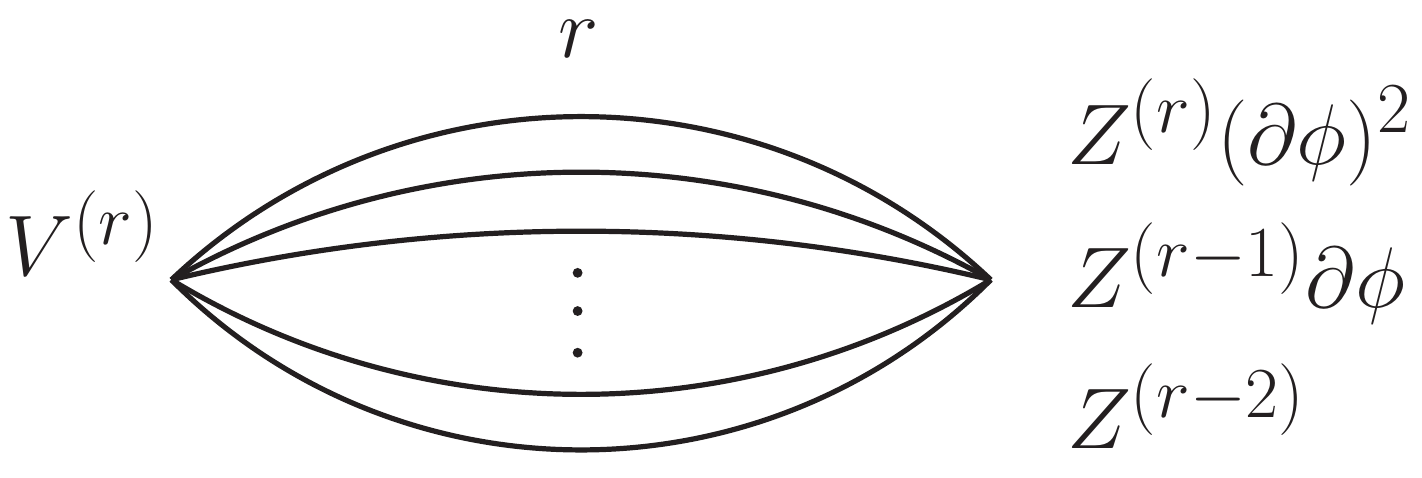} \label{zmelon}
\end{center}
\caption{General melon diagram with $r$ propagators, a vertex from $V$ and a vertex from $Z$. Each of the $Z$ vertices can appear in three ways.}
\label{zmelon}
\end{figure}\noindent
with $r$ number of propagators, which is left unspecified for the moment. The $Z$ vertex can appear in three different ways depending on whether one, two or non of the fields in $(\partial\phi)^2$ is involved in the propagators. These are
\be 
\frac{1}{2}\frac{1}{r!}Z^{(r)}(\partial\phi)^2, \qquad
\frac{1}{(r-1)!}Z^{(r-1)}\partial\phi, \qquad
\frac{1}{2}\frac{1}{(r-2)!}Z^{(r-2)}.
\ee
Using this the evaluation of the diagram is 
\be \label{vz}
\int_{x,y}\left[\frac{1}{2}\frac{1}{r!}Z_x^{(r)}(\partial_x\phi_x)^2\,G^r_{xy}\,V_y^{(r)}
+\frac{1}{r!}Z_x^{(r-1)}\partial_x\phi_x\,\partial_x G^r_{xy}\,V_y^{(r)}
+\frac{1}{2}\frac{1}{(r\!-\!2)!}Z_x^{(r-2)}\,G^{r-2}_{xy}(\partial_x G_{xy})^2\,V_y^{(r)}\right].
\ee
The quantity $G^r_{xy}$ in the first two terms of the above expression is given in Eq.\eqref{Grx}. The quantity $G^{r-2}_{xy}(\partial_x G_{xy})^2$ can be calculated in a similar way by using \eqref{G} to write ($G_x \equiv G_{x0}$)
\be 
(\partial_\mu G_x)^2 G^{r-2}_x  = \frac{4\delta^2}{(4\pi)^{rk}\Gamma(k)^r}\,\frac{\Gamma(\delta)^r}{\pi^{r\delta}} \frac{1}{|x|^{2r\delta +2}}
\ee
and transforming this to momentum space
\be  \label{mel}
\int_x \!e^{ip\cdot x}\, (\partial_x G_x)^2 G^{r-2}_x = \frac{1}{(4\pi)^{(r-1)d/2}\Gamma(k)^r}\,\frac{\delta}{r}\,\frac{\Gamma(\delta)^r}{\Gamma(r\delta)}\;\Gamma(k-(r-1)\delta-1)\,(p^2)^{(r-1)\delta+1-k} 
\ee
The  useful coordinate space representation can now be obtained by replacing $p^2$ with $-\Box$ in the above expression.

\section{Quadratic counter-terms of the form $Z^{(a)}Z^{(b)}$} \label{s:zzmelon}

The melon diagram with two $Z$ vertices gives the quadratic contribution in $Z$. The general diagram of this form with $r$ propagators is 
\begin{figure}[H]
\begin{center}
\includegraphics[width=0.49\textwidth]{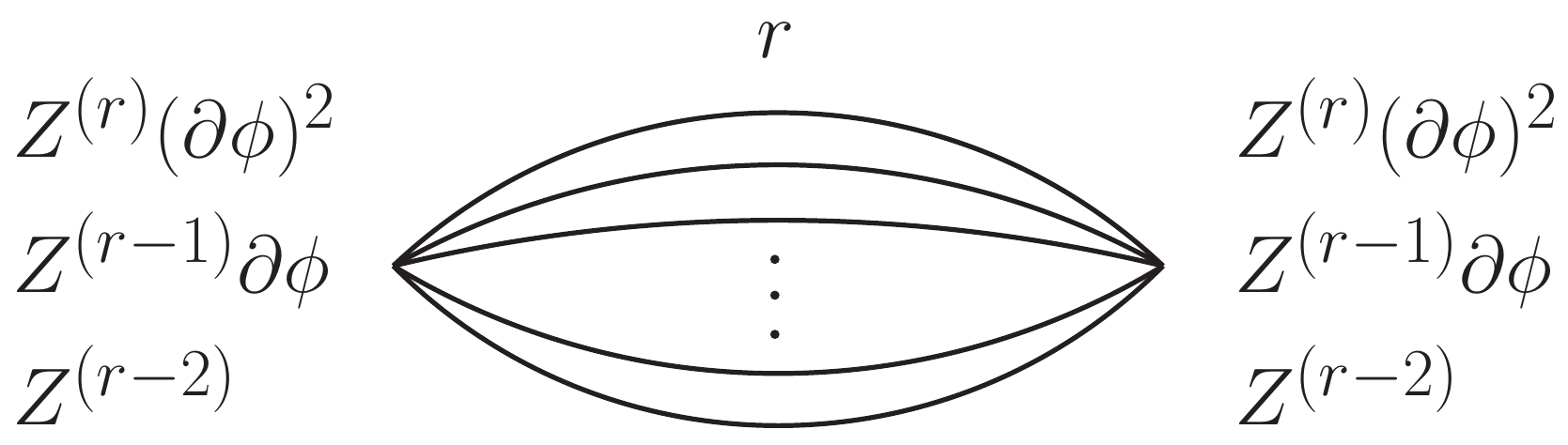} \label{zzmelon}
\end{center}
\caption{General melon diagram with $r$ propagators and with vertices from $Z$. The $Z$ vertex can appear in three ways.}
\label{zzmelon}
\end{figure}\noindent
As seen in the previous section and demonstrated in the above diagram, the $Z$ function appears in three ways in the vertex, so the expression for this diagram can be written as a sum of six terms 
{\setlength\arraycolsep{1pt}
\bea  \label{zz}
&& \hspace{-8.8mm}\int_{x,y}\left[\frac{1}{2}r! \left(\frac{1}{2}\frac{1}{r!}Z^{(r)}(\partial\phi)^2\right)_x G^{r} \left(\frac{1}{2}\frac{1}{r!}Z^{(r)}(\partial\phi)^2\right)_y \right. \nn\\[10pt]
&+& \left. \frac{1}{2}2r!\left(\frac{1}{2}\frac{1}{r!}Z^{(r)}(\partial\phi)^2\right)_x \left(-\frac{1}{r}\partial^\mu G^{r}\right)\left(\frac{1}{(r-1)!}Z^{(r-1)}\partial_\mu\phi\right)_y \right. \nn\\[10pt]
&+& \left. \frac{1}{2}\left(\frac{1}{(r-1)!}Z^{(r-1)}\partial_\mu\phi\right)_x \left(-\frac{(r-1)!}{r}\partial^\mu\partial^\nu G^{r}\right) \left(\frac{1}{(r-1)!}Z^{(r-1)}\partial_\nu\phi\right)_y  \right. \nn\\[10pt]
&+& \left. \frac{1}{2}\left(\frac{1}{2}\frac{1}{(r-2)!}Z^{(r-2)}\right)_x\Big(2(r-2)!(\partial_\mu\partial_\nu G)^2G^{r-2}+4(r-2)(r-2)!(\partial_\mu\partial_\nu G)\partial^\mu G \partial^\nu G G^{r-3} \right. \nn\\  
&+& \left. (r-2)(r-3)(r-2)!(\partial G)^4 G^{r-4}\Big)\left(\frac{1}{2}\frac{1}{(r-2)!}Z^{(r-2)}\right)_y \right. \nn\\[10pt]
&+& \left. \frac{1}{2}2r! \left(\frac{1}{2}\frac{1}{r!}Z^{(r)}(\partial\phi)^2\right)_x (\partial G)^2 G^{r-2}\left(\frac{1}{2}\frac{1}{(r-2)!}Z^{(r-2)}\right)_y \right. \nn\\[10pt]
&+& \left. \frac{1}{2}2\left(\frac{1}{(r-1)!}Z^{(r-1)}\partial^\mu\phi\right)_x \,(r-1)!\,\partial_\mu\!\left[(\partial G)^2 G^{r-2}\right]\, \left(\frac{1}{2}\frac{1}{(r-2)!}Z^{(r-2)}\right)_y \right]
\eea}%
The indices $x,y$ show the point at which the expressions inside the parentheses are evaluated at. Also, the spacetime indices in the propagators $G_{xy}$ are omitted for simplicity, and (here and in what follows) the partial derivatives on the propagators are understood to be evaluated at the spacetime point $x$. Next we need to know the combination of the propagators and their derivatives that appear in the middle parts of each of the six terms in the above expression. The useful representation for these quantities can be found by moving to momentum space and then replacing $p^2$ by $-\Box$. In the first three lines all we need is $G^r$ which is given in Eq.\eqref{Grx}. In the fourth term we need to calculate three new quantities. Using \eqref{G} these can be written (dropping the spacetime index $x$ in the propagators and partial derivatives) as 
\be 
(\partial_\mu\partial_\nu G)^2G^{r-2} = \frac{8\delta^2(2\delta^2+3\delta+k) c^{r}}{|x|^{2r\delta+4}},
\ee
\be   
\partial_\mu\partial_\nu G\,\partial^\mu G\,\partial^\nu G\, G^{r-3} = \frac{8\delta^3(2\delta+1) c^{r}}{|x|^{2r\delta+4}},
\ee
\be 
(\partial G)^4 G^{r-4} = \frac{16\delta^4c^{r}}{|x|^{2r\delta+4}}.
\ee
In momentum space these are respectively given as
\be    \label{r-2}
\int_x \!e^{ip\cdot x} \; (\partial_\mu\partial_\nu G)^2G^{r-2} =  \frac{1}{2(4\pi)^{2r-1}}\,\delta^2(2\delta^2+3\delta+k)\;\frac{\Gamma(\delta)^{r}\Gamma(-(r-1)\delta)}{\Gamma(r\delta +2)}\,(p^2)^{(r-1)\delta},
\ee
\be   \label{r-3}
\int_x \!e^{ip\cdot x} \; \partial_\mu\partial_\nu G\,\partial^\mu G\,\partial^\nu G \, G^{r-3} =  \frac{1}{2(4\pi)^{2r-1}}\,\delta^3(2\delta +1)\;\frac{\Gamma(\delta)^{r}\Gamma(-(r-1)\delta)}{\Gamma(r\delta +2)}\,(p^2)^{(r-1)\delta},
\ee
\be   \label{r-4}
\int_x \!e^{ip\cdot x} \; (\partial G)^4 G^{r-4} = \frac{1}{(4\pi)^{2r-1}}\,\delta^4\;\frac{\Gamma(\delta)^{r}\Gamma(-(r-1)\delta)}{\Gamma(r\delta +2)}\,(p^2)^{(r-1)\delta}.
\ee
Finally the combination of propagators in the last two terms is already reported in equation \eqref{mel}. The divergent parts of these quantities can be evaluated after fixing the number of propagators $r$ which is done case by case in the examples considered throughout the text.

\section{Computational details for a sample triangle diagram}  \label{s:vvz.sample}

In this appendix we give the details of the computations for a sample triangle diagram in $\Box^2$ theory with $n=3$. The ones cubic in the potential have already been discussed in Sect.\ref{ss:tri}. We therefore concentrate here on an example involving a $Z$ coupling. Out of the four diagrams of this kind discussed in Sect.\ref{s:type2.exmpl} let us choose the one corresponding to the divergence \eqref{113z4uv} which is slightly more involved than the others. As discussed in the text the $Z$ function can appear in three different ways in a vertex. In the case of the example considered here these are
\be \label{z-verts}
\frac{1}{2}\frac{1}{2!}Z^{(2)}, \qquad 
\frac{1}{3!}Z^{(3)}\partial_\mu\phi = \frac{1}{3!}\partial_\mu Z^{(2)} , \qquad
\frac{1}{2}\frac{1}{4!}Z^{(4)}(\partial\phi)^2.
\ee    
Here we distinguish the three diagrams with the $Z$ vertex being any of the above. In the first case on the left there are two derivatives acting on the propagators emanating from the $Z$ vertex in all possible ways. This can happen in two ways: either both derivatives act on the propagators connecting $Z$ to $V^{(4)}$ or one of the derivatives acts on the propagator connecting $Z$ to $V^{(2)}$. The diagrams representing these two cases are shown below
\begin{figure}[H]
\begin{center}
\includegraphics[width=0.43\textwidth]{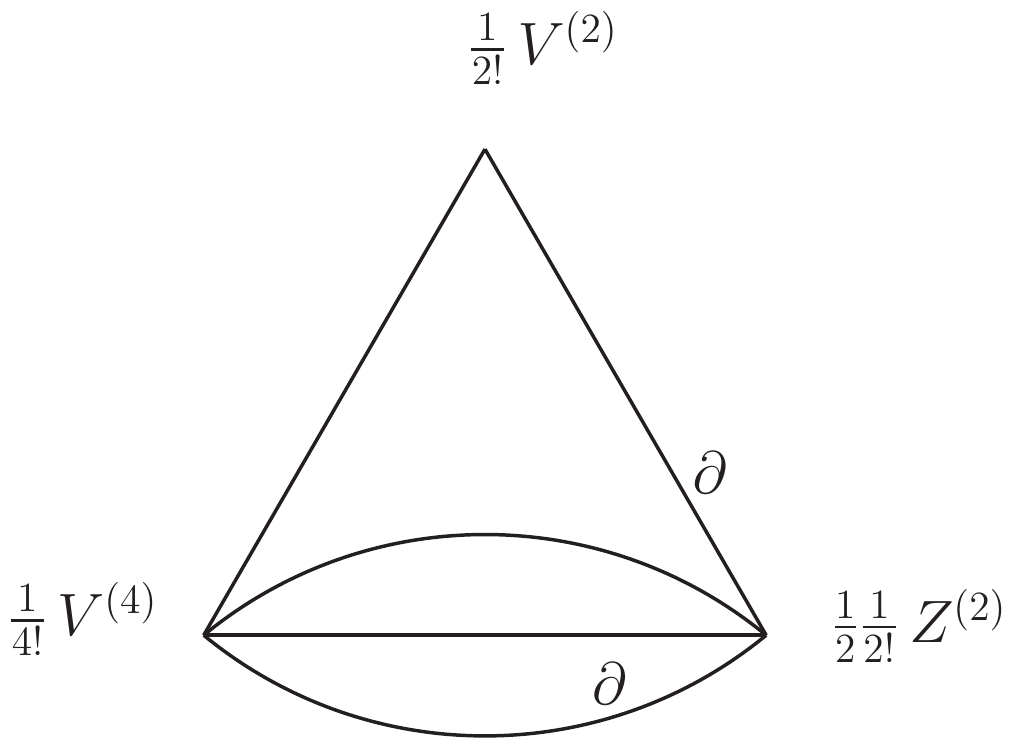} \label{sample21} \qquad\quad
\includegraphics[width=0.43\textwidth]{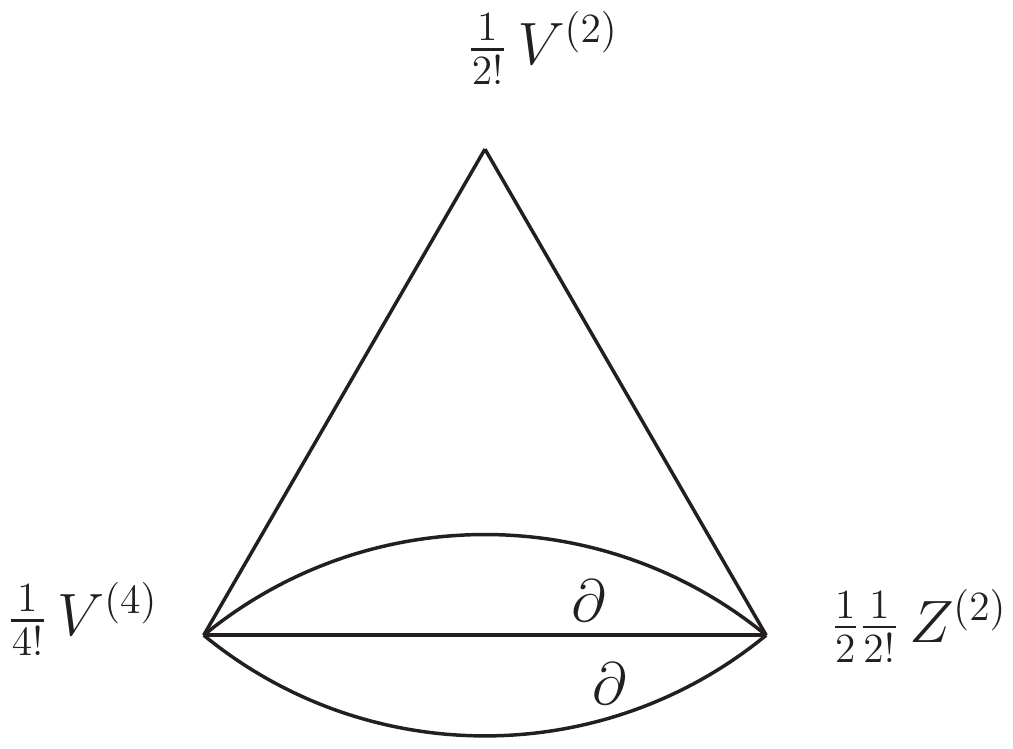} \label{sample22}
\end{center}
\caption{Contributions to the sample triangle diagram from the leftmost term in \eqref{z-verts}.}\label{samplel}
\end{figure}\noindent
We have made explicit the factors accompanying each function and the number of field derivatives on them. Both of these diagrams appear with a coefficient
\be \label{coeff2} 
-\frac{1}{2!}\frac{1}{4!}\left(\frac{1}{2}\frac{1}{2!}\right)\,\cdot\,2\times 4\times 2 \,\cdot\, 3! = -\frac{1}{2}.
\ee 
The minus sign comes from the three minus signs accompanying each function. The next three factors are those shown in the figure which, apart from the one half factor in $Z$, result from the Taylor expansion of these functions. The subsequent three factors show respectively the possible ways the lines emanating from $V^{(2)}$, $V^{(4)}$ and $Z^{(2)}$ are distributed among the other two vertices. Finally the $3!$ is the number of ways the three propagators from $Z^{(2)}$ and $V^{(4)}$ are connected to each other. Putting aside the coefficient \eqref{coeff2} the diagram on the left is given in coordinate space \nolinebreak as
\be 
\int_{x,y,z} \!Z^{(2)}_x\,V^{(4)}_{y}\,V^{(2)}_{z}\, G^2_{xy}\partial_\mu G_{xy}\, \partial^\mu G_{xz}\, G_{yz}, \qquad  G^2_{xy}\partial_\mu G_{xy} = {\textstyle{\frac{1}{3}}}\partial_\mu G^3_{xy} 
\ee
where the partial derivatives are taken at the space-time point $x$. To proceed further it is convenient to move to momentum space. Let us assume that $p_1$, $p_2$ and $p_3$ are the three incoming momenta into the vertices $Z^{(2)}$, $V^{(2)}$ and $V^{(4)}$ respectively.
Then using \eqref{Grp} this diagram reduces to the following expression in momentum space
\be 
\left(\!-\frac{i}{3}\right)i\frac{1}{(4\pi)^6}\frac{\Gamma(\delta)^3}{\Gamma(3\delta)}\Gamma(2-2\delta)\int_p \frac{(p+k_1)\!\cdot\!(p+k_2)}{[(p+k_1)^2]^{2-2\delta}[(p+k_2)^2]^2[(p+k_3)^2]^2},
\ee 
where $p$ is the loop momentum integrated over and we have defined $p_1=k_2-k_1$, $p_2=k_3-k_2$ and $p_3=k_1-k_3$. Expressing the numerator in the integrand in terms of complete squares this can be written as a sum of three terms
{\setlength\arraycolsep{0.5pt}
\bea  
\frac{1}{6}\frac{1}{(4\pi)^6}\frac{\Gamma(\delta)^3}{\Gamma(3\delta)}\Gamma(2-2\delta)\int_p &\bigg[&\frac{1}{[(p+k_1)^2]^{1-2\delta}[(p+k_2)^2]^2[(p+k_3)^2]^2} \nn\\
&+& \frac{1}{[(p+k_1)^2]^{2-2\delta}[(p+k_2)^2][(p+k_3)^2]^2} \nn\\
&-& \frac{p_1^2}{[(p+k_1)^2]^{2-2\delta}[(p+k_2)^2]^2[(p+k_3)^2]^2}\, \bigg].
\eea}%
Using Feynman parameterization \eqref{feynp} and the shorthand definition of integrals over Feynman parameters given in \eqref{shnfeyn}, this can be rewritten as
{\setlength\arraycolsep{0.5pt}
\bea  
\frac{1}{6}\frac{1}{(4\pi)^6}\frac{\Gamma(\delta)^3}{\Gamma(3\delta)}\Gamma(2-2\delta)\int_{P,\vec a} &\bigg[& a^{-2\delta}\,b\,c\,\frac{\Gamma(5-2\delta)}{\Gamma(1-2\delta)}\,\frac{1}{(P^2+\Delta)^{5-2\delta}} + a^{1-2\delta}\,c\,\frac{\Gamma(5-2\delta)}{\Gamma(2-2\delta)}\,\frac{1}{(P^2+\Delta)^{5-2\delta}} \nn\\
&-& a^{1-2\delta}\,b\,c\,\frac{\Gamma(6-2\delta)}{\Gamma(2-2\delta)}\,\frac{p_1^2}{(P^2+\Delta)^{6-2\delta}}\, \bigg] 
\eea}%
where $\vec a = (a,b,c)$ are the three Feynman parameters, $P = p +ak_1+bk_2+ck_3$ is the shifted momentum and
\be 
\Delta = ab\, p_1^2 + bc\, p_2^2 + ac\, p_3^2.
\ee
Performing the momentum integral reduces this expression to
\be   \label{expression}
\frac{1}{6}\frac{1}{(4\pi)^9}\frac{\Gamma(\delta)^3}{\Gamma(3\delta)}\int_{\vec a} \bigg[a^{-2\delta}\,b\,c\,\Gamma(3-3\delta)\,\frac{1-2\delta}{\Delta^{3-3\delta}} + a^{1-2\delta}\,c\,\Gamma(3-3\delta)\,\frac{1}{\Delta^{3-3\delta}} - a^{1-2\delta}\,b\,c\,\Gamma(4-3\delta)\,\frac{p_1^2}{\Delta^{4-3\delta}}\bigg].
\ee
We will now evaluate each term in turn using the Mellin-Barnes representation and the notation introduced in Appendix.\ref{s:not} for the contour integrals in the complex plane. The first term is  
\be   
\frac{1}{6}\frac{1}{(4\pi)^9}\frac{\Gamma(\delta)^3}{\Gamma(3\delta)}(1-2\delta)\int_{\vec a,y,z} a^{-2\delta}\,b\,c\,\Gamma(-y)\Gamma(-z)\Gamma(y+z+3-3\delta)\,(abp_1^2)^y(bcp_2^2)^{3\delta-3-y-z}(acp_3^2)^z,
\ee
which after integrating over the Feynman parameters $a,b,c$, using the general formula \eqref{intfeyn}, becomes
{\setlength\arraycolsep{0.5pt}
\bea  \label{int}
&& \frac{1}{6}\frac{1}{(4\pi)^9}\frac{\Gamma(\delta)^3}{\Gamma(3\delta)}\frac{(1-2\delta)}{\Gamma(4\delta-1)}\int_{y,z} \Gamma(-y)\Gamma(-z)\Gamma(y+z+3-3\delta)\nn\\
&& \phantom{\frac{1}{6}\frac{1}{(4\pi)^9}\frac{\Gamma(\delta)^3}{\Gamma(3\delta)}\frac{1-2\delta}{\Gamma(4\delta-1)}}\;\;\; \times\Gamma(y+z+1-2\delta)\Gamma(3\delta-1-z)\Gamma(3\delta-1-y) \left(\frac{p_1^2}{p_2^2}\right)^y \left(\frac{p_3^2}{p_2^2}\right)^z.
\eea}%
In this expression the relevant poles in $y,z$ that lead to a divergence in $\epsilon$ are $(y,z)=(0,0)$, $(y,z)=(1,0)$ and $(y,z)=(0,1)$. It might be instructive to have a closer look at this integral and analyze it carefully. In order not to digress from the main steps we move this analysis to the next appendix. Picking these poles and extracting the $\epsilon$ divergence gives
\be 
\frac{1}{(4\pi)^9}\,\frac{1}{36}\left(\frac{1}{\epsilon^2} + \frac{9-6\gamma}{4\epsilon}\right) + \frac{1}{(4\pi)^9}\,\frac{1}{24\epsilon}\left(\frac{p_1^2}{p_2^2} + \frac{p_3^2}{p_2^2}\right),
\ee
where the first term which is local comes from the first pole in $y,z$ while the second term which is non-local results from the last two $y,z$ poles. Let us now move to the second term in \eqref{expression}. In a similar way this term can be expressed in the Mellin-Barnes representation as
{\setlength\arraycolsep{0.5pt}
\bea  
&& \frac{1}{6}\frac{1}{(4\pi)^9}\frac{\Gamma(\delta)^3}{\Gamma(3\delta)}\frac{1}{\Gamma(4\delta-1)}\int_{y,z} \Gamma(-y)\Gamma(-z)\Gamma(y+z+3-3\delta)\nn\\
&& \phantom{\frac{1}{6}\frac{1}{(4\pi)^9}\frac{\Gamma(\delta)^3}{\Gamma(3\delta)}\frac{(1-2\delta)}{\Gamma(4\delta-1)}}\;\;\; \times\Gamma(y+z+2-2\delta)\Gamma(3\delta-2-z)\Gamma(3\delta-1-y) \left(\frac{p_1^2}{p_2^2}\right)^y \left(\frac{p_3^2}{p_2^2}\right)^z.
\eea}%
This time the only pole in $y,z$ that leads to an $\epsilon$ divergence is $(y,z)=(0,0)$ giving 
\be 
\frac{1}{(4\pi)^9}\,\frac{1}{36}\left(\frac{1}{\epsilon^2} + \frac{15-6\gamma}{4\epsilon}\right).
\ee
Finally the third term is evaluated in a similar way
{\setlength\arraycolsep{0.5pt}
\bea  
&-& \frac{1}{6}\frac{1}{(4\pi)^9}\frac{\Gamma(\delta)^3}{\Gamma(3\delta)}\frac{1}{\Gamma(4\delta\!-\!2)}\int_{y,z}\!\! \Gamma(-y)\Gamma(-z)\Gamma(y+z+4-3\delta) \\
&& \phantom{\frac{1}{6}\frac{1}{(4\pi)^9}\frac{\Gamma(\delta)^3}{\Gamma(3\delta)}\frac{(1-2\delta)}{\Gamma(4\delta-1)}} \times\Gamma(y+z+2-2\delta)\Gamma(3\delta-2-z)\Gamma(3\delta-2-y) \left(\frac{p_1^2}{p_2^2}\right)^{\!y+1}\!\! \left(\frac{p_3^2}{p_2^2}\right)^z, \nn
\eea}%
giving a divergence at the pole $(y,z)=(0,0)$
\be 
-\frac{1}{(4\pi)^9}\,\frac{1}{12\epsilon}\,\frac{p_1^2}{p_2^2}.
\ee
Summing up all three contributions we finally get the UV divergence of the diagram on the left
\be \label{sample21uv}
\frac{1}{(4\pi)^9}\,\frac{1}{18}\left(\frac{1}{\epsilon^2} + \frac{6-3\gamma}{2\epsilon}\right) + \frac{1}{(4\pi)^9}\,\frac{1}{24\epsilon}\left(\frac{p_3^2}{p_2^2} - \frac{p_1^2}{p_2^2}\right).
\ee
We will now follow similar steps to evaluate the diagram on the right. In momentum space this is 
{\setlength\arraycolsep{0.5pt}
\bea  
&& \frac{1}{(4\pi)^6}\frac{\delta^2\Gamma(\delta)^3}{\Gamma(3\delta+1)}\Gamma(1-2\delta)\int_p \frac{1}{[(p+k_1)^2]^{1-2\delta}[(p+k_2)^2]^2[(p+k_3)^2]^2} \nn\\
&=& \frac{1}{(4\pi)^6}\frac{\delta^2\Gamma(\delta)^3}{\Gamma(3\delta+1)}\Gamma(5-2\delta) \int_{P,\vec a} a^{-2\delta}\,b\,c\,\frac{1}{(P^2+\Delta)^{5-2\delta}} \nn\\
&=& \frac{1}{(4\pi)^9}\frac{\delta^2\Gamma(\delta)^3}{\Gamma(3\delta+1)}\Gamma(3-3\delta) \int_{\vec a} a^{-2\delta}\,b\,c\,\frac{1}{\Delta^{3-3\delta}} \nn\\
&=& \frac{1}{(4\pi)^9}\frac{\delta^2\Gamma(\delta)^3}{\Gamma(3\delta+1)}\int_{\vec a,y,z} \!\! a^{-2\delta}\,b\,c\,\Gamma(-y)\Gamma(-z)\Gamma(y+z+3-3\delta)\,(abp_1^2)^y(bcp_2^2)^{3\delta-3-y-z}(acp_3^2)^z \nn\\
&=& \frac{1}{(4\pi)^9}\frac{\delta^2\Gamma(\delta)^3}{\Gamma(3\delta+1)} \frac{1}{\Gamma(4\delta-1)}\int_{y,z} \Gamma(-y)\Gamma(-z)\Gamma(y+z+3-3\delta)\nn\\
&& \phantom{\frac{1}{6}\frac{1}{(4\pi)^9}\frac{\Gamma(\delta)^3}{\Gamma(3\delta)}\frac{1-2\delta}{\Gamma(4\delta-1)}}\;\;\; \times\Gamma(y+z+1-2\delta)\Gamma(3\delta-1-z)\Gamma(3\delta-1-y) \left(\frac{p_1^2}{p_2^2}\right)^y \!\!\left(\frac{p_3^2}{p_2^2}\right)^z,
\eea}%
where in the second line we have used Feynman parameterization and in the third line we have integrated over the momenta. This is then expressed in Mellin-Barnes representation in the fourth line and finally in the fifth line integrated over Feynman parameters. Just like the previous case the relevant poles in $y,z$ are $(y,z)=(0,0)$, $(y,z)=(1,0)$ and $(y,z)=(0,1)$. The UV divergence is
\be \label{sample22uv}
-\frac{1}{(4\pi)^9}\,\frac{1}{18}\left(\frac{1}{\epsilon^2} + \frac{11-6\gamma}{4\epsilon}\right) - \frac{1}{(4\pi)^9}\,\frac{1}{12\epsilon}\left(\frac{p_1^2}{p_2^2} + \frac{p_3^2}{p_2^2}\right).
\ee
Summing up the two contributions \eqref{sample21uv} and \eqref{sample22uv} from the diagrams in Fig.\eqref{samplel},  multiplying by the coefficient \eqref{coeff2}, and moving to coordinate space we get
\be \label{sample2uv}
-\frac{1}{(4\pi)^9}\,\frac{1}{144\epsilon}\,Z^{(2)}V^{(2)}V^{(4)} + \frac{1}{(4\pi)^9}\,\frac{1}{16\epsilon}\,\Box Z^{(2)}\frac{1}{\Box}V^{(2)}\,V^{(4)} + \frac{1}{(4\pi)^9}\,\frac{1}{48\epsilon}\,Z^{(2)}\frac{1}{\Box}V^{(2)}\Box V^{(4)}. 
\ee
We are now left with two diagrams in one of which one of the derivatives at the $Z$ vertex acts on the propagators, while in the other both derivatives are external.
\begin{figure}[H] 
\begin{center}
\includegraphics[width=0.43\textwidth]{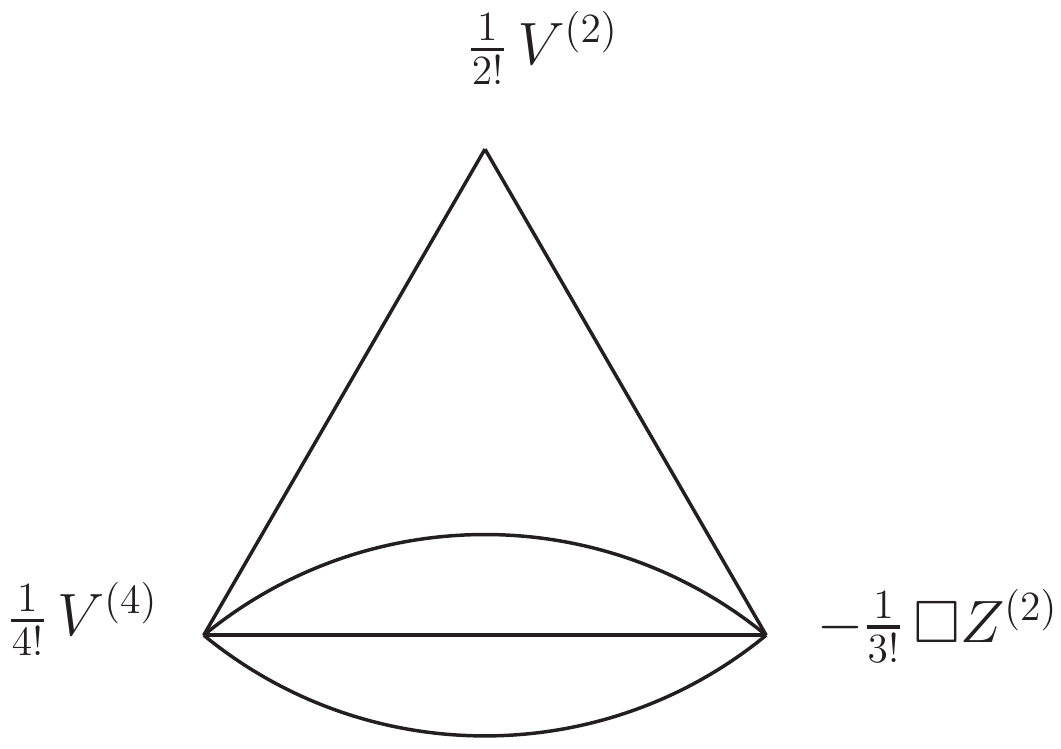} \label{sample3} \qquad\quad
\includegraphics[width=0.43\textwidth]{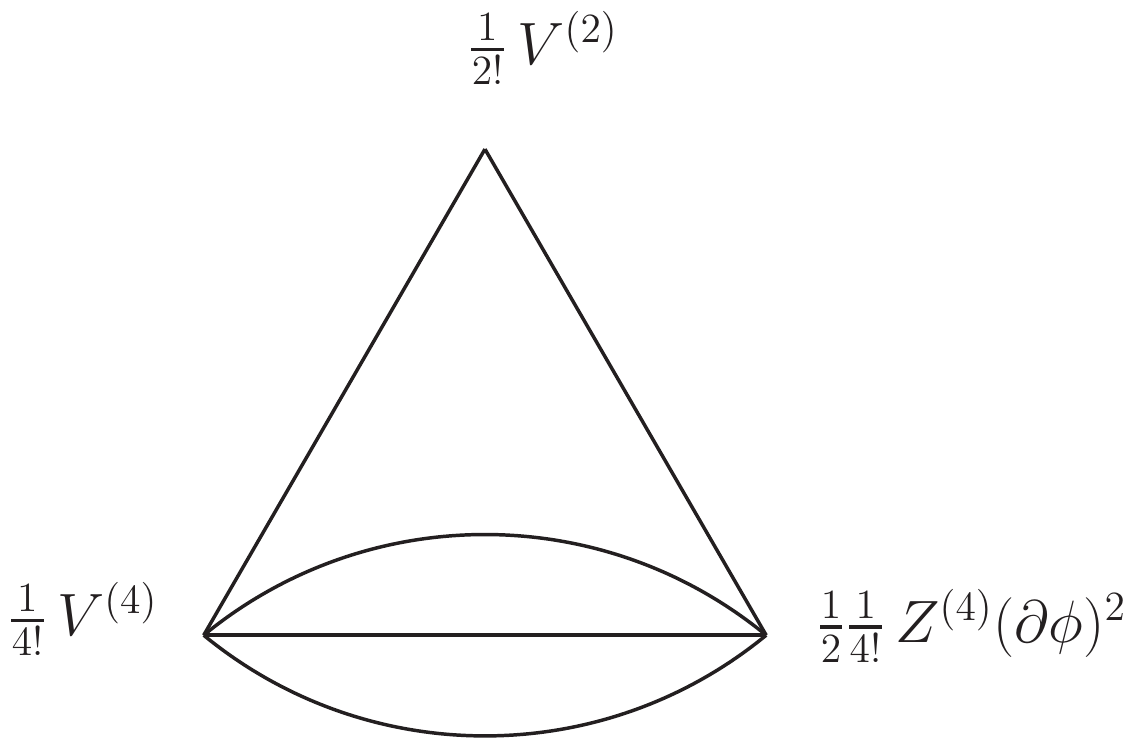} \label{sample4}
\end{center}
\caption{Left diagram shows the contribution from the middle term in \eqref{z-verts} to the sample triangle diagram, while the diagram on the right shows the contribution from the rightmost term in \eqref{z-verts}.}\label{samplecr}
\end{figure}\noindent
In the first case the $Z$ vertex is the middle term in \eqref{z-verts}. 
This diagram has an overall factor 
\be \label{coeff3}
-\frac{1}{3!}\frac{1}{2!}\frac{1}{4!}\cdot 4\times 2\times 3! = -\frac{1}{3!}
\ee
which is multiplied by the coordinate space expression 
\be 
\int_{x,y,z} \![\partial_\mu Z^{(2)}]_x\;\partial_x^\mu G^3_{xy}\, V^{(4)}_{y}\, G_{yz}\, V^{(2)}_{z}\, G_{xz} + \int_{x,y,z} \![\partial_\mu Z^{(2)}]_x\, G^3_{xy}\, V^{(4)}_{y}\,G_{yz}\, V^{(2)}_{z}\,\partial_x^\mu G_{xz}.
\ee
Using integration by parts this can be rewritten as
\be \label{3uv}
-\int_{x,y,z} \![\Box Z^{(2)}]_x\,V^{(4)}_{y}\,V^{(2)}_{z}\, G^3_{xy}\, G_{xz}\, G_{yz}.
\ee
Next, we consider the diagram in which the $Z$ vertex is the last term in \eqref{z-verts}. This term has an overall factor
\be  \label{coeff4}
-\frac{1}{2}\frac{1}{4!}\frac{1}{2!}\frac{1}{4!}\cdot 2\times 4^2\times 3! = -\frac{1}{12},
\ee
to be multiplied by 
\be \label{4uv}
\int_{x,y,z} \![Z^{(4)}(\partial\phi)^2]_x\,V^{(4)}_{y}\,V^{(2)}_{z}\, G^3_{xy}\, G_{xz}\, G_{yz}.
\ee
In \eqref{3uv} and \eqref{4uv} the combination of propagators is the same and there are no derivatives on the propagators. Therefore we do not need a separate computation for each of these expressions. All we need to compute is a triangle diagram of which two of the edges include 1 propagator and the third edge includes 3 propagators and with no derivatives acting on the internal lines. Using \eqref{Grp} this is computed as follows

{\setlength\arraycolsep{0.5pt}
\bea  
&& \frac{1}{(4\pi)^6}\frac{\Gamma(\delta)^3}{\Gamma(3\delta)}\Gamma(2-2\delta)\int_p \frac{1}{[(p+k_1)^2]^{2-2\delta}[(p+k_2)^2]^2[(p+k_3)^2]^2} \nn\\
&=& \frac{1}{(4\pi)^6}\frac{\Gamma(\delta)^3}{\Gamma(3\delta)}\Gamma(6-2\delta) \int_{p,\vec a} a^{1-2\delta}\,b\,c\,\frac{1}{(P^2+\Delta)^{6-2\delta}} \nn\\
&=& \frac{1}{(4\pi)^9}\frac{\Gamma(\delta)^3}{\Gamma(3\delta)}\Gamma(4-3\delta) \int_{p,\vec a} a^{1-2\delta}\,b\,c\,\frac{1}{\Delta^{4-3\delta}} \nn\\
&=& \frac{1}{(4\pi)^9}\frac{\Gamma(\delta)^3}{\Gamma(3\delta)}\int_{\vec a,y,z} \!\! a^{1-2\delta}\,b\,c\,\Gamma(-y)\Gamma(-z)\Gamma(y+z+4-3\delta)\,(abp_1^2)^y(bcp_2^2)^{3\delta-4-y-z}(acp_3^2)^z \nn\\
&=& \frac{1}{(4\pi)^9}\frac{\Gamma(\delta)^3}{\Gamma(3\delta)} \frac{1}{\Gamma(4\delta-2)}\int_{y,z} \Gamma(-y)\Gamma(-z)\Gamma(y+z+4-3\delta) \\
&& \phantom{\frac{1}{6}\frac{1}{(4\pi)^9}\frac{\Gamma(\delta)^3}{\Gamma(3\delta)}\frac{1-2\delta}{\Gamma(4\delta-1)}} \times\Gamma(y+z+2-2\delta)\Gamma(3\delta-2-z)\Gamma(3\delta-2-y) \left(\frac{p_1^2}{p_2^2}\right)^y \!\!\left(\frac{p_3^2}{p_2^2}\right)^z\frac{1}{p_2^2}.\nn
\eea}%
The only pole in $y,z$ that gives rise to an $\epsilon$ divergence is $(y,z)=(0,0)$. This divergence is 
\be 
\frac{1}{(4\pi)^9}\,\frac{1}{2\epsilon}\,\frac{1}{p_2^2}.
\ee
For the left diagram in Fig.\ref{samplecr} this has to be multiplied by \eqref{coeff3} and the functions in \eqref{3uv}. In coordinate space this gives
\be \label{sample3uv}
-\frac{1}{(4\pi)^9}\,\frac{1}{12\epsilon}\,\Box Z^{(2)}\frac{1}{\Box}V^{(2)}\,V^{(4)}.
\ee 
For the right diagram multiplying by the factor \eqref{coeff4} and the functions in \eqref{4uv} and moving to coordinate space we get
\be  \label{sample4uv}
\frac{1}{(4\pi)^9}\,\frac{1}{24\epsilon}\, Z^{(4)}(\partial\phi)^2\,\frac{1}{\Box}V^{(2)}\,V^{(4)}.
\ee
Summing up the three contributions \eqref{sample2uv}, \eqref{sample3uv} and \eqref{sample4uv} we obtain the final result in \eqref{113z4uv}.

\section{A closer look at a sample Mellin-Barnes integral}

Throughout the paper, we repeatedly encounter Mellin-Barnes integrals involving two contour integrations and extract their poles in $\epsilon$.  
In this section we analyze carefully, as a nontrivial example, the contour integral of Eq.\eqref{int}. Similar arguments can be used for other contour integrals discussed in this work. Dropping the overall coefficient in Eq.\eqref{int}, which we are not interested in, we are left with the integral
\be 
\int_{y,z} \Gamma(-y)\Gamma(-z)\Gamma(y+z+3-3\delta) \Gamma(y+z+1-2\delta)\Gamma(3\delta-1-z)\Gamma(3\delta-1-y) \left(\frac{p_1^2}{p_2^2}\right)^y \left(\frac{p_3^2}{p_2^2}\right)^z.
\ee
The value of $p_2$ is taken to be larger than $p_1,p_3$, so that the contour can be closed at infinity in the right half plane. The integral therefore picks $y,z$ poles lying in this region. These include $y,z=0,1,\cdots$ from the first two gamma functions on the left, as well as $y,z=3\delta -1,3\delta,3\delta+1,\cdots$ which come from the last two gamma functions. Using the residue theorem to pick the contribution from all these poles we can write the integral as
{\setlength\arraycolsep{1pt}
\bea 
&& \sum_{m,n=0}\! \frac{(-1)^{m+n}}{m!n!} \Gamma(m\!+\!n\!+\!3\!-\!3\delta)\Gamma(3\delta\!-\!1\!-\!m)\Gamma(3\delta\!-\!1\!-\!n)\Gamma(m\!+\!n\!+\!1\!-\!2\delta) \left(\frac{p_2^2}{p_1^2}\right)^m\left(\frac{p_3^2}{p_1^2}\right)^n \nn\\
&+& \sum_{m,n=0}\! \frac{(-1)^{m+n}}{m!n!} \Gamma(-m\!-\!3\delta\!+\!1)\Gamma(-n\!-\!3\delta\!+\!1)\Gamma(m\!+\!n\!+\!3\delta+1)\Gamma(m\!+\!n\!+\!4\delta\!-\!1) \left(\frac{p_2^2}{p_1^2}\right)^{\!m+2}\!\!\left(\frac{p_3^2}{p_1^2}\right)^{n+2} \nn\\
&+& \sum_{m,n=0}\! \frac{(-1)^{m+n}}{m!n!} \Gamma(-n\!-\!3\delta\!+\!1)\Gamma(m\!+\!n\!+\!2)\Gamma(3\delta\!-\!1\!-\!m)\Gamma(m\!+\!n\!+\!\delta)\left[\left(\frac{p_2^2}{p_1^2}\right)^{\!\!m}\!\!\!\left(\frac{p_3^2}{p_1^2}\right)^{\!\!n+2}\!\!\!\!\!\!+\left(\frac{p_2^2}{p_1^2}\right)^{\!n+2}\!\!\left(\frac{p_3^2}{p_1^2}\right)^{\!m}\right] \nn\\
\eea}%
In the first line we separate the summation in $m,n$ into six parts given by the pairs below:
\be 
\ba{|c|c|cc|c|cc|cc|c|}
\hline
m & 0 & 1 & 0 & 1 & 0 & \geq 2 & 1 & \geq 2 & \geq 2 \\ \hline
n & 0 & 0 & 1 & 1 & \geq 2 & 0 & \geq 2 & 1 & \geq 2 \\ \hline 
\ea
\ee
and in the third line we separate the sum into three terms $m=0, 1, \geq 2$. The result is
{\setlength\arraycolsep{1pt}
\bea 
&& \Gamma(3-3\delta)\Gamma(3\delta-1)^2\Gamma(1-2\delta) \nn\\
&-& \Gamma(4-3\delta)\Gamma(3\delta-1)\Gamma(3\delta-2)\Gamma(2-2\delta) \left[\left(\frac{p_2^2}{p_1^2}\right)+\left(\frac{p_3^2}{p_1^2}\right)\right] \nn\\
&+& \Gamma(5-3\delta)\Gamma(3\delta-2)\Gamma(3\delta-2)\Gamma(3-2\delta) \left(\frac{p_2^2}{p_1^2}\right)\left(\frac{p_3^2}{p_1^2}\right) \nn\\
&+& \sum_{n=0} \frac{(-1)^{n}}{(n+2)!} \Gamma(n+5-3\delta)\Gamma(3\delta-1)\Gamma(3\delta-3-n)\Gamma(n+3-2\delta) \left[\left(\frac{p_2^2}{p_1^2}\right)^{n+2}+\left(\frac{p_3^2}{p_1^2}\right)^{n+2}\right] \nn\\
&-& \sum_{n=0} \frac{(-1)^{n}}{(n+2)!} \Gamma(n+6-3\delta)\Gamma(3\delta-2)\Gamma(3\delta-3-n)\Gamma(n+4-2\delta) \left[\left(\frac{p_2^2}{p_1^2}\right)^{n+2}\left(\frac{p_3^2}{p_1^2}\right) +\left(\frac{p_2^2}{p_1^2}\right)\left(\frac{p_3^2}{p_1^2}\right)^{n+2}\right] \nn\\
\eea}%
{\setlength\arraycolsep{1pt}
\bea
&+& \sum_{m,n=0} \frac{(-1)^{m+n}}{(m+2)!(n+2)!} \Gamma(m+n+7-3\delta)\Gamma(3\delta-3-m)\Gamma(3\delta-3-n) \nn\\
&& \times \; \Gamma(m+n+5-2\delta) \left(\frac{p_2^2}{p_1^2}\right)^{m+2}\left(\frac{p_3^2}{p_1^2}\right)^{n+2} \nn\\[3mm]
&+& \sum_{m,n=0} \frac{(-1)^{m+n}}{m!n!} \Gamma(-m-3\delta+1)\Gamma(-n-3\delta+1)\Gamma(m+n+3\delta+1) \nn\\
&& \times \; \Gamma(m+n+4\delta-1) \left(\frac{p_2^2}{p_1^2}\right)^{m+2}\left(\frac{p_3^2}{p_1^2}\right)^{n+2} \nn\\[3mm]
&+& \sum_{n=0} \frac{(-1)^{n}}{n!} \Gamma(-n-3\delta+1)\Gamma(n+2)\Gamma(3\delta-1)\Gamma(n+\delta)\left[\left(\frac{p_3^2}{p_1^2}\right)^{\!n+2}\!\!\!\!+\left(\frac{p_2^2}{p_1^2}\right)^{\!n+2}\right] \nn\\[3mm]
&-& \sum_{n=0} \frac{(-1)^{n}}{n!} \Gamma(-n-3\delta+1)\Gamma(n+3)\Gamma(3\delta-2)\Gamma(n+1+\delta)\left[\left(\frac{p_2^2}{p_1^2}\right)\!\!\left(\frac{p_3^2}{p_1^2}\right)^{\!n+2}\!\!\!\!+\left(\frac{p_2^2}{p_1^2}\right)^{\!n+2}\!\!\left(\frac{p_3^2}{p_1^2}\right)\right] \nn\\[3mm]
&+& \sum_{m,n=0} \frac{(-1)^{m+n}}{(m+2)!n!} \Gamma(-n-3\delta+1)\Gamma(m+n+4)\Gamma(3\delta-3-m) \nn\\ 
&& \times \; \Gamma(m+n+2+\delta)\left[\left(\frac{p_2^2}{p_1^2}\right)^{\!m+2}\!\!\left(\frac{p_3^2}{p_1^2}\right)^{\!n+2}\!\!\!\!+\left(\frac{p_2^2}{p_1^2}\right)^{\!n+2}\!\!\left(\frac{p_3^2}{p_1^2}\right)^{\!m+2}\right].
\eea}%
The first and second lines are divergent while the third is finite. The remaining terms are also divergent but these divergences cancel among each other. In fact the fourth and eighth terms sum up to a finite term, the sum of the fifth and the ninth terms is finite, and the divergences of the sixth, seventh and tenth terms cancel out.

\section{Some notation and useful formulas} \label{s:not}

Here we collect some shorthand notation that we have used throughout the text along with some useful formulas. We have abbreviated the notation for integrals over spacetime coordinates, momentum space and integrals over contours in the complex plane. These are respectively given \nolinebreak as
\be  \label{shn}
\int_x = \int d^dx, \qquad
\int_p = \int \frac{d^dp}{(2\pi)^d}, \qquad
\int_z = \frac{1}{2\pi i}\int dz.
\ee
Also, we have written the integrals over Feynman parameters $\vec a = (a_1,a_2,\cdots)$ in a compact way as
\be  \label{shnfeyn}
\int_{\vec a} = \int_0^1 \prod da_i\; \delta\left(\sum a_i-1\right).
\ee
We have used the Feynman parameterization formula
\be \label{feynp}
\frac{1}{\prod A_i^{\rho_i}} = \int_{\vec a} \;\frac{\prod a^{\rho_i-1}_i}{[\sum a_iA_i]^{\sum \rho_i}}\;\frac{\Gamma(\sum \rho_i)}{\prod \Gamma(\rho_i)},
\ee
and the following formula regarding integration over Feynamn parameters
\be \label{intfeyn}
\int_{\vec a} \;\prod a^{\alpha_i-1}_i = \;\frac{\prod \Gamma(\alpha_i)}{\Gamma(\sum \alpha_i)}.
\ee

%
%
%

\end{document}